\documentclass[twocolumn]{aastex6}
\usepackage{apjfonts,color,amsmath,textcomp,url,graphicx,subfigure}

\newcommand{\ms}{\hbox{m~s$^{-1}$}}
\newcommand{\msyr}{\hbox{m~s$^{-1}$~yr$^{-1}$}}
\newcommand{\kms}{\hbox{km\,s$^{-1}$}}
\newcommand{\rad}{$\mathrm{rad}$}
\newcommand{\MEarth}{$M_{\Earth}$}
\newcommand{\MJup}{$M_{\mathrm{Jup}}$}

\newcommand{\Msol}{$M_{\odot}$}
\newcommand{\masyr}{$\mathrm{mas}\,\mathrm{yr}^{-1}$}

\DeclareMathOperator\erf{erf}
\DeclareMathOperator\ierf{erf^{-1}}

\shorttitle{A High-Precision NIR Survey for RV variable low-mass stars}
\shortauthors{Gagn\'e et al.}

\begin{document}

\title{A HIGH-PRECISION NEAR-INFRARED SURVEY FOR RADIAL VELOCITY VARIABLE LOW-MASS STARS USING CSHELL AND A METHANE GAS CELL}

\author{Jonathan Gagn\'e\altaffilmark{1,2}, Peter Plavchan\altaffilmark{3}, Peter Gao\altaffilmark{4}, Guillem Anglada-Escude\altaffilmark{5,6}, Elise Furlan\altaffilmark{7}, Cassy Davison\altaffilmark{8}, Angelle Tanner\altaffilmark{9}, Todd J. Henry\altaffilmark{8}, Adric R. Riedel\altaffilmark{10}, Carolyn Brinkworth\altaffilmark{7,11}, David Latham\altaffilmark{14}, Michael Bottom\altaffilmark{15}, Russel White\altaffilmark{8}, Sean Mills\altaffilmark{12}, Chas Beichman\altaffilmark{13}, John A. Johnson\altaffilmark{14}, David R. Ciardi\altaffilmark{7}, Kent Wallace\altaffilmark{16}, Bertrand Mennesson\altaffilmark{16}, Kaspar von Braun\altaffilmark{17}, Gautam Vasisht\altaffilmark{16}, Lisa Prato\altaffilmark{17}, Stephen R. Kane\altaffilmark{18}, Eric E. Mamajek\altaffilmark{19}, Bernie Walp\altaffilmark{20}, Timothy J. Crawford\altaffilmark{16}, Rapha\"{e}l Rougeot\altaffilmark{21}, Claire S. Geneser\altaffilmark{3}, Joseph Catanzarite\altaffilmark{22}}

\affil{\altaffilmark{1} Carnegie Institution of Washington DTM, 5241 Broad Branch Road NW, Washington, DC~20015, USA; jgagne@carnegiescience.edu\\
\altaffilmark{2} NASA Sagan Fellow\\
\altaffilmark{3} Department of Physics, Missouri State University, 901 S National Ave, Springfield, MO~65897, USA; peterplavchan@missouristate.edu\\
\altaffilmark{4} Division of Geological and Planetary Sciences, California Institute of Technology, Pasadena, CA 91125, USA\\
\altaffilmark{5} School of Physics and Astronomy, Queen Mary University of London, 327 Mile End Rd, E1 4NS, London, UK\\
\altaffilmark{6} Centre for Astrophysics Research, University of Hertfordshire, College Lane, AL10 9AB, Hatfield, UK\\
\altaffilmark{7} NASA Exoplanet Science Institute, California Institute of Technology, 770 S. Wilson Ave., Pasadena, CA 91125, USA\\
\altaffilmark{8} Department of Physics and Astronomy, Georgia State University, Atlanta, GA 30303, USA\\
\altaffilmark{9} Mississippi State University, Department of Physics \& Astronomy, Hilbun Hall, Starkville, MS 39762, USA\\
\altaffilmark{10} Division of Physics, Mathematics and Astronomy, California Institute of Technology, Pasadena, CA 91125, USA\\
\altaffilmark{11} National Center for Atmospheric Research, P.O. Box 3000, Boulder, CO 80307, USA\\
\altaffilmark{12} Department of Astronomy and Astrophysics, University of Chicago, 5640 S. Ellis Ave, Chicago, IL 60637, USA\\
\altaffilmark{13} NASA Exoplanet Science Institute, California Institute of Technology, Pasadena, CA 91125, USA\\
\altaffilmark{14} Institute for Theory and Computation, Harvard-Smithsonian Center for Astrophysics, 60 Garden Street, Cambridge, Massachusetts 02138 USA\\
\altaffilmark{15} Department of Astronomy, California Institute of Technology, Pasadena, CA 91125, USA\\
\altaffilmark{16} Jet Propulsion Laboratory, California Institute of Technology, 4800 Oak Grove Drive, Pasadena, CA 91125, USA\\
\altaffilmark{17} Lowell Observatory, West Mars Hill Road, Flagstaff, AZ 86001, USA\\
\altaffilmark{18} Department of Physics \& Astronomy, San Francisco State University, 1600 Holloway Avenue, San Francisco, CA 94132, USA\\
\altaffilmark{19}  Department of Physics \& Astronomy, University of Rochester, Rochester, NY 14627, USA\\
\altaffilmark{20} Stratospheric Observatory for Infrared Astronomy, NASA Dryden Flight Research Center, Mail Stop DAOF, S233, P.O. Box 273 Edwards, CA 93523, USA\\
\altaffilmark{21} ESA, European Space Research and Technology Centre\\
\altaffilmark{22} NASA Ames Research Center, MS 245-3, P.O. Box 1, Moffett Field, CA 94035-0001, USA}

\begin{abstract}

We present the results of a precise near-infrared (NIR) radial velocity (RV) survey of 32 low-mass stars with spectral types K2--M4 using CSHELL at the\deleted{3\,m}NASA IRTF in the $K$-band with an isotopologue methane gas cell to achieve wavelength calibration and a novel iterative RV extraction method. We surveyed 14 members of young ($\approx$\,25--150\,Myr) moving groups, the young field star \replaced{$\epsilon$~Eri}{$\varepsilon$~Eridani} as well as 18 nearby ($<$\,25\,pc) low-mass stars and achieved typical single-measurement\deleted{ RV}precisions of 8--15\,\ms\ \replaced{within a single night and}{with a long-term stability of} 15--50\,\ms\ over longer baselines. \added{We obtain the best NIR RV constraints to date on 27 targets in our sample, 19 of which were never followed by high-precision RV surveys.\deleted{ We find no significant statistical difference in RV variability between the nearby and young low-mass stars in our sample.}Our results indicate that very active stars can display long-term RV variations as low as $\sim$\,25--50\,\ms\ at $\approx$\,2.3125\,$\mu$m, thus constraining the effect of jitter at these wavelengths.}\deleted{ These results represent a significant improvement over previous NIR surveys using this nearly $\sim$\,25\,yr-old detector, which is limited by a significant number of bad pixels and its low spectral grasp of $\approx$\,5.6\,nm at 2.3125\,$\mu$m. Additionally, non-detections allowed the determination of upper limits on the projected masses of close-in companions to several nearby, young and/or active stars.}We provide the first multi-wavelength confirmation of \replaced{GJ~876~b}{GJ~876~bc} and independently retrieve orbital parameters consistent with previous studies. We recovered RV variability for HD~160934~AB and GJ~725~AB that are consistent with \replaced{the binary orbits reported in the literature. Nine}{their known binary orbits, and nine} other targets are candidate RV variables \replaced{at}{with} a statistical significance of 3--5$\sigma$.\deleted{ LO~Peg is recovered as a statistically significant RV variable ($\sim$\,6$\sigma$) although the rotational broadening due to its fast rotation alone could explain this.}\replaced{These results demonstrate the potential of our data extraction pipeline and the methane gas cell, and suggest that using the same method on}{Our method combined with} the new iSHELL spectrograph\deleted{ at IRTF} will yield long-term\deleted{precision }RV \replaced{measurements}{precisions} of $\lesssim$\,5\,\ms\ in the NIR, which will allow the detection of Super-Earths near the habitable zone of mid-M dwarfs.

\end{abstract}

\clearpage
\section{INTRODUCTION}

The method of Doppler radial velocity (RV) variations has proven itself fruitful in the last decades both for the identification of new exoplanets (e.g., \citealp{1995Natur.378..355M,1998ApJ...505L.147M,1998AA...338L..67D,2001ApJ...556..296M,2002DPS....34.4202C,2003AJ....126.3099E,2004ApJ...617..580B,2005ApJ...634..625R,2010ApJ...719..890R,2011ApJ...727..117M,2012Natur.491..207D,2014ApJ...781...28M,2014MNRAS.441.1545T}) and the confirmation of exoplanets detected by the method of transit (e.g. Kepler--78b; \citealp{2013ApJ...774...54S,2013Natur.503..377P,2013PASP..125..989A}). Recent developments have shown that cool ($\lesssim$\,3800\,K) stellar hosts in the M spectral class represent valuable targets for the identification of new Earth-mass planetary companions in the habitable zone with the RV method due to their smaller mass and significantly larger population \citep{2006AJ....132.2360H}. 

However, it becomes gradually harder to obtain sufficient signal-to-noise (S/N) ratios at optical wavelengths \replaced{for}{at} decreasing effective temperatures \citep{2010ApJ...710..432R,2013PASP..125..240B}. To worsen the case, late-type M dwarfs are on average more active and display more stellar spots (e.g., \citealp{2009ApJ...699..649S,2010MNRAS.407.2269M,2012ApJ...758...56S,2014ApJ...788...81M,2015AJ....149..158S}), which can induce RV signals very similar to those of planetary companions (e.g., \citealp{2001AA...379..279Q,2005ESASP.560..871P}).  \cite{2014Sci...345..440R} demonstrated the importance of a careful consideration of stellar activity in exoplanet searches by demonstrating that the purported GJ~581~d habitable-zone exoplanet \citep{2007AA...469L..43U} was most likely a false-positive signal caused by stellar spots. Overcoming this limitation is especially important in the search for very low-mass companions that induce RV variations of low amplitudes (e.g. a few \ms) comparable to stellar activity jitter.

The study of RV variations in the regime of near-infrared (NIR) wavelengths addresses both these issues. First, a larger fraction of the flux of cooler ($\lesssim$\,3850\,K), later-type ($\gtrsim$\,M0; \citealt{2013ApJS..208....9P}) host stars is emitted at these wavelengths, although care must be used in choosing the observed NIR wavelength range for M0--M4 dwarfs as their lack of spectral features can counterbalance the brightness advantage \citep{2010ApJ...710..432R}.

Secondly and perhaps more importantly, the RV signal induced by stellar spots \replaced{has}{is expected to have} an $\approx \lambda^{-1}$ dependence for modest spot contrast temperatures, where $\lambda$ is the wavelength \citep{2010ApJ...710..432R,2012PASP..124..586A,2013Natur.503..381H,2013Natur.503..377P,2015arXiv150301770P,2015ApJ...798...63M}, which means that the effect of stellar spots on NIR RV measurements is less important than it is at visible wavelengths by a factor $\sim$\,4. \added{However, there have also been predictions that the Zeeman effect resulting from the strong magnetic activity in young stars could cause jitter to increase as a function of wavelength \citep{2013AA...552A.103R}. }Furthermore, the RV signal induced on a stellar host by a substellar companion via the Doppler effect is wavelength-independent, hence a multi-wavelengths RV follow-up opens the possibility of rejecting exoplanet candidates that are caused by other astrophysical phenomena that would cause RV variations of different amplitudes in the optical and NIR regimes.

RV surveys in the NIR are still trailing behind \deleted{those of }their optical counterparts in terms of long-term RV stability, with best reported NIR results at $\approx$\,5\,\ms\ \citep{2010ApJ...713..410B} using 8\,m-class telescopes, or $\approx$\,45--60\,\ms\ \citep{2010ApJ...723..684B,2011ApJ...735...78C,2012ApJS..203...10T,2012ApJ...749...16B} using smaller facilities, versus 0.8--15\,\ms\ in the optical \citep{1994Ap&SS.212..281C,1994Msngr..76...51K,2006ApJ...649..436E,2008PhST..130a4010M,2010Sci...330..653H,2012Natur.491..207D}. This is mainly due to \added{the presence of stronger telluric absorption features in the NIR \citep{2010AA...524A..11S,2010ApJ...723..684B,2011PASP..123.1302B,2015arXiv150301770P}}, and technical challenges \replaced{and to the fact}{, given} that NIR instrumentation and observing methods have only been developed relatively recently. As an example, iodine gas cells, which have been used extensively as wavelength calibrators in optical RV surveys, do not offer a sufficient density of absorption lines in the NIR domain. For this reason, most existing NIR RV studies have used telluric lines to achieve wavelength calibration \citep{2010ApJ...723..684B,2011ApJ...735...78C,2012ApJ...749...16B,2012ApJS..203...10T,2015AJ....149..106D}, with the exception of \cite{2010ApJ...713..410B} who used an ammonia gas cell with CRIRES at the VLT to obtain unprecedented long-term precisions of $\approx$\,5--10\,\ms\ in the NIR on targets with $K_S$-band magnitudes between 4.4 and 8.0.

Our team has recently developed a methane isotopologue gas cell that offers a high absorption line density in the NIR regime to achieve RV measurements of the order of a few \ms\ with the limited spectral grasp of CSHELL at the NASA InfraRed Telescope Facility (IRTF; \citealp{2012PASP..124..586A,2013SPIE.8864E..1JP}), as well as an iterative algorithm that allows for the simultaneous solving of the wavelength solution, the construction of an empirical stellar spectrum and the measurement of stellar RVs (P.~Gao et al., submitted to PASP). In this paper, we present the results of a NIR RV survey of 32 late-type, nearby stars using CSHELL at the IRTF using this new RV extraction pipeline. We achieve long-term single-measurement high-S/N RV precisions of $\approx$\,8\,\ms\ within a single night and $\approx$\,15\,\ms\ over long-term baselines up to several years, which represents a substantial improvement over previously reported single-measurements precisions using CSHELL and no gas cell (e.g., 58\,\ms\ within a single night; \citealt{2011ApJ...735...78C}; or 90\,\ms\ on baselines of several years; \citealt{2015AJ....149..106D}).

In Section~\ref{sec:sample}, we present the method by which we constructed our target sample. The observing setup and strategy is then detailed in Section~\ref{sec:obs}. We summarize the spectral extraction method and the RV measurement algorithm in Sections~\ref{sec:ext} and \ref{sec:rvpip}, respectively. The method by which we combine individual RV measurements is presented in Section~\ref{sec:rvcomb}. Our global survey results are then presented in Section~\ref{sec:results}, and we discuss individual targets in Section~\ref{sec:indivresults}. We finally present our conclusions in Section~\ref{sec:conclusion}.

\section{SAMPLE SELECTION}\label{sec:sample}

\begin{figure}
	\centering
	\includegraphics[width=0.488\textwidth]{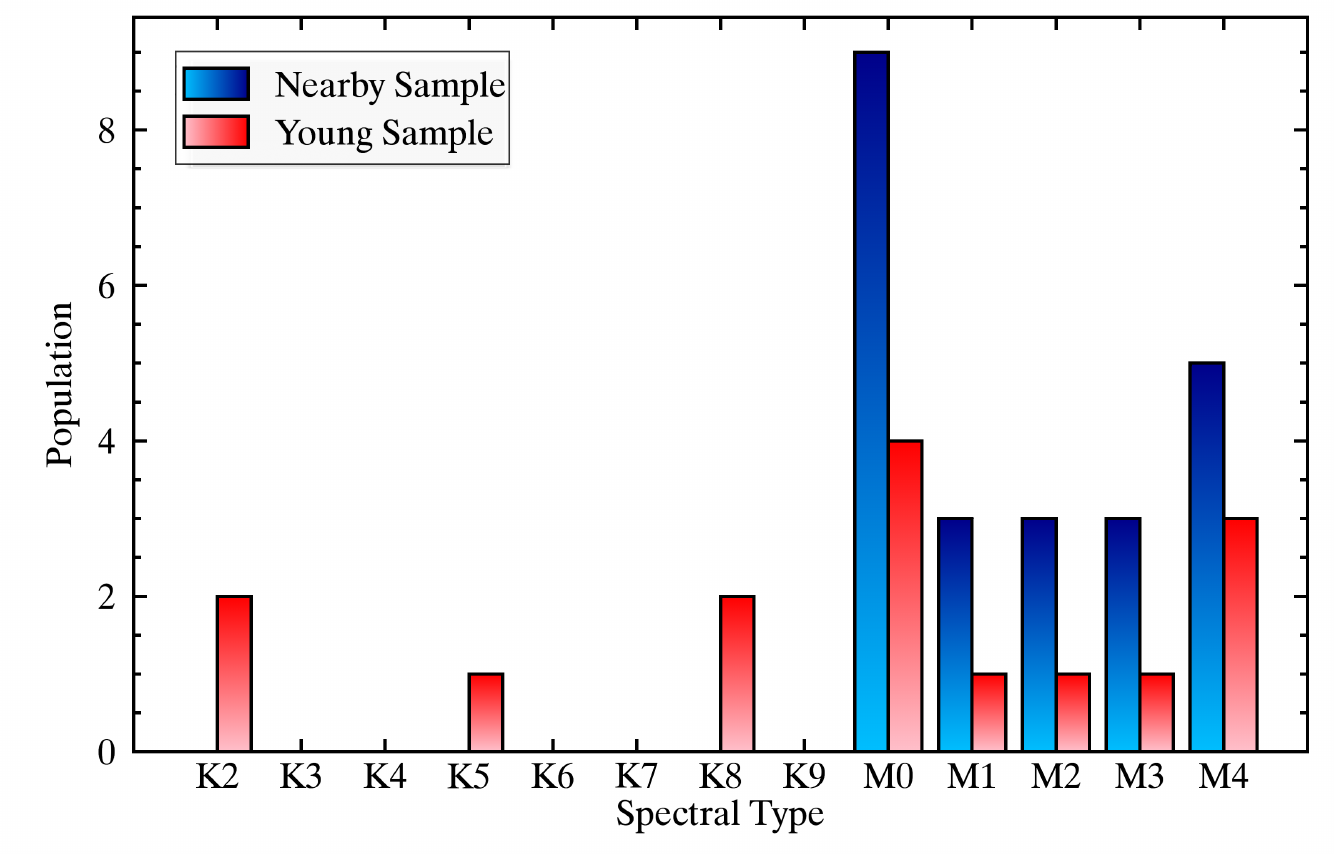}
	\caption{Spectral type histogram of the nearby and young target samples that were studied in this work. Targets that fall into both samples are shown in both histograms (and thus appear twice). For more details, see Section~\ref{sec:sample}}
	\label{fig:spthist}
\end{figure}

Our survey sample is composed of two parts; the first consists of nearby, young stars mostly selected from known members of young moving groups, and the second consists of nearby ($<$\,25\,pc) stars of any age. Comparing our results with those of optical RV surveys for the young sample will eventually allow us to characterize how the signature of stellar activity on RV curves differ between these two wavelength regimes, whereas completing the census of giant close-in planets in a volume-limited sample will be useful to derive population statistics in the near future. In this Section, we describe the process by which these samples were constructed. The complete survey sample is presented in Table~\ref{tab:sample} and a spectral types histogram of it is presented in Figure~\ref{fig:spthist}.

\subsection{The Young Sample}

The young sample was constructed by selecting K\added{-} to mid M-type young stars in the solar neighborhood, mainly from known members of young moving groups\added{,} such as $\beta$~Pictoris ($24 \pm 3$\,Myr; \citealp{2001ApJ...562L..87Z,2015arXiv150805955B}), AB~Doradus ($149^{+51}_{-19}$\,Myr; \citealp{2004ApJ...613L..65Z,2015arXiv150805955B}) and the Octans-Near association ($\approx$\,30--100\,Myr; \citealt{2013ApJ...778....5Z}), with \emph{2MASS} $K_S$-band magnitudes brighter than $7.1$ (the brightest target \replaced{$\epsilon$~Eri}{$\varepsilon$~Eridani} has $K_S = 1.8$) so that a S/N of $\sim$\,80 could be achieved within approximately an hour. \added{The young age of all stars in this sample is strongly constrained by kinematic membership to these young associations, typically in addition to lithium abundance measurements and/or comparison to isochrones (e.g., \citealt{2008hsf2.book..757T}), except for the star \replaced{$\epsilon$~Eri}{$\varepsilon$~Eridani}, which is not a member of any known young moving group. In this case, the age is constrained to 400--800\,Myr by \cite{2008ApJ...687.1264M}}.

No selection cut was applied on projected rotational velocities; all stars in this sample have measured $v \sin i$ values\added{ in the literature} ranging from 2 to 70\,\kms\deleted{\ in the literature}. All $<$\,2\textquotedbl\ binaries were rejected from the sample, except for GJ~3305~AB (0\farcs093; \citealt{2007AA...472..321K}) and HD~160934~AB (0\farcs12; \citealt{2006Ap&SS.304...59G}), which were not known to be binary stars at the time \replaced{where}{when} the young sample was assembled. A total of 15 targets were selected and are listed in Table~\ref{tab:sample}, with spectral types ranging from K2 to M4.

Six targets in the young sample had never benefitted from a precise RV follow-up. Six \replaced{targets}{others} were already followed either at optical or NIR wavelengths albeit at a $\gtrsim$\,50\,\ms\ precision (AG~Tri, AT~Mic~A, AT~Mic~B, AU~Mic, BD+01~2447, V1005~Ori; \citealp{2012ApJ...749...16B,2006PASP..118..706P}), and three\deleted{ others} targets benefitted from a precise RV follow-up at optical wavelengths (BD+20~1790, \replaced{$\epsilon$~Eri}{$\varepsilon$~Eridani}, GJ~3305~AB; e.g., \citealp{2010AA...513L...8F,2015AA...576A..66H,1988ApJ...331..902C,2014AA...568A..26E}).

\added{There are a total of 3/15 (20\%) targets in the young sample that have a projected rotational velocity smaller than $\sim$\,3\,\kms, which seems unexpected for young stars (i.e. such slow rotators have been shown to be typically older than a few hundred Myr; \citealp{2011ApJ...727...56I,2012ApJ...746...43R}). We suggest that projection effects could explain this, e.g. 20\% of stars with a random inclination $i$ would have a projected rotational velocities three times smaller or less than their actual rotational velocities (i.e., $\sin i \leq 0.31$)}.

\subsection{The Nearby Sample}\label{sec:samplenearby}

The nearby sample was constructed by selecting all M dwarfs from the Reserach Consortium on Nearby Stars (RECONS; \citealp{2014AAS...22412026H}) and the L\'epine and Shara Proper Motion (LSPM) catalog \citep{2005AJ....130.1680L} with a  trigonometric distance measurement that places them within 25\,pc of the Sun. We avoided including targets that were already part of precise RV follow-up programs, such as the Hobble-Eberly Telescope survey (HET; \citealp{2003AJ....126.3099E,2006ApJ...649..436E}), the California Planet Survey (CPS; \citealp{2010ApJ...721.1467H,2014ApJ...781...28M}), and the Ultraviolet and Visual Echelle Spectrograph re-analysis of \cite{2014MNRAS.441.1545T}. We selected the targets that are easily accessible from the IRTF ($-35$\textdegree\,$<$\,DEC\,$< 65$\textdegree) with apparent \emph{2MASS} magnitudes $K_S < 6.4$. We used a more conservative $K_S$-band cut in this sample to achieve a S/N of at least 200 per pixel within a few hours (see Section~\ref{sec:obs}).

All targets with a known stellar companion or a background star at $<$\,2\textquotedbl\ were rejected from the sample. We obtained projected rotational velocities ($v \sin i$) from the literature when available and rejected targets with $v \sin i \geq 20$\,\kms. From this initial list of targets, we followed 21 low-mass stars with spectral types in the M0--M4 range, which are listed in Table~\ref{tab:sample}. All of these targets have rotational velocity measurements in the literature, which range from 3 to 16\,\kms. \replaced{We note}{It can be noted} that 4 targets are present \replaced{both in}{in both} the nearby and young samples.

Fourteen targets in the nearby sample \replaced{had never benefitted from a precise RV follow-up}{never had precise RV follow-up observations}. Five targets were already followed at NIR wavelengths albeit at a $\gtrsim$\,50\,\ms\ precision (AT~Mic~A, AT~Mic~B, AU~Mic, EV~Lac and GJ~725~A; \citealt{2012ApJ...749...16B}), and two other targets benefitted from a precise RV follow-up at optical wavelengths (GJ~15~A and GJ~876; e.g., \citealp{1998AA...338L..67D,1998ApJ...505L.147M,2010ApJ...719..890R,2014ApJ...794...51H}).

\begin{deluxetable*}{llllccccccccc}[p]
\tabletypesize{\scriptsize}
\tablecolumns{13}
\tablecaption{Precise NIR RVs Follow-Up Sample \label{tab:sample}}
\tablehead{\colhead{Common} & \colhead{RA J2000} & \colhead{DEC J2000} &  \colhead{Sp.} & \colhead{Ref.} & \colhead{2MASS} & \colhead{$v \sin i$} & \colhead{Ref.} & \colhead{Binary} & \colhead{Ref.} & \colhead{$\log R^\prime_{HK}$} & \colhead{Activity\tablenotemark{a}} & \colhead{Ref.} \\
\colhead{Name} & \colhead{(hh:mm:ss)} & \colhead{(dd:mm:ss)} &  \colhead{Type} & \colhead{} & \colhead{$K_S$} & \colhead{(\kms)} & \colhead{} & \colhead{Sep. (\textquotedbl)} & \colhead{} & \colhead{}}
\startdata
\sidehead{\textbf{Nearby, Young Sample}}
AT~Mic~B & 20:41:51.147 & -32:26:10.22 & M4 & 27 & 4.94\tablenotemark{b} & $15.8 \pm 1.2$ & 12 & 3.6 & $\cdots$ & $\cdots$ & $\cdots$ & $\cdots$ \\
AT~Mic~A & 20:41:51.156 & -32:26:06.58 & M4 & 27 & 4.94\tablenotemark{b} & $10.1 \pm 1.2$ & 12 & 3.6 & $\cdots$ & $\cdots$ & $\cdots$ & $\cdots$ \\
AU~Mic & 20:45:09.492 & -31:20:26.66 & M1 & 27 & 4.53 & $9.3 \pm 1.2$ & 12 & $\cdots$ & $\cdots$ & $-4.053 \pm 0.057$ & VA & 11 \\
EQ~Peg~A & 23:31:52.087 & 19:56:14.22 & M3.5 & 28 & 5.33 & $12.8 \pm 5.0$ & 10 & $5.0$ & 14 & ($-5.162 \pm 0.057$) & (VI) & 6 \\
\sidehead{\textbf{Young Sample}}
AG~Tri & 02:27:29.254 & 30:58:24.61 & K8 & 27 & 7.08 & $5 \pm 2$ & 1 & $22.2$ & 2 & $\cdots$ & $\cdots$ & $\cdots$ \\
$\varepsilon$~Eridani & 03:32:55.911 & -09:27:29.86 & K2 & 29 & 1.78 & $2.45 \pm 0.5$ & 3 & $\cdots$ & $\cdots$ & $-4.478 \pm 0.031$ & A & 4 \\
V577~Per & 03:33:13.491 & 46:15:26.53 & K2 & 5 & 6.37 & $8.9 \pm 0.9$ & 38 & $\cdots$ & $\cdots$ & $-4.109 \pm 0.057$ & VA & 6 \\
GJ~3305~AB & 04:37:37.467 & -02:29:28.45 & M0 & 30 & 6.41 & $6.50 \pm 0.5$ & 7 & 0.093 & $\cdots$ & $\cdots$ & $\cdots$ & $\cdots$ \\
TYC~5899--26--1 & 04:52:24.407 & -16:49:21.97 & M3 & 27 & 6.89 & $< 3$ & 5 & $\cdots$ & $\cdots$ & $\cdots$ & $\cdots$ & $\cdots$ \\
V1005~Ori & 04:59:34.831 & 01:47:00.68 & M0 & 27 & 6.26 & $8.7 \pm 0.9$ & 8 & $\cdots$ & $\cdots$ & $-4.080 \pm 0.151$ & VA & 8 \\
BD+20~1790 & 07:23:43.592 & 20:24:58.66 & K5 & 31 & 6.88 & $15.70 \pm 3.16$ & 9 & $\cdots$ & $\cdots$ & $-3.800 \pm 0.0037$ & VA & 8 \\
BD+01~2447 & 10:28:55.551 & 00:50:27.62 & M2 & 27 & 5.31 & $0.1 \pm 0.2$ & 10 & $\cdots$ & $\cdots$ & $-5.029 \pm 0.057$ & I & 11 \\
HD~160934~AB & 17:38:39.634 & 61:14:16.03 & M0 & 27 & 6.81 & $19.1 \pm 0.6$ & 8 & 0.12 & $\cdots$ & $-3.87 \pm 0.120$ & VA & 8 \\
LO~Peg & 21:31:01.711 & 23:20:07.47 & K8 & 27 & 6.38 & $70 \pm 10$ & 10 & $\cdots$ & $\cdots$ & $-3.906 \pm 0.050$ & VA & 13 \\
BD--13~6424 & 23:32:30.864 & -12:15:51.43 & M0 & 27 & 6.57 & $8.8 \pm 1.2$ & 12 & $\cdots$ & $\cdots$ & $\cdots$ & $\cdots$ & $\cdots$ \\
\sidehead{\textbf{Nearby Sample}}
GJ~15~A & 00:18:22.885 & 44:01:22.63 & M2 & 32 & 4.02 & $1.43 \pm 0.5$ & 7 & $31.9$ & 2 & $-5.27 \pm 0.04$ & VI & 8 \\
GJ~169 & 04:29:00.138 & 21:55:21.48 & M0.5 & 33 & 4.88 & $3.8 \pm 1.8$ & 10 & $\cdots$ & $\cdots$ & $-4.813 \pm 0.154$ & I & 8 \\
LHS~26 & 04:31:11.479 & 58:58:37.57 & M4 & 32 & 5.72 & $< 3$ & 15 & $8.9$ & 2 & $\cdots$ & $\cdots$ & $\cdots$ \\
GJ~338~A & 09:14:22.982 & 52:41:12.53 & M0 & 32 & 3.99 & $2.9 \pm 1.2$ & 16 & $18.1$ & 17 & $-4.652 \pm 0.166$ & A & 8 \\
GJ~338~B & 09:14:24.856 & 52:41:11.84 & M0 & 32 & 4.14 & $2.8 \pm 1.2$ & 16 & $18.1$ & 17 & $-4.420 \pm 0.050$ & A & 18 \\
GJ~458~A & 12:12:20.847 & 54:29:08.69 & M0 & 31 & 6.06 & $5.4 \pm 1$ & 8 & $14.7$ & 20 & $-4.970 \pm 0.040$ & I & 8 \\
GJ~537~B & 14:02:33.128 & 46:20:23.92 & M0 & 34 & 5.39 & $< 4$ & 21 & $3.0$ & $\cdots$ & ($-4.971 \pm 0.057$) & (I) & 4 \\
GJ~537~A & 14:02:33.240 & 46:20:26.64 & M2 & 34 & 5.43 & $< 4$ & 21 & $3.0$ & 2 & $-4.733 \pm 0.057$ & A & 4 \\
LHS~371 & 14:25:43.496 & 23:37:01.06 & M0 & 34 & 5.97 & $< 4$ & 21 & $45.4$ & 20 & $\cdots$ & $\cdots$ & $\cdots$ \\
LHS~372 & 14:25:46.671 & 23:37:13.31 & M1 & 34 & 6.09 & $< 4$ & 21 & $45.3$ & 22 & $\cdots$ & $\cdots$ & $\cdots$ \\
LHS~374 & 14:30:47.794 & -08:38:46.57 & M0 & 34 & 5.77 & $< 3$ & 10 & $50.0$ & 23 & $-4.896 \pm 0.057$ & I & 24 \\
GJ~9520 & 15:21:52.919 & 20:58:39.48 & M1.5 & 31 & 5.76 & $6.5 \pm 3$ & 21 & $\cdots$ & $\cdots$ & $\cdots$ & $\cdots$ & $\cdots$ \\
GJ~3942 & 16:09:03.097 & 52:56:37.95 & M0 & 35 & 6.33 & $< 4$ & 21 & $\cdots$ & $\cdots$ & $\cdots$ & $\cdots$ & $\cdots$ \\
GJ~725~A & 18:42:46.679 & 59:37:49.47 & M3 & 32 & 4.43 & $< 2.5$ & 26 & 13.3 & $\cdots$ & $-5.267 \pm 0.229$ & VI & 11 \\
GJ~740 & 18:58:00.140 & 05:54:29.70 & M0.5 & 21 & 5.36 & $3.8 \pm 2.8$ & 8 & $\cdots$ & $\cdots$ & $-4.825 \pm 0.175$ & I & 8 \\
EV~Lac & 22:46:49.807 & 44:20:03.10 & M3.5 & 36 & 5.30 & $6.9 \pm 0.8$ & 8 & $10.0$ & 2 & $-3.972 \pm 0.147$ & VA & 8 \\
GJ~876 & 22:53:16.722 & -14:15:48.91 & M4 & 37 & 5.01 & $2.8 \pm 2.2$ & 25 & $\cdots$ & $\cdots$ & ($-5.146 \pm 0.057$) & (VI) & 11 \\
\enddata
\tablenotetext{a}{VI: Very Inactive; I: Inactive; A: Active; VA: Very active; $\cdots$: Information not available in the literature (see Section~\ref{sec:samplenearby} for more details).}
\tablenotetext{b}{Unresolved photometry.}
\tablecomments{See Section~\ref{sec:sample} for more details}
\tablerefs{(1)~\citealt{2000AA...364..205C}; (2)~\citealt{2001AJ....122.3466M}; (3)~\citealt{2006ApJ...646..505B}; (4)~\citealt{1991ApJS...76..383D}; (5)~\citealt{2010AJ....140..119S}; (6)~\citealt{2013AA...551L...8P}; (7)~\citealt{2010MNRAS.407.1657H}; (8)~\citealt{2012AA...537A.147H}; (9)~\citealt{2007AJ....133.2524W}; (10)~\citealt{2005ESASP.560..571G}; (11)~\citealt{2010ApJ...725..875I}; (12)~\citealt{2006AA...460..695T}; (13)~\citealt{2003AJ....126.2048G}; (14)~\citealt{2002AA...384..180F}; (15)~\citealt{2003ApJ...583..451M}; (16)~\citealt{1998AA...331..581D}; (17)~\citealt{2012RMxAA..48..177O}; (18)~\citealt{2013AA...555A..11E}; (19)~\citealt{2013MNRAS.431.2063S}; (20)~\citealt{2007AJ....133..889L}; (21)~\citealt{2012AJ....143...93R}; (22)~\citealt{2003ApJ...582.1011S}; (23)~\citealt{1997AAS..125..523W}; (24)~\citealt{2011ApJ...734...70A}; (25)~\citealt{1992ApJ...390..550M}; (26)~\citealt{2010AJ....139..504B}; (27)~\citealt{2013ApJ...762...88M}; (28)~\citealt{2015AJ....149..106D}; (29)~\citealt{1989ApJS...71..245K}; (30)~\citealt{2007AA...472..321K}; (31)~\citealt{2004AJ....128..463R}; (32)~\citealt{2009ApJ...704..975J}; (33)~\citealt{2010MNRAS.403.1949K}; (34)~\citealt{2014MNRAS.443.2561G}; (35)~\citealt{1956AJ.....61..201V}; (36)~\citealt{1997AJ....113.1458H}; (37)~\citealt{2007ApJ...670.1367L}; (38)~\citealt{2014AJ....148...70M}.}
\end{deluxetable*}

Both the HET and CPS catalogs used similar selection criteria with the exception that young and/or chromospherically active systems were rejected. Thus, we are subject to a bias towards these systems despite our rejection of high-$v \sin i$ targets. One target in the sample (GJ~740) has been followed as part of the HARPS survey \citep{2013AA...549A.109B}, but no results were published at the time the target list was assembled, hence we have \replaced{not excluded it from}{included it in} the sample.

We flag chromospheric activity in Table~\ref{tab:sample} based on \deleted{available} $\log{R^\prime_{HK}}$ \replaced{coefficients in}{index values from} the literature, when available (e.g., see \citealt{1968SoPh....3..164Z,1984ApJ...287..769N}.) \added{We assume a measurement error of 5\% in the case of LO~Peg, since \cite{2003AJ....126.2048G} do not report one. Nine objects have no available $\log{R^\prime_{HK}}$ index but have a measurement for the alternative $S_{HK}$ index that also traces chromospheric activity. In these cases, we used the relation of \cite{1982AA...107...31M} and \cite{1984ApJ...287..769N} to translate it into a value for $\log{R^\prime_{HK}}$. Using Figure~1 of \cite{1982AA...107...31M}\footnote{Using the WebPlotDigitizer tool available at \url{http://arohatgi.info/WebPlotDigitizer/}}, we calculated that this relation is associated with an uncertainty of $\sim 0.057$\,dex, which is larger than typical measurement errors on $S_{HK}$ themselves. In a few cases where the $B-V$ color of a target is redder than the color range where the relation has been tested ($0.45 \leq B-V \leq 1.5$), we display the value between parentheses in Table~\ref{tab:sample}. We use the classification of \cite{2003AJ....126.2048G} to categorize targets with $\log{R^\prime_{HK}} > -4.2$ as very active (VA), those with $-4.2 > \log{R^\prime_{HK}} > -4.75$ as active (A), those with $-4.75 > \log{R^\prime_{HK}} > -5.1$ as inactive (I), and those with $\log{R^\prime_{HK}} < -5.1$ as very inactive (VI).}

\section{OBSERVATIONS}\label{sec:obs}

We obtained our data using the NIR high resolution single-order echelle spectrograph CSHELL \citep{1993SPIE.1946..313G,1990SPIE.1235..131T} at the \replaced{3-meters}{3\,m} IRTF from \replaced{October 2010}{2010 October} to \replaced{January 2015}{2015 January}. The survey presented here includes a total of 3794 individual spectra obtained \added{with}in a total of 65 observing nights. We used the 0\farcs5 slit and the continuous variable filter\footnote{\added{More information is available at }\url{http://irtfweb.ifa.hawaii.edu/~cshell/rpt\_cvf.html}}, yielding a resolving power of $R = 46\,000$ (\citealp{2011ApJ...735...78C,2015ApJ...808...12P}) with a spectral grasp of $\approx$\,5.55\,nm at 2.3125\,$\mu$m. It is noteworthy to mention that \citealt{2015AJ....149..106D} measured a resolving power as high as $R = 57\,000$ with the 0\farcs5 slit from modelling observed absorption telluric lines and suggested that this could be due to the slit being slightly narrower than its designation. The spectrograph uses a $\approx$\,25 yr-old 256x256 pixels InSb detector that has a significant number of bad pixels compared to more recent NIR detectors.

A $^{13}$CH$_4$ isotopologue methane gas cell \citep{2012PASP..124..586A,2013SPIE.8864E..1JP} was inserted in the shutter beam to provide a wavelength reference in the NIR. \added{The gas cell temperature is stabilized at $10.0 \pm 0.1$\degr\,Celsius, which is adequate to keep it stable at better than the $\sim$\,1\,\ms\ level \citep{2012PASP..124..586A,2013SPIE.8864E..1JP}}. The grating angle was set once at the beginning of each night by observing an A-type star and ensuring that the deep methane absorption line of the gas cell at $\approx$\,2.31355\,$\mu$m was located on column 179 of the detector to achieve identical wavelength coverage, as the position of the deep methane feature typically varied by $\sim$\,1--2 detector pixels between observing nights. This setup typically corresponded to central wavelength values of $\approx$\,2.31255\,$\mu$m.

We obtained science exposures of 300\,s or less to avoid saturation or significant background variations, and obtained several exposures (typically $\sim$\,5--60) to achieve a combined S/N of $\sim$\,70 to 200 depending on the survey sample. We did not obtain pair-subtracted spectra as our targets are significantly brighter than background sky emission, and all spectra were obtained with the methane gas cell in beam.

Starting in 2014, we systematically observed one bright A-type star (either Sirius, Castor, Vega, Alphecca, $\beta$~Arietis or 32~Pegasi depending on their altitude) at the beginning of every night to obtain a S/N\,$\approx 200$ spectrum of the gas cell \replaced{to}{and} characterize any potential long-term variation. Fifteen flat-field exposures of \replaced{300\,s}{15\,s} were initially obtained at the beginning of every night, however starting from 2014 April 2, we obtained fifteen \replaced{300\,s}{15\,s} flat-field exposures \replaced{before}{after} every science target to minimize systematic instrumental effects. We did not apply dark frame calibrations as we found that they did not improve the quality of our results. A log of all observations obtained in this work is presented in Table~\ref{tab:obslog}, where S/N values are calculated by assuming that the spectra are photon-noise limited.

\begin{deluxetable}{lccc}
\tablecolumns{4}
\tablecaption{Log of All Science Star Observations \label{tab:obslog}}
\tablehead{\colhead{Target} & \colhead{UT Date} & \colhead{Num. of} &  \colhead{Median}\\
\colhead{Name} & \colhead{(YYMMDD)} & \colhead{Spectra} & \colhead{S/N}}
\startdata
\textbf{AG~Tri} & 101009 & 6 & 23\\
& 101011 & 15 & 16\\
& 101012 & 12 & 23\\
& 101122 & 12 & 22\\
& 101123 & 12 & 26\\
& 101124 & 12 & 20\\
& 110216 & 9 & 22\\
& 110219 & 10 & 25\\
& 110716 & 10 & 20\\
& 110819 & 14 & 28\\
\hline
\textbf{AT~Mic~A} & 101010 & 6 & 39\\
& 101011 & 11 & 35\\
& 101013 & 10 & 28\\
& 101123 & 2 & 26\\
& 101124 & 6 & 43\\
\enddata
\tablecomments{See Section~\ref{sec:obs} for more details. \added{Table~\ref{tab:obslog} is published in its entirety in the machine-readable format. A portion is shown here for guidance regarding its form and content.}}
\end{deluxetable}

\section{SPECTRAL EXTRACTION}\label{sec:ext}

Due to several challenges in extracting RVs out of CSHELL spectra, we constructed our own \replaced{unique}{custom} data reduction pipeline to extract the 3794 spectra obtained in this work in a consistent way. We summarize below this data extraction pipeline, which is \replaced{detailed}{described in more details} in P.~Gao et al. (submitted to PASP).

Per-target flat fields were generated by median-combining  data sets that each consist of 15 individual files. 2D sinusoidal fringing with amplitude $\sim$\,0.2--0.6\% was found to be present in these flat fields due to the CSHELL circular variable filter, which limited our \added{long-term }RV precision to $>$\,55\,\ms\ when left uncorrected. Fringing subtraction was thus achieved by median-combining a large number of per-target flat fields spanning several years to obtain a master flat field. This averaged out the fringing due to its spatial and spectral variations over time. Individual per-target flat fields were then normalized by the master flat field to bring out their fringing pattern, which was fitted and corrected individually\deleted{ (for more details, see P.~Gao et al., submitted to PASP)}.

Fringing is also present in the science observations, but similar efforts \replaced{as the flat field}{to correct} fringing\deleted{ removal} were not possible due to low illumination of most of the image aside from the \replaced{spectrum}{spectral trace} of the target. Thus, we removed this fringing \replaced{during}{in} the RV extraction process.\deleted{ (P.~Gao et al., submitted to PASP).}

Spectral extraction was performed by first dividing the fringing-corrected, per-target flat fields from the science observations, followed by correcting for a linear tilt in the target trace on the detector. A Moffat profile was then fit to the trace in the spatial direction, and the 1D spectrum was extracted using the median trace profile with an optimal extraction procedure \citep{1986PASP...98..609H,2013pss2.book...35M}. Following this, a synthetic spectral trace was constructed from the extracted spectrum and the spatial profile and was \deleted{is }divided out from the observed spectral trace. The 2D residuals resulting from this operation were combed for large deviations from the median, which were flagged as bad pixels. A 1D spectrum was extracted again from the spectral trace after having masked these bad pixels. Additional bad pixels were flagged by noting any major deviations from the continuum in the\added{ final} 1D spectra, \replaced{so that they were then}{in order for them to be} ignored by the RV extraction pipeline. 

\section{RADIAL VELOCITY EXTRACTION}\label{sec:rvpip}

We used \replaced{the}{a novel forward modelling} MATLAB RV pipeline \deleted{described by P.~Gao et al. (submitted to PASP)} to compute the relative RV of every individual spectrum\replaced{.}{, which we summarize in this section. The pipeline is described in detail by P.~Gao et al. (submitted to PASP), who demonstrate that it allows for the construction of a stellar template simultaneously to gas cell observations, and properly accounts for the significant telluric features in the NIR. Using this pipeline with CSHELL/IRTF data, they achieve a\deleted{n} RV precision of $\sim$\,3\,\ms\ over a few days using photon-noise limited observations of the M-type giant SV~Peg.}

The pipeline extracts the RVs associated with a spectrum by fitting a forward spectral model to the observed data. The forward model consists of: an empirical telluric transmission spectrum $T_\lambda$ \citep{1991aass.book.....L} where $\lambda$ is the wavelength measured at rest and in vacuum; a measured isotopologue methane gas cell spectrum $G_\lambda$ \citep{2013SPIE.8864E..1JP} that has been obtained from \replaced{JPL}{the NASA Jet Propulsion Laboratory (JPL)} and corrected using high-S/N observations of A-type telluric standards; a line spread function (LSF) $L$ that models broadening and distortions in the spectral line profile due to both instrumental and atmospheric effects and that is represented by a weighted sum of the first 5 terms of the Hermite function (i.e., a normal distribution multiplied by Hermite polynomials, see \citealt{Arfken:2012uf}) where the weights are free parameters; a quadratic curve $B_\lambda$ that models the instrumental Blaze function and variations in the NIR sky background; and an instrumental fringing term $\phi_\lambda$ that is represented by a multiplicative sinusoid function. The forward model $F_{\lambda}$ is then given by :
\begin{align}
	F_{\lambda} = \left( G_\lambda \cdot T_\lambda \cdot S_\lambda \cdot B_\lambda \cdot \phi_\lambda \right) \otimes L,
\end{align}
\noindent where $S_\lambda$ is an estimated stellar spectrum of the target in the absence of a gas cell, and $\otimes$ represents a convolution. The forward model $F_{\lambda}$ must then be Doppler-shifted to account for the relative line-of-sight velocity of the target with respect to an observer on the Earth. This is done by applying a linear shift to $F_{\lambda}$ by a factor \replaced{$\exp{\left(W/c\right)}$}{$\left(W/c\right)$} in logarithmic wavelength space, where $c$ is the speed of light and $W$ is the combined contributions from the RV signal of the target and the motion of the Earth relative to the target. $F_{\lambda}$ is finally mapped to the detector pixels using a two-degrees polynomial mapping relation. The forward model is fit to the data using a Nelder-Mead downhill simplex algorithm \citep{Nelder:1965in}, which is especially useful in this situation where the number of free parameters (17) is relatively large.

A difficulty in using this method is the apparent need of measuring the stellar spectrum $S_\lambda$, since it would require obtaining high-S/N spectra for every target with the methane gas cell out of beam. This would be problematic, as moving the methane gas cell in and out of beam will affect the wavelength solution of the observed spectrum. The stellar template thus needs to be measured simultaneously as the RV data is collected. The algorithm addresses this need, using only observations obtained with the gas cell in beam, by: (1) performing a first fit with $S_\lambda = 1\ \forall\ \lambda$; (2) identifying deep CO lines in the residuals of the best fit; (3) repeating step 1 with the CO lines masked to obtain a better solution given that CO lines are not yet included in the stellar template $S_\lambda$; (4) constructing a stellar spectrum from a weighted linear combination of the de-convolved fitting residuals of all individual spectra for a given science target in the stellar rest frame; and (5) repeating step 4 up to 20 times, adding the residuals back into the best estimated stellar spectrum at every iteration. The deconvolution of the residuals is computer-intensive and only significantly benefits the first iteration, hence it was only performed at the first iteration. It is important to note that all RVs measured with this algorithm are \emph{relative}, since they are measured with respect to the stellar template that was constructed from all of the data \replaced{itself}{themselves}.

\replaced{We note}{It can be noted} that combining all observations to create the stellar template does not account for any variability of the star within the wavelength range that we observe. This is not problematic for the targets in our sample since they do not show significant temporal variations in their spectral morphology, however stars that vary significantly such as SV~Peg (e.g., M.~Bottom et al., in preparation) would need to be reduced in several individual steps of smaller temporal coverage to account for the varying stellar spectrum.

The quality of the fit typically increased for a few iterations (typically $\lesssim$\,10), until noise \replaced{began}{started} to dominate the residuals that are added to the stellar spectrum. The best-fit parameters of the iteration where the RV scatter is minimized were then preserved as the true solution to the fit, and the barycentric-corrected RVs were calculated from $W$ using \added{barycentric corrections that were computed a priori for every science exposure with} the \emph{barycentric\_vel.pro} IDL routine.\deleted{ We refer the reader to P.~Gao et al. (submitted to PASP) for a detailed explanation and characterization of the algorithm described in this Section.} Since the stellar spectrum is not modelled, this analysis does not allow for a measurement of the projected stellar rotational velocity.

\section{THE COMBINATION OF RV MEASUREMENTS}\label{sec:rvcomb}

The algorithm described in Section~\ref{sec:rvpip} yields an individual RV measurement $\nu_i$ for every spectrum $i$ that corresponds to a single exposure on a given science target. These RV measurements are relative in the sense that they are computed with respect to the stellar template, which is built from the data \replaced{itself}{themselves}. The Nelder-Mead downhill simplex algorithm does not provide error measurements on the best-fit parameters, however the general quality of the forward model can be assessed from $\sigma_{R_i}$, the standard deviation of the residuals ($R_i$) of the best fit to the spectrum in question, with bad pixels ignored.

Since we aim at reaching high S/N ratios and at detecting variability on timescales of more than several hours, we combined all science exposures obtained within \replaced{the same}{a given} night (typically $\sim$\,5--60), using a weighted mean that is designed to minimize the impact of low-quality data:
\begin{align}
	\bar{\nu}_k = \frac{\sum_i w_i\,\nu_i}{\sum_i w_i},
\end{align}
\noindent where the weight factors are defined as $w_i = \sigma_{R_i}^{-2}$. $\bar{\nu}_k$ thus represents the mean RV within observing night $k$. We define the measurement error that is associated with this quantity as a weighted standard deviation of the individual RV measurements obtained in this night:
\begin{align}
	\sigma_k^2 = \frac{1}{N_{exp}} \frac{\sum_i w_i\,\left(\nu_i-\bar{\nu}_k\right)^2}{\sum_i w_i},
\end{align}
\noindent where $N_{exp}$ is the total number of exposures obtained within the observing night.

There are a few quantities that are interesting to calculate for every RV curve, in the sense that they can shed light on the typical\deleted{ instrumental} precision and on the possibility that a target is an RV variable source. The most straightforward of those is the weighted standard deviation of the per-night RVs:
\begin{align}
	\varsigma^2 = \frac{\sum_k w_k^\prime\,\bar{\nu}_k^2}{\sum_k w_k^\prime},
\end{align}
\noindent where we use the $\varsigma$ symbol to distinguish this quantity from $\sigma_k$. We use the optimal weight factors $w_k^\prime$ that correspond to the inverse of the variance of a per-night RV measurement:
\begin{align}
	w_k^\prime = \sigma_k^{-2}.
\end{align}

In order to avoid an artificial over-weighting of RV points that have a very small $\sigma_k$, which happens from time to time when the number of exposures is very low, we define a maximum weight of $w^\prime_{k,max} = \left(15\,\mbox{\ms}\right)^{-2}$, which corresponds to the weight that one RV data point would have at the typical single-measurement precision values that we obtain for high-S/N observations.

Another quantity of interest is the reduced chi-square $\chi^2_r$ of an RV curve with respect to a zero-variation curve, given by:
\begin{align}
	\chi^2_r = \frac{1}{N_k-1}\sum_{k=1}^{N_k} \frac{\bar{\nu}_k^2}{\sigma_k^2}
\end{align}
\noindent where $N_k$ is the number of nights where a given target was observed, and the denominator $N_k-1$ corresponds to the number of degrees of freedom to the RV curve (we subtract one fitted parameter corresponding to the floating relative RV). Targets with a higher $\chi^2_r$ value will be more likely to be true RV variable sources.

In an ideal case where all per-night RV measurements have the same intrinsic RV precision $\sigma_k = S\ \forall\ k$, it can be shown from the previous equations that:
\begin{align}
	S^2=\frac{\varsigma^2}{\chi^2_r}\,\left(\frac{N_k}{N_k-1}\right).
\end{align}

We will refer to this quantity $S$ as the single-measurement precision; it is not affected by the fact that a given source is an RV variable as long as it is only variable on timescales larger than a few hours (otherwise $\sigma_k$ would be partly composed of a variability term). For the sake of being able to compare all of our targets in a single $\varsigma^2$--$\chi_r^2$ plane, we will make the following approximation:
\begin{align}
	S^2 \approx \frac{\varsigma^2}{\chi^2_r},
\end{align}
\noindent which would normally only be valid for large values of $N_k$. We can bring out another interesting measurement by making the supposition that the scatter $\varsigma$ of a given RV curve is due to only two uncorrelated sources: a physical RV variability $V$ and an instrumental error term corresponding to the single-measurement precision $S$. It immediately follows that:
\begin{align}
	\varsigma = \sqrt{S^2 + V^2}.
\end{align}

We define another quantity $N_\varsigma$:
\begin{align}
	N_\varsigma = \frac{V}{S} = \sqrt{\left(\varsigma/S\right)^2 - 1} \approx \sqrt{\chi^2_r-1},
\end{align}
\noindent which will be used in the following sections to assess the statistical significance of the RV variability $V$ of our targets.

One last quantity of interest is $V_{\rm max}$, the maximal admissible RV variability on the timescales that we probed given our RV measurements,\deleted{ which we set} at a statistical significance of \replaced{3$\sigma$}{$N_\varsigma$}. If we assume that the probability density function \replaced{$P(\nu)$}{$\mathcal{N}(\nu)$} associated with the RV variability measurement that we obtained for a target is a normal distribution centered on $V$ and with a characteristic width $S$, a simplistic estimate for the maximal admissible RV variability would be \replaced{$V_{\rm max} = V + 3S$}{$V_{\rm max} = V + N_\varsigma S$} such that:
\begin{align}
	f=\int_{-\infty}^{V_{\rm max}\left(N_\varsigma\right)} \mathcal{N}(\nu)\,\mathrm{d}\nu = \erf{\left(N_\varsigma/\sqrt{2}\right)} \approx 0.9973,\label{eqn:fnsig}
\end{align}
\noindent where $\erf{\left(x\right)}$ is the error function. However, since negative RV variabilities are unphysical, the normal probability density function must be set to zero for all negative RV values. We must thus identify the value for $V_{\rm max}$ that ensures:
\begin{align}
	\int_0^{V_{\rm max}\left(N_\varsigma\right)} \mathcal{N}(\nu)\,\mathrm{d}\nu = f\int_0^{\infty} \mathcal{N}(\nu)\,\mathrm{d}\nu.
\end{align}

Solving for $V_{\rm max}$ yields:
\begin{align}
	V_{\rm max}\big(N_\varsigma\big) &= V + S\,\ierf{\Big(f-\left(1-f\right)\erf{\left(V/S\right)}\Big)},\label{eqn:vmax}\\
	f &= \erf{\left(N_\varsigma/\sqrt{2}\right)},
\end{align}
\noindent where $\ierf{\left(x\right)}$ is the inverse error function.\explain{A few changes were made to the Equations above to generalize the case to any value of $N_\varsigma$}

\section{SURVEY RESULTS}\label{sec:results}

We present in this section the resulting NIR RV measurements that we obtained for the two survey samples described in Section~\ref{sec:sample} using our novel RV extraction method. Global results of this survey are described in Section~\ref{sec:ensemble}\deleted{, and individual science targets are discussed in Section~\ref{sec:indivresults}}\added{, and upper limits on the projected masses of possible companions to our targets are derived in Section~\ref{sec:nondet}. We discuss the effect of rotational velocity and stellar activity in Section~\ref{sec:rotvel}, and discuss\deleted{ on} bi-sector measurements in Section~\ref{sec:bisector}}.

\subsection{Ensemble Results}\label{sec:ensemble}

\begin{figure*}
	\centering
	\subfigure[AT~Mic~B]{\includegraphics[width=0.49\textwidth]{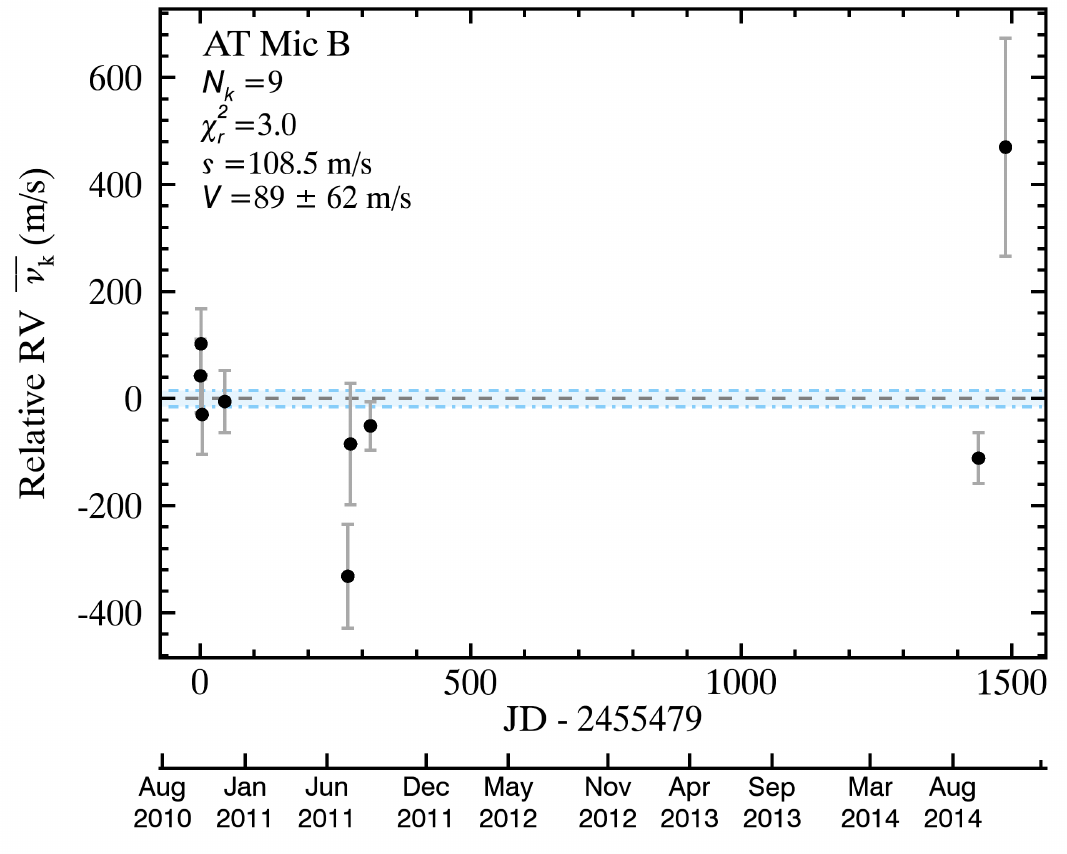}}
	\subfigure[AT~Mic~A]{\includegraphics[width=0.49\textwidth]{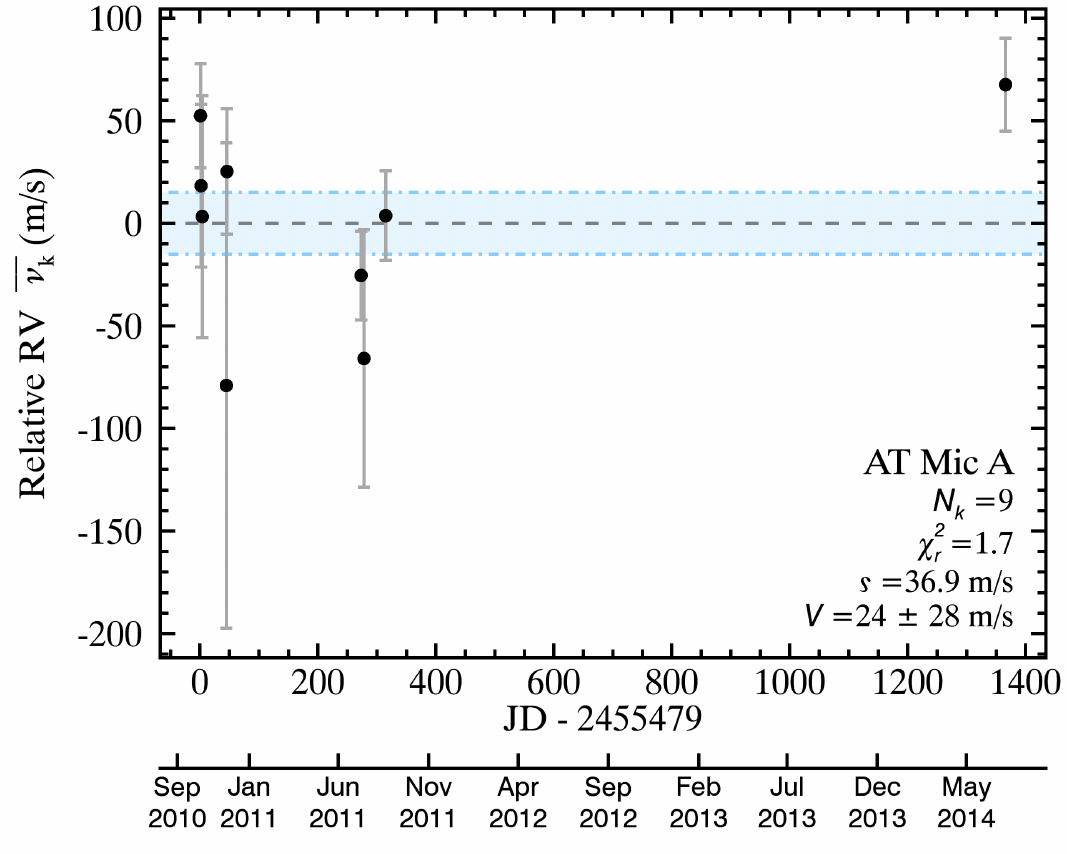}}
	\subfigure[AU~Mic]{\includegraphics[width=0.49\textwidth]{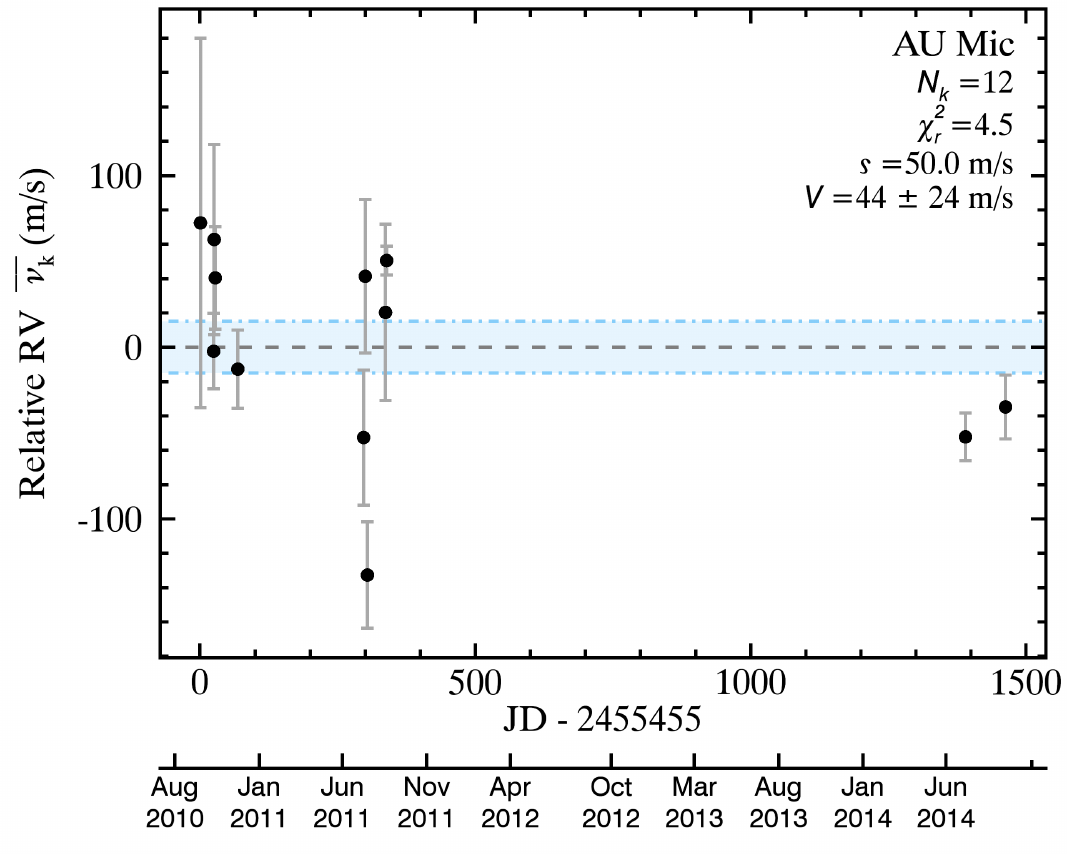}}
	\subfigure[EQ~Peg~A]{\includegraphics[width=0.49\textwidth]{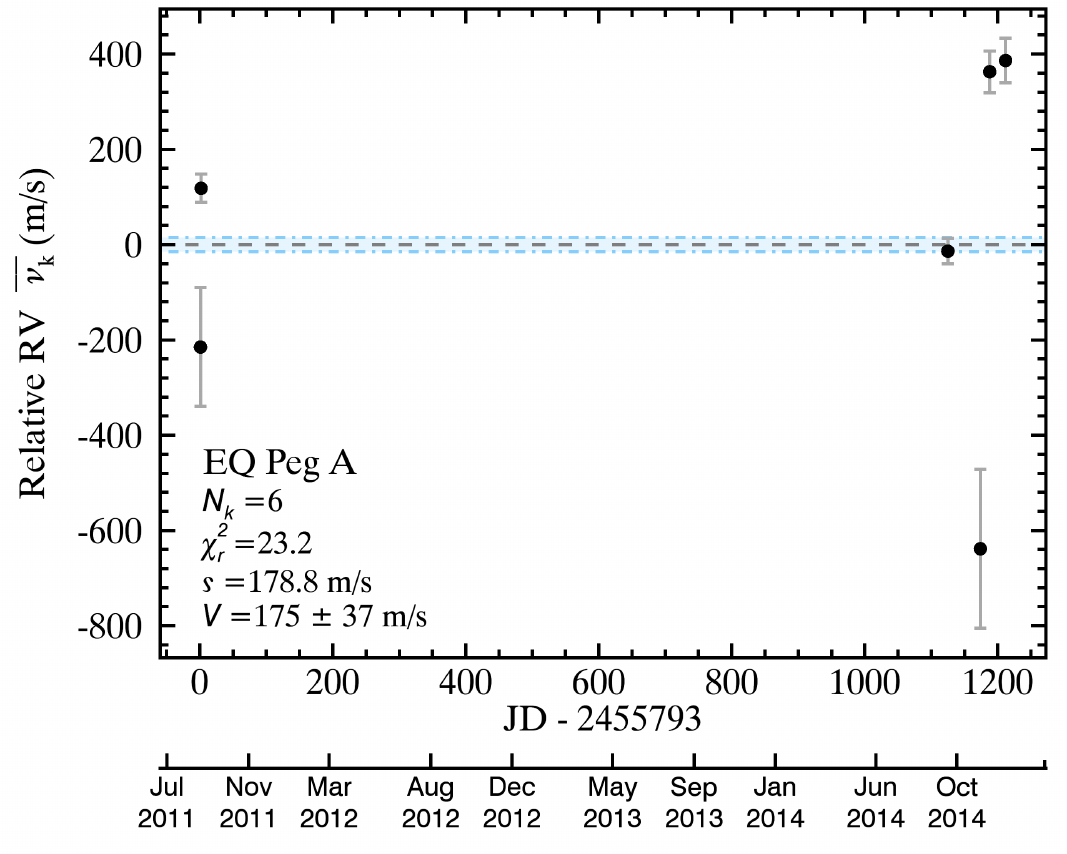}}
		\caption{RV curves of the targets that are present in both the young and nearby samples. The grey dashed line corresponds to the relative zero-level, whereas the light blue region delimited by dash-dotted lines corresponds to a scatter of 15\,\ms, which is similar to the best average single-measurement precisions that we achieve. Target names are displayed in the legends as well as the total number of measurements $N_k$, the error-weighted reduced chi-square $\chi_r^2$ with respect to a zero-variation curve, the error-weighted standard deviation $\varsigma$ and the expected RV variability $V$ (see Section~\ref{sec:rvcomb} for detail). See Section~\ref{sec:results} for a discussion of the global survey results, and Section~\ref{sec:indivresults} for a discussion of individual targets}
	\label{fig:RV_Curves_1}
\end{figure*}

\begin{figure*}
	\centering
	\subfigure[AG~Tri]{\includegraphics[width=0.325\textwidth]{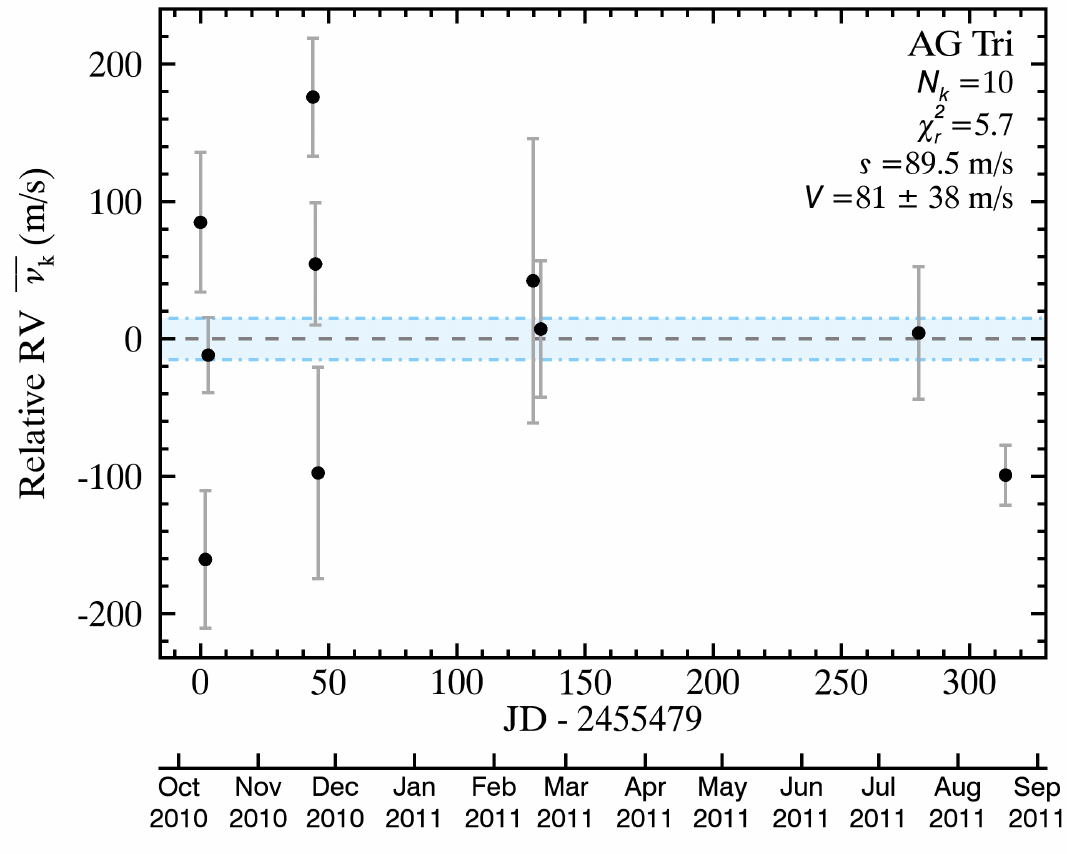}}
	\subfigure[$\varepsilon$~Eridani]{\includegraphics[width=0.325\textwidth]{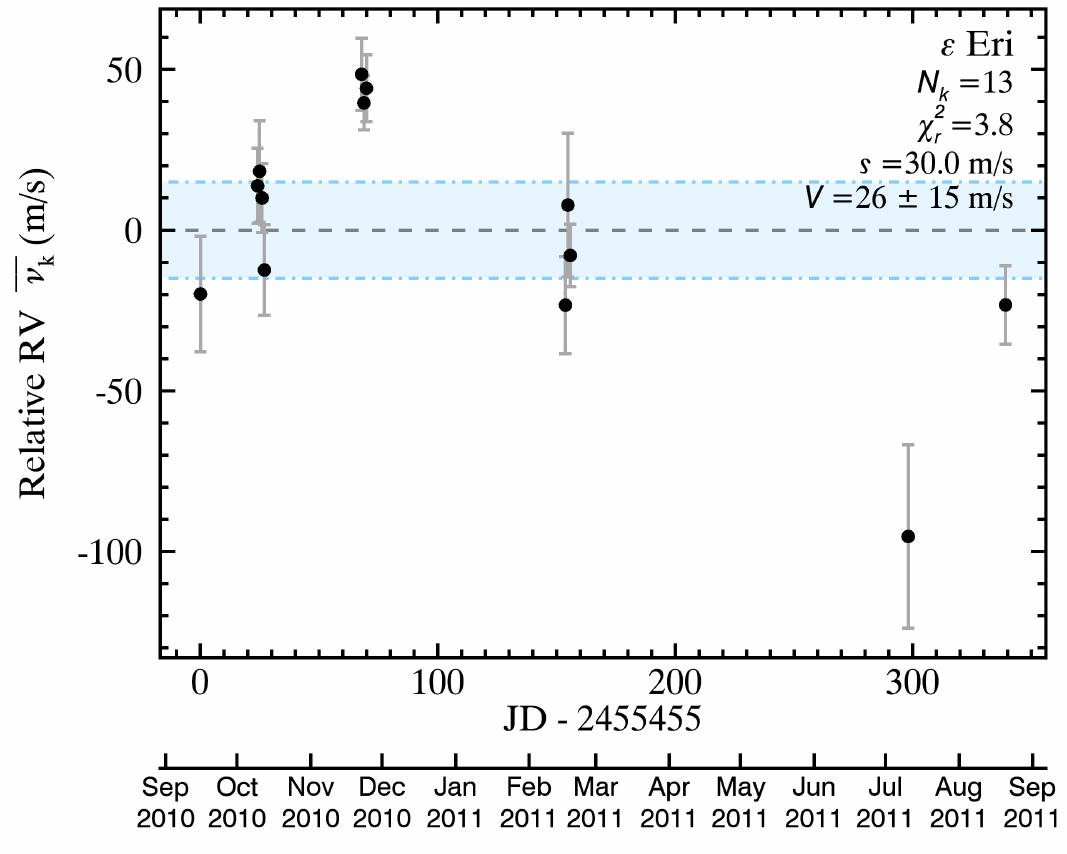}}
	\subfigure[V577~Per]{\includegraphics[width=0.325\textwidth]{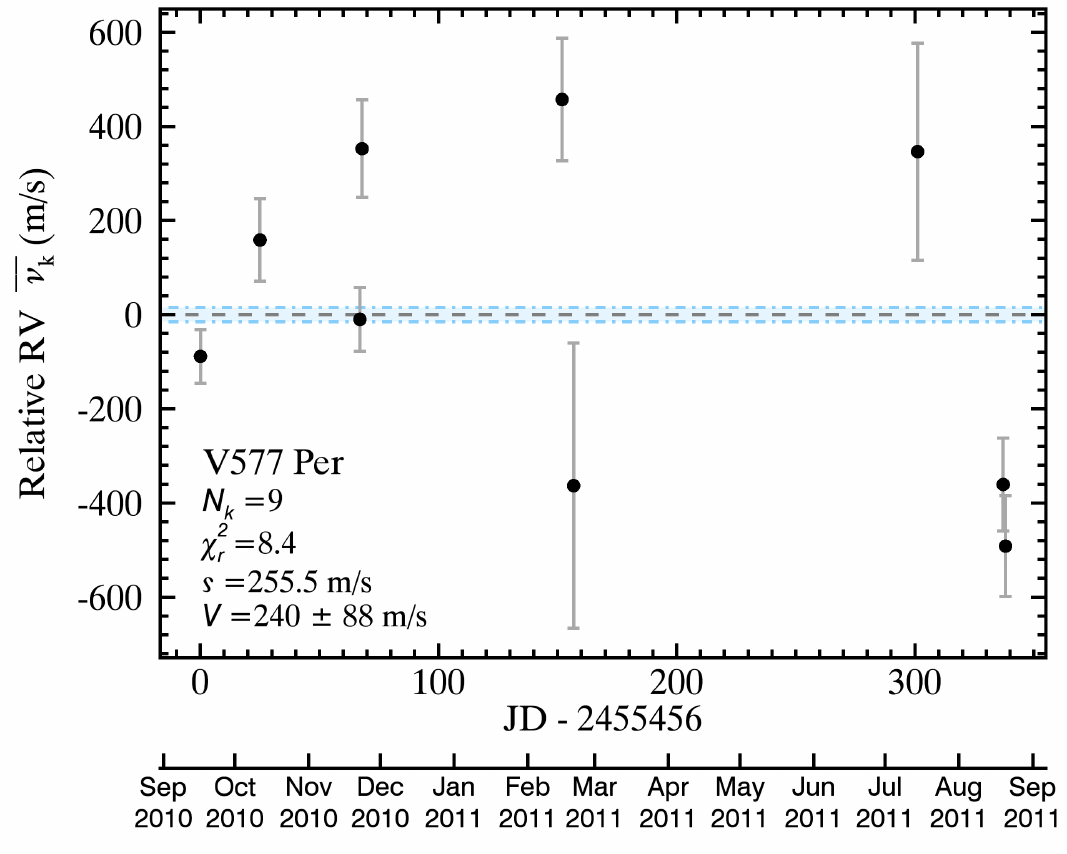}}
	\subfigure[GJ~3305~AB]{\includegraphics[width=0.325\textwidth]{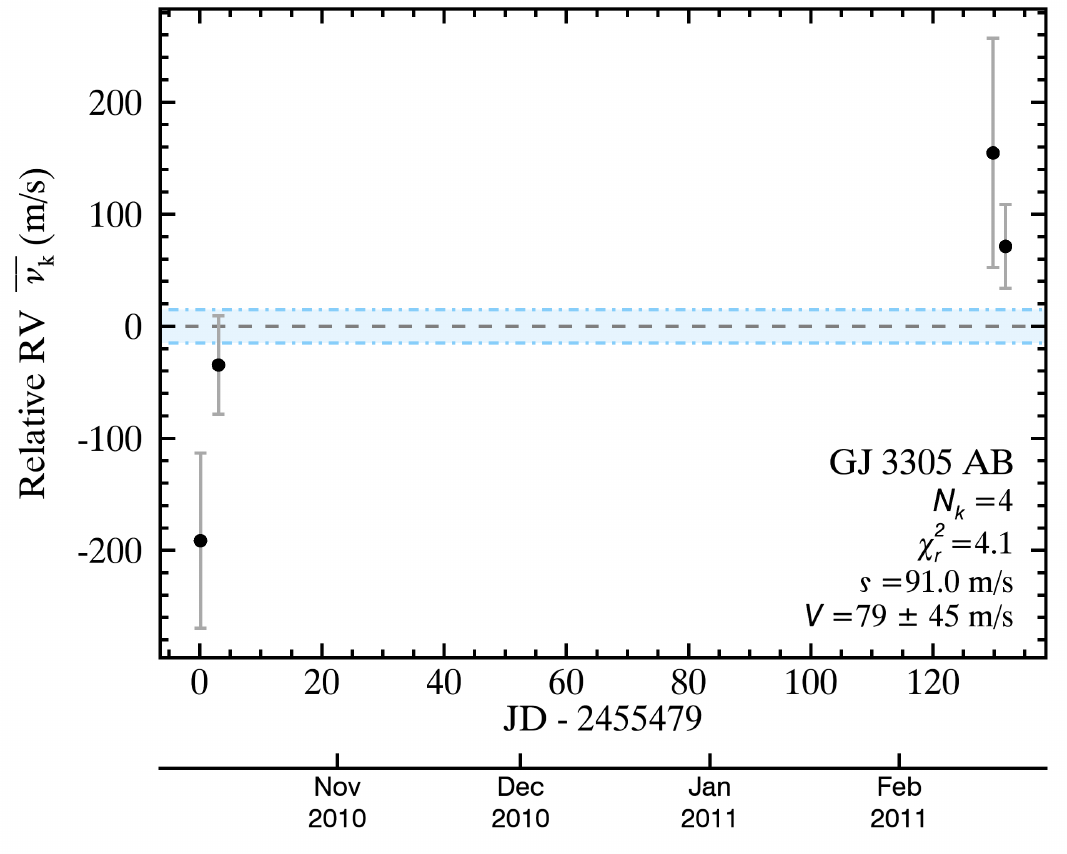}\label{fig:RV_Curves_GJ3305}}
	\subfigure[TYC~5899--26--1]{\includegraphics[width=0.325\textwidth]{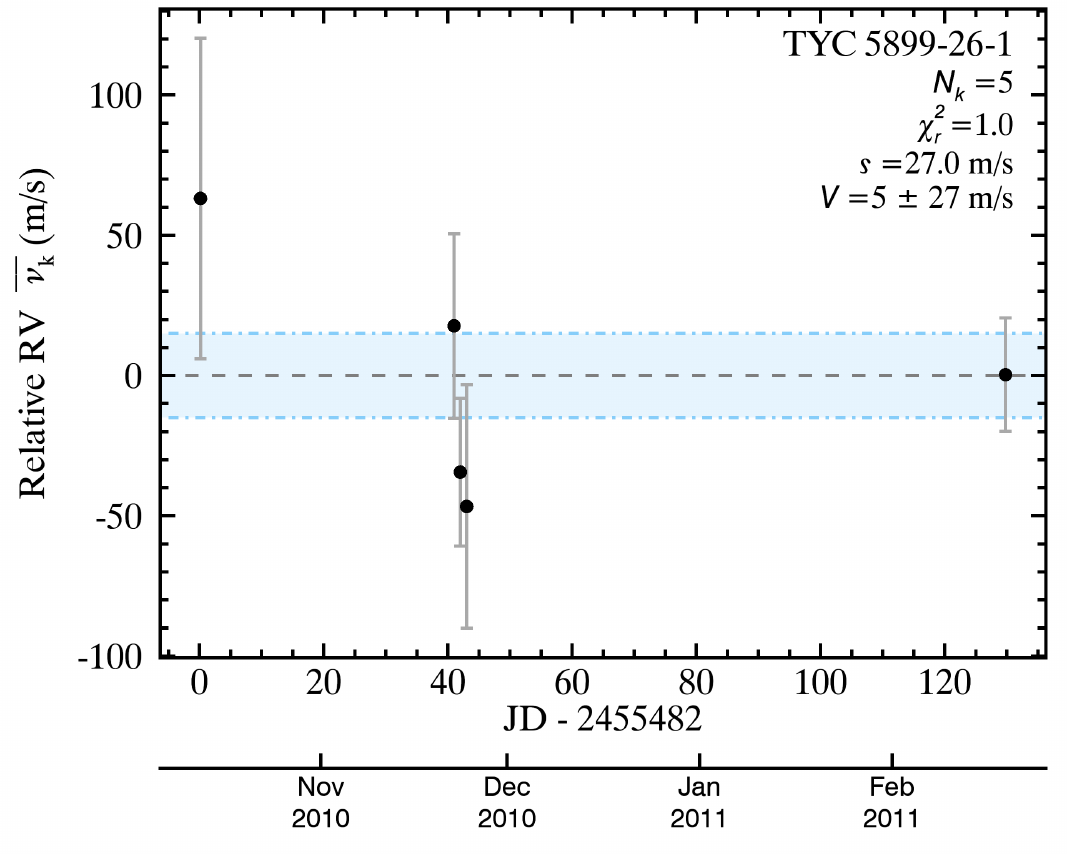}}
	\subfigure[V1005~Ori]{\includegraphics[width=0.325\textwidth]{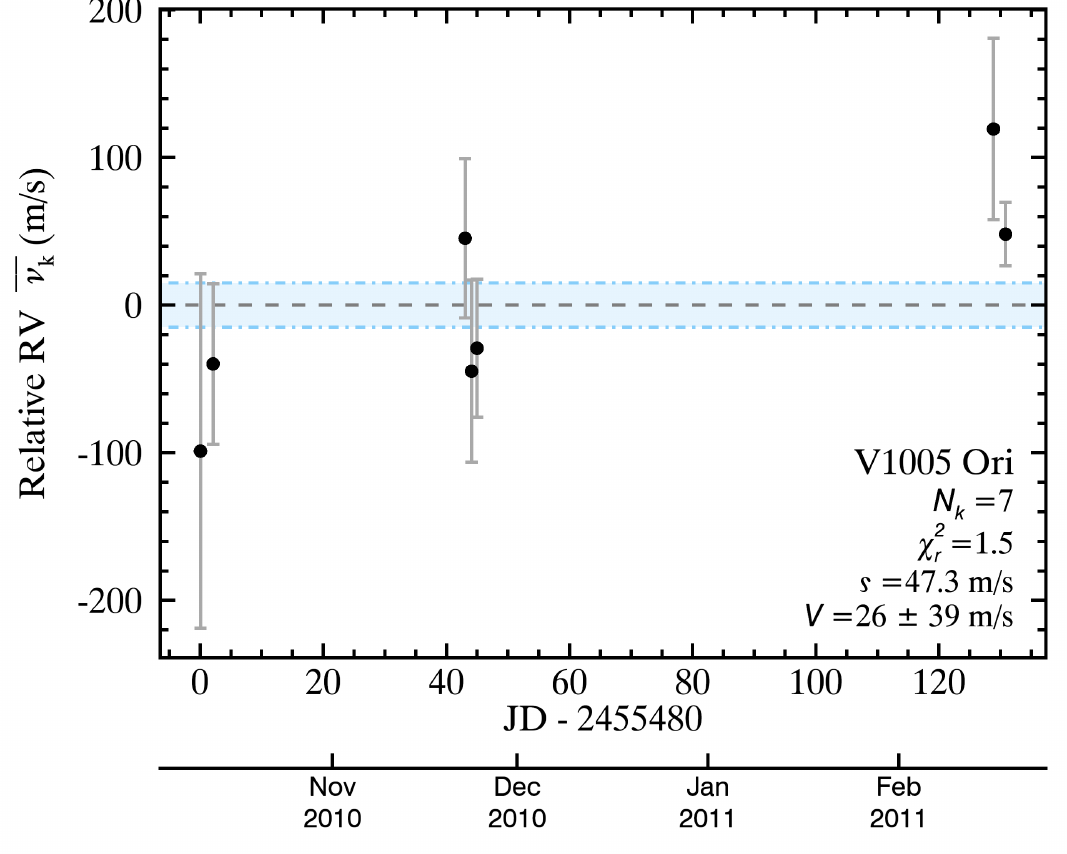}}
	\subfigure[BD+20~1790]{\includegraphics[width=0.325\textwidth]{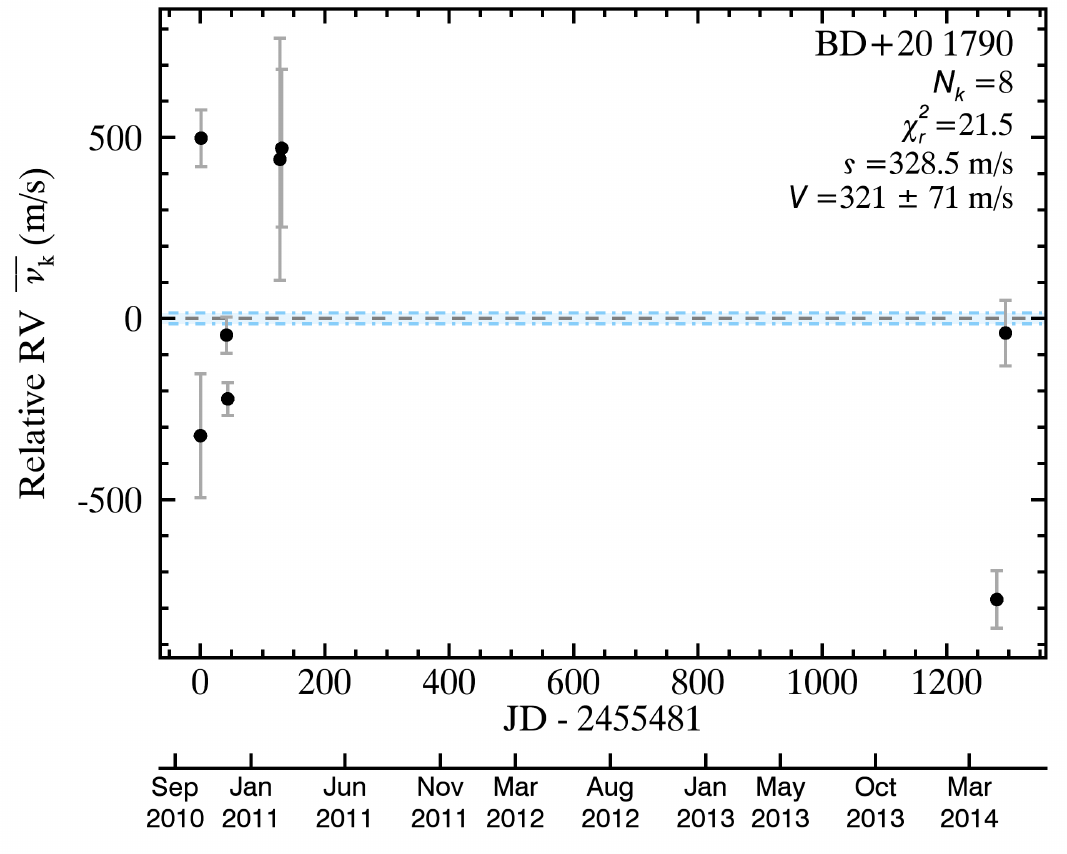}\label{fig:BD2017}}
	\subfigure[BD+01~2447]{\includegraphics[width=0.325\textwidth]{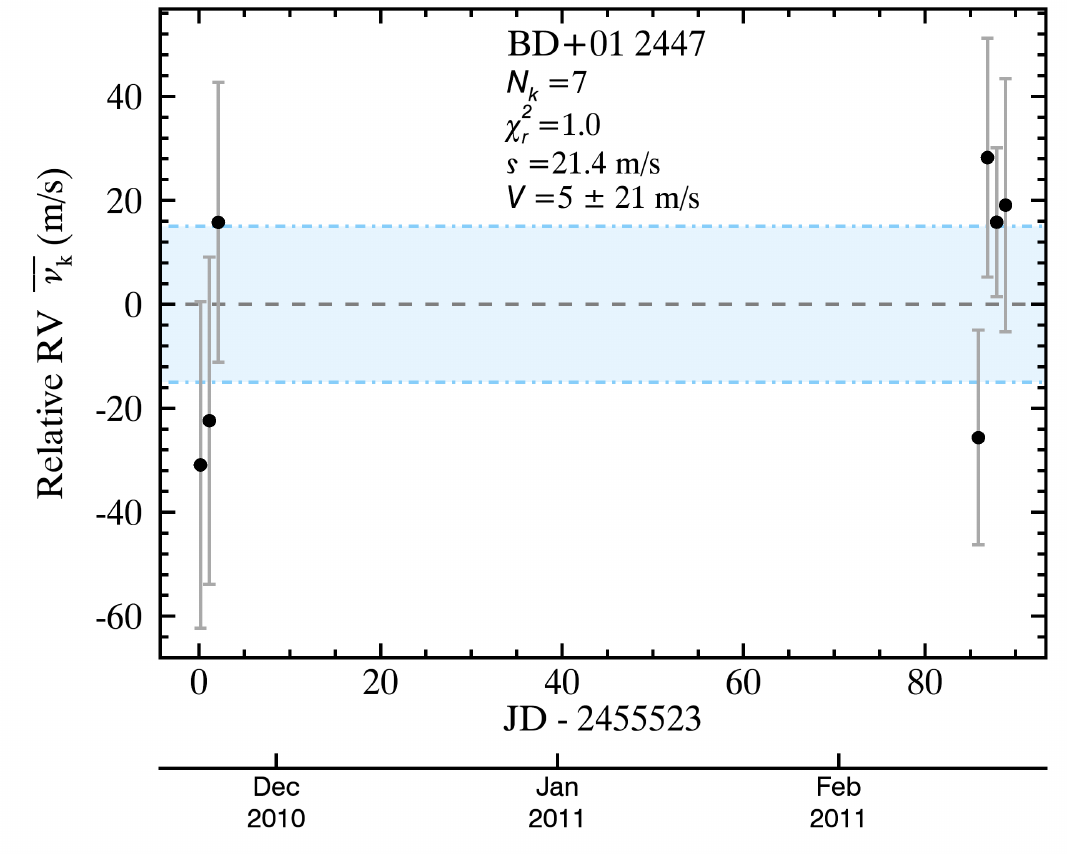}}
	\subfigure[HD~160934~AB]{\includegraphics[width=0.325\textwidth]{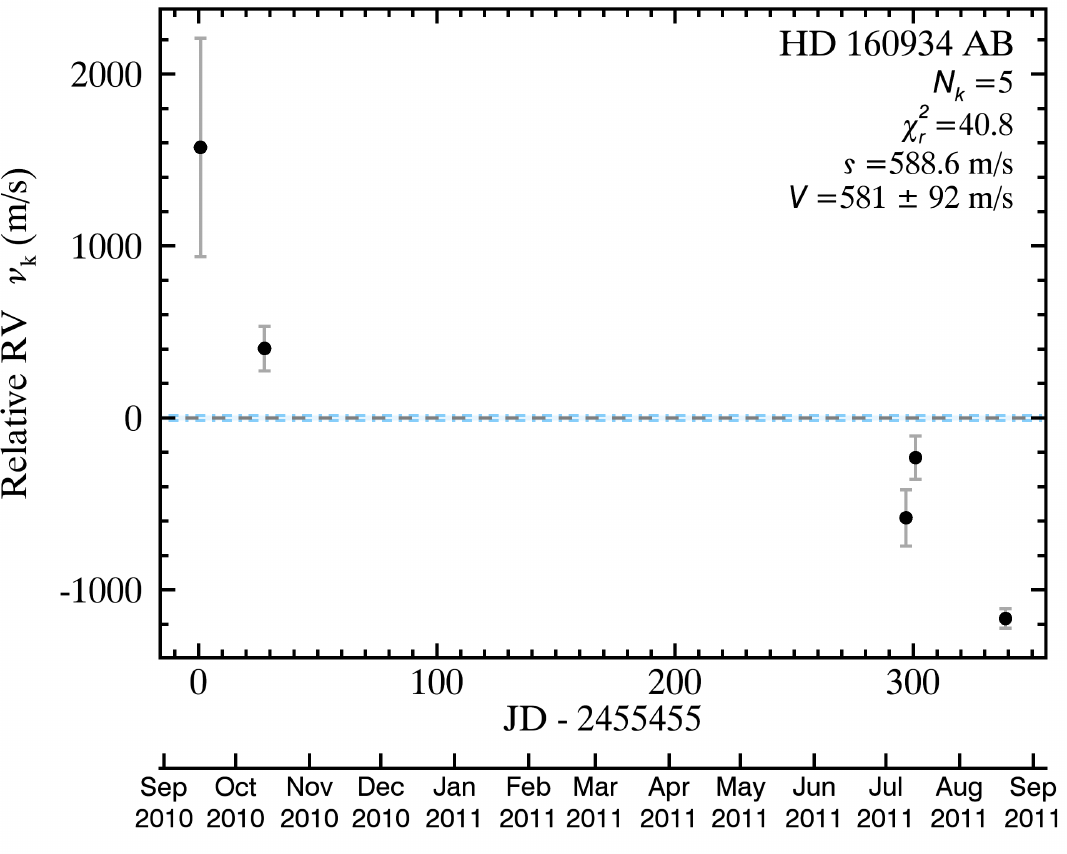}}
	\subfigure[LO~Peg]{\includegraphics[width=0.325\textwidth]{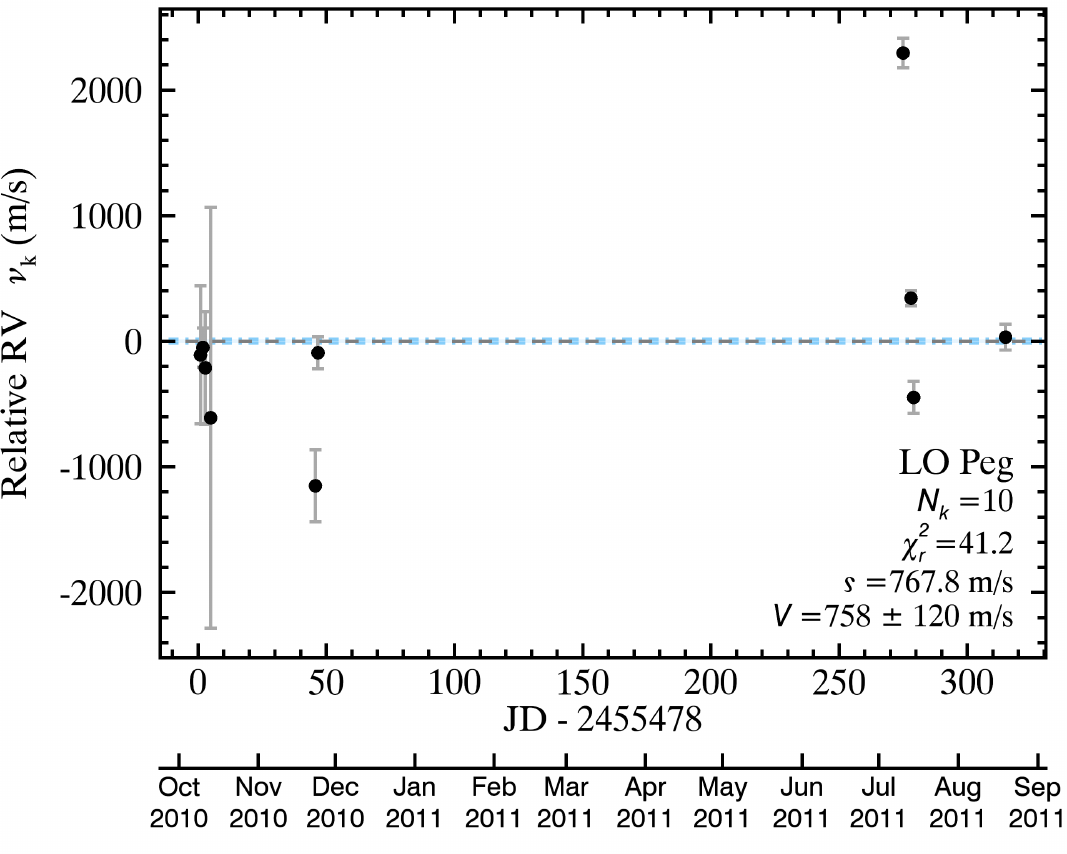}}
	\subfigure[BD--13~6424]{\includegraphics[width=0.325\textwidth]{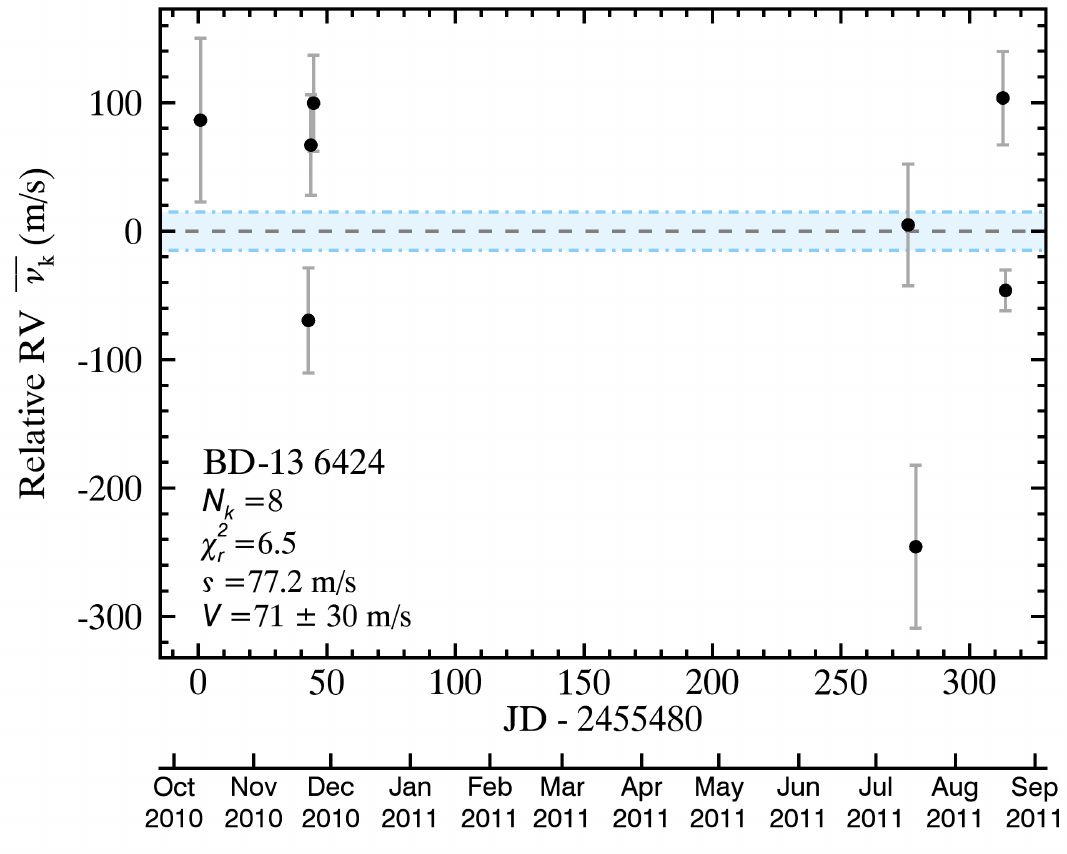}}
	\caption{RV curves of the young target sample. The formatting is identical to that of Figure~\ref{fig:RV_Curves_1}. See Section~\ref{sec:results} for a discussion of the global survey results, and Section~\ref{sec:indivresults} for a discussion of individual targets}
	\label{fig:RV_Curves_2}
\end{figure*}

\begin{figure*}
	\centering
	\subfigure[GJ~15~A]{\includegraphics[width=0.325\textwidth]{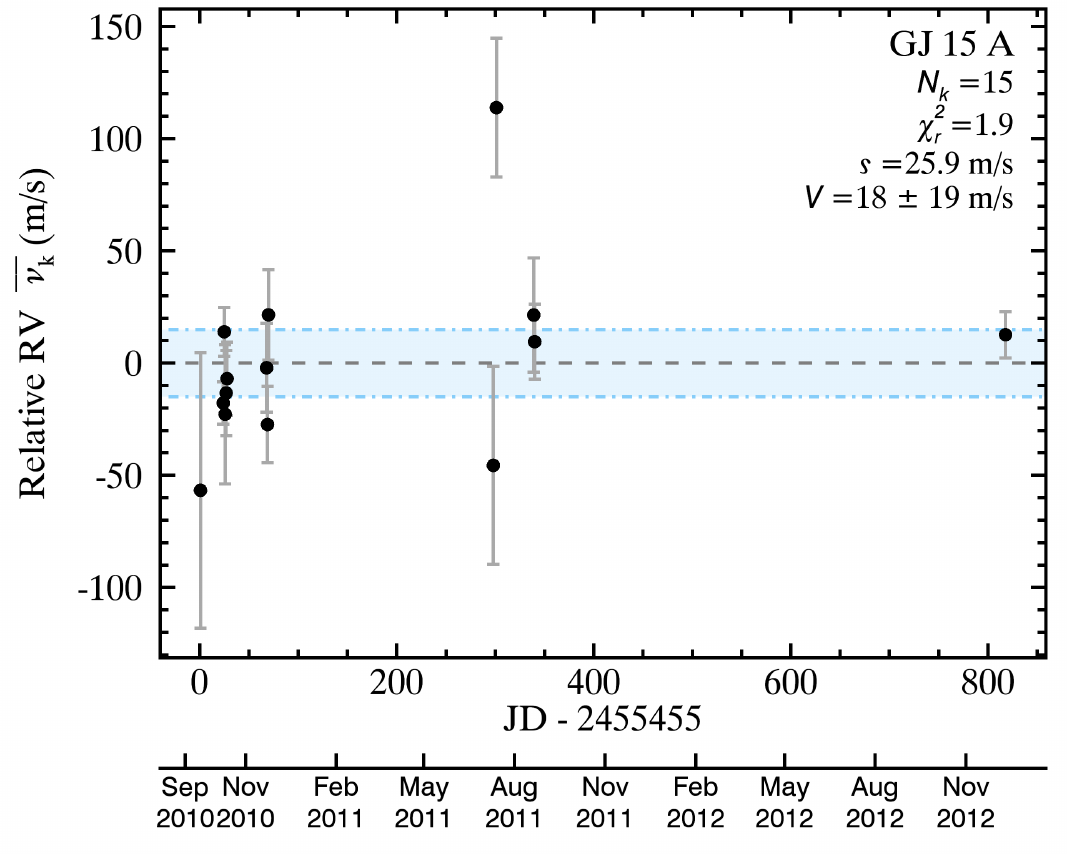}}
	\subfigure[GJ~169]{\includegraphics[width=0.325\textwidth]{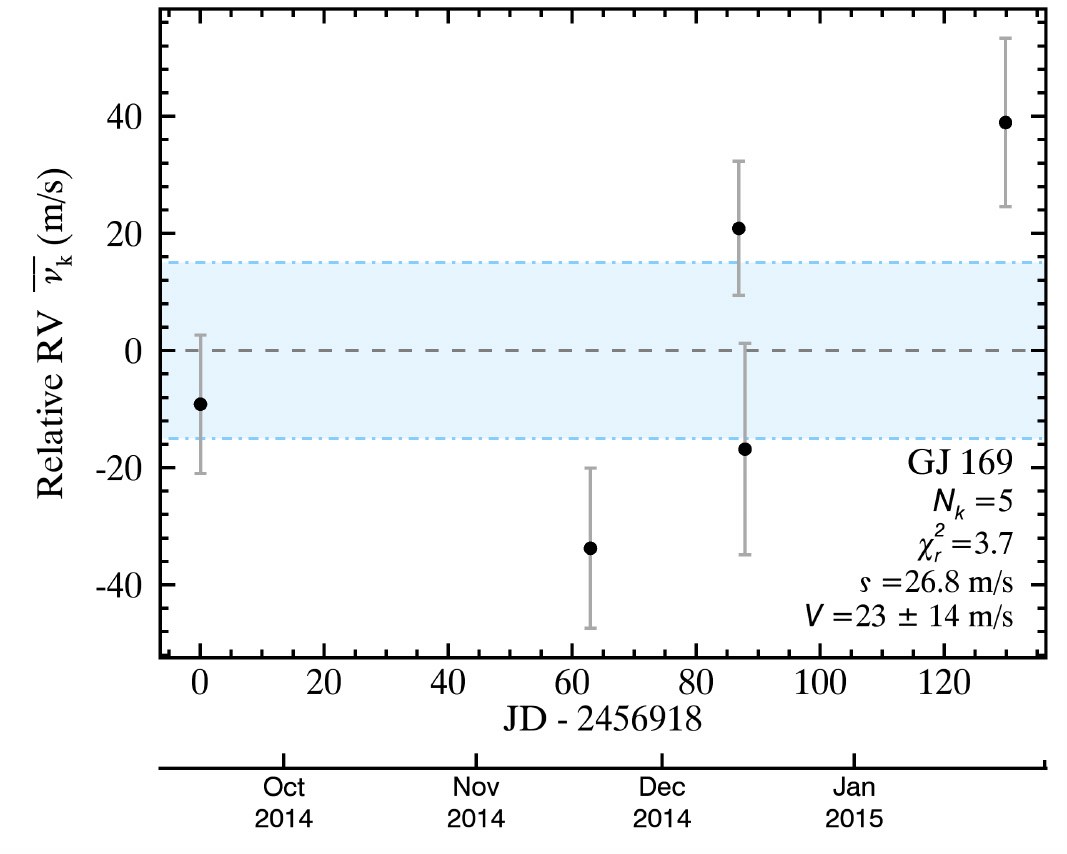}}
	\subfigure[LHS~26]{\includegraphics[width=0.325\textwidth]{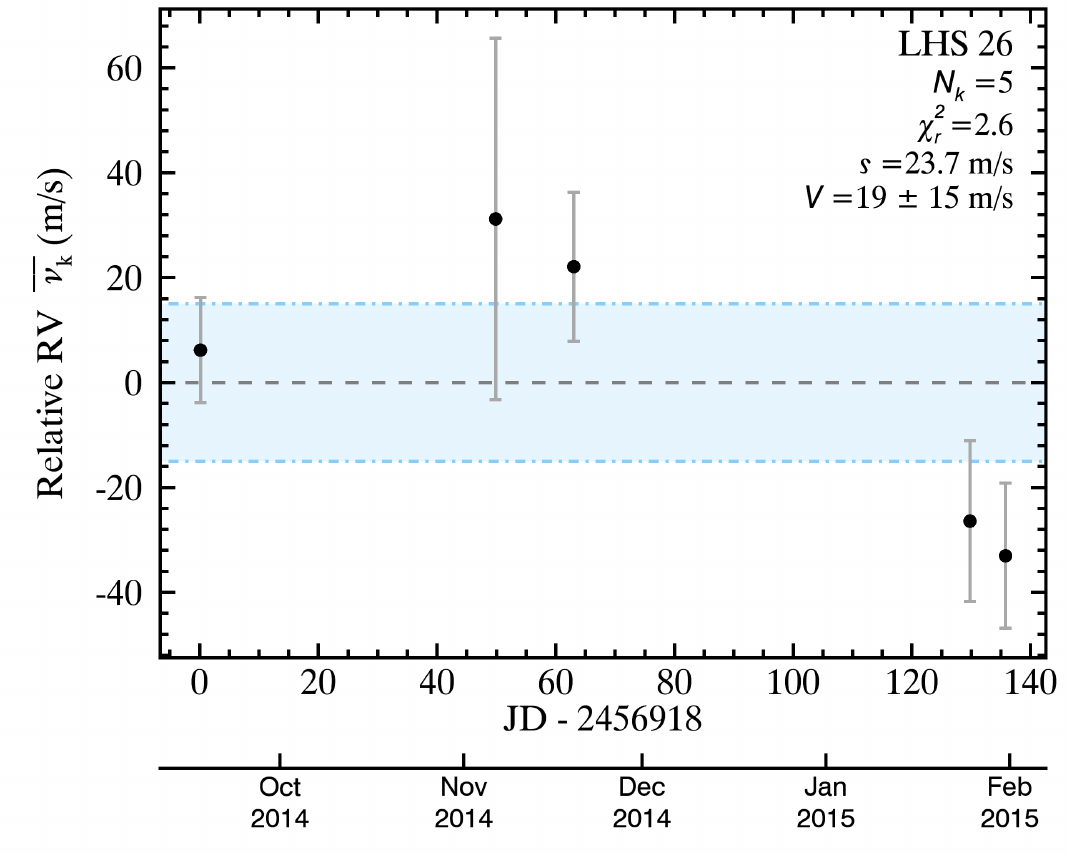}}
	\subfigure[GJ~338~A]{\includegraphics[width=0.325\textwidth]{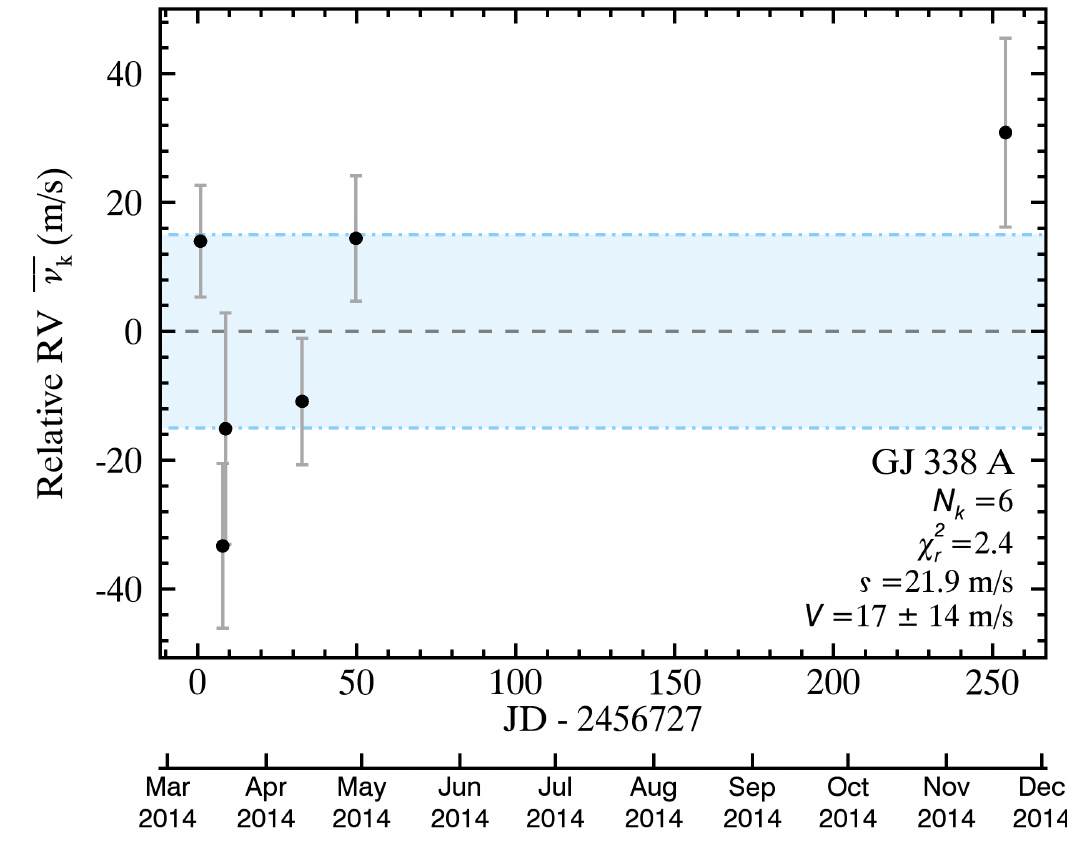}}
	\subfigure[GJ~338~B]{\includegraphics[width=0.325\textwidth]{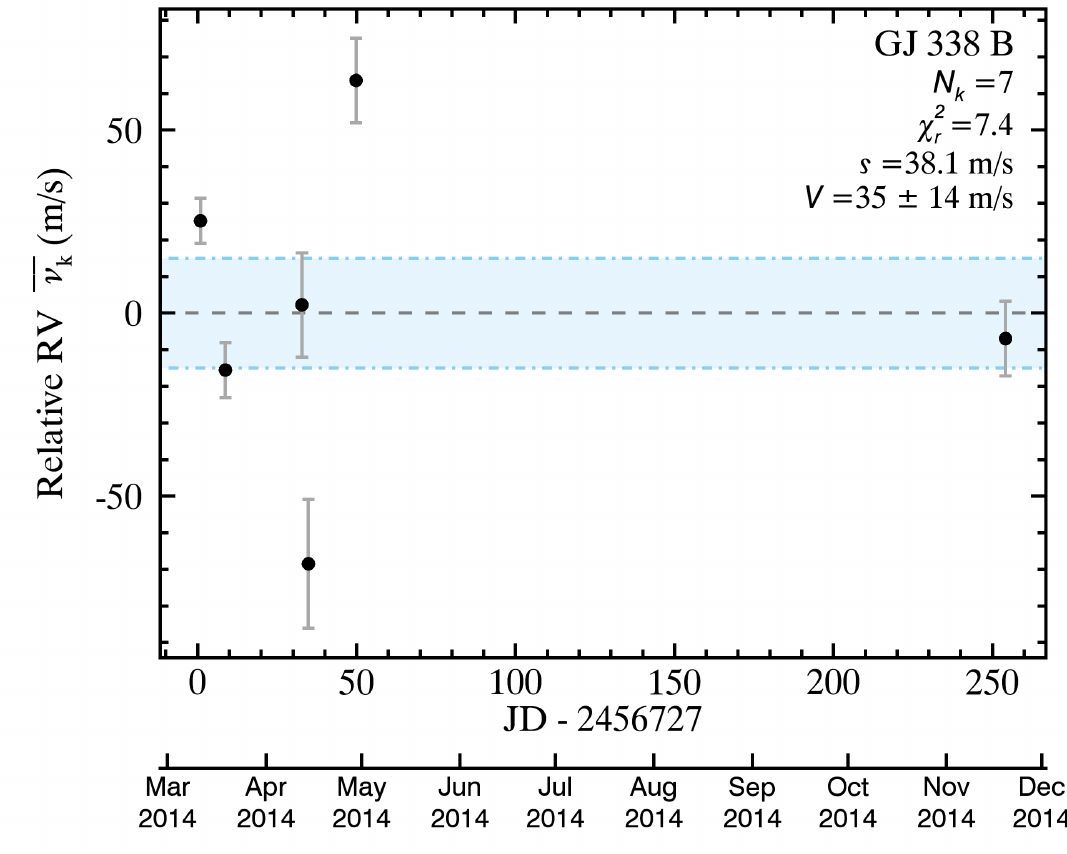}}
	\subfigure[GJ~458~A]{\includegraphics[width=0.325\textwidth]{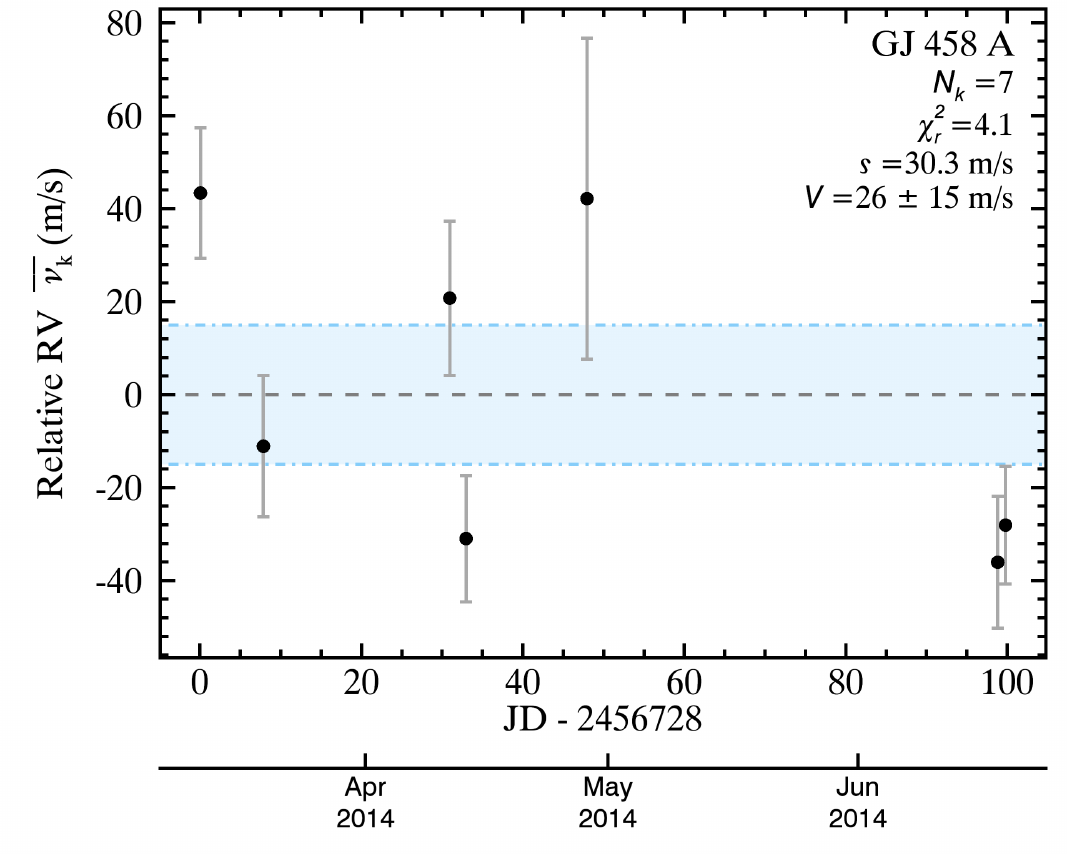}\label{fig:458A}}
	\subfigure[GJ~537~B]{\includegraphics[width=0.325\textwidth]{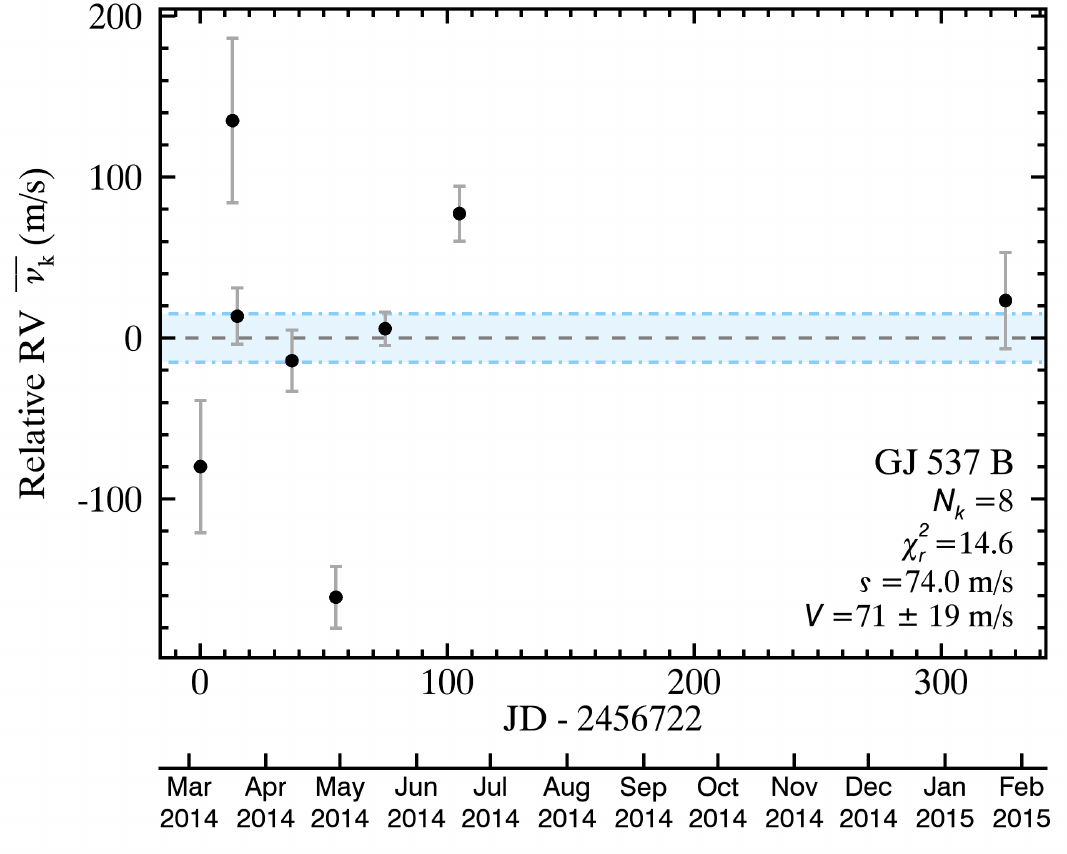}}
	\subfigure[GJ~537~A]{\includegraphics[width=0.325\textwidth]{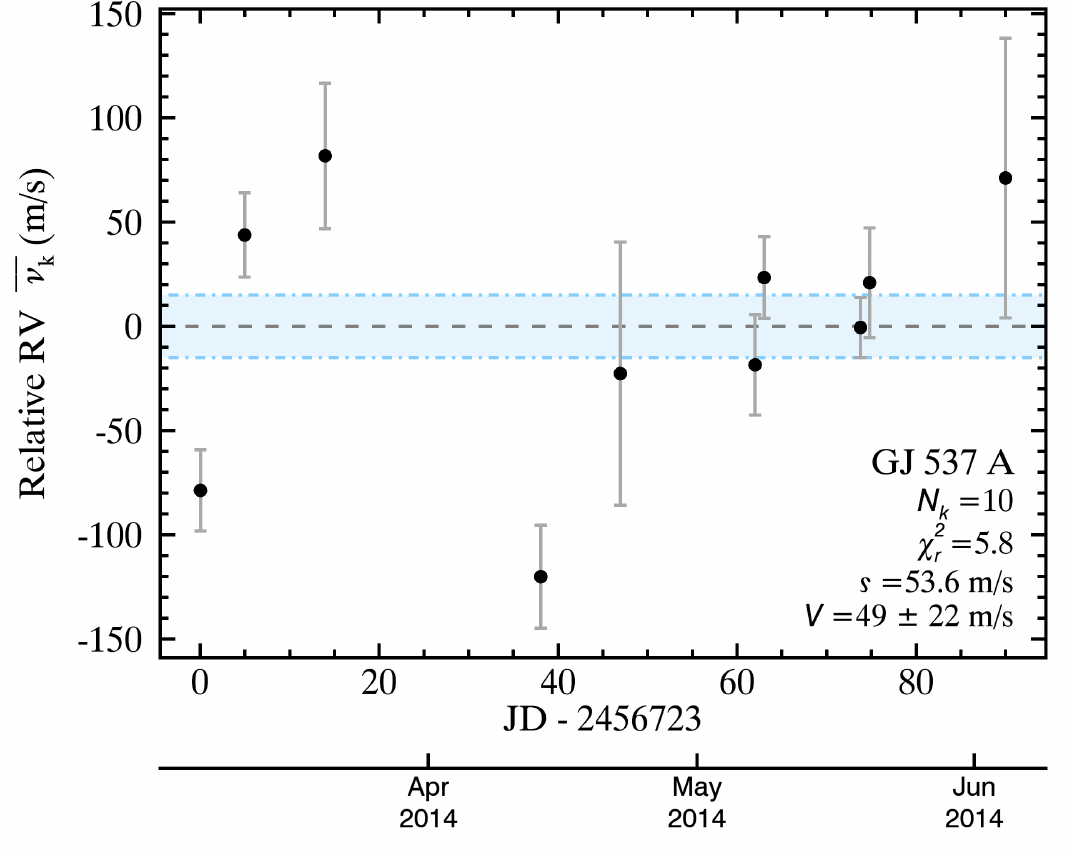}}
	\subfigure[LHS~371]{\includegraphics[width=0.325\textwidth]{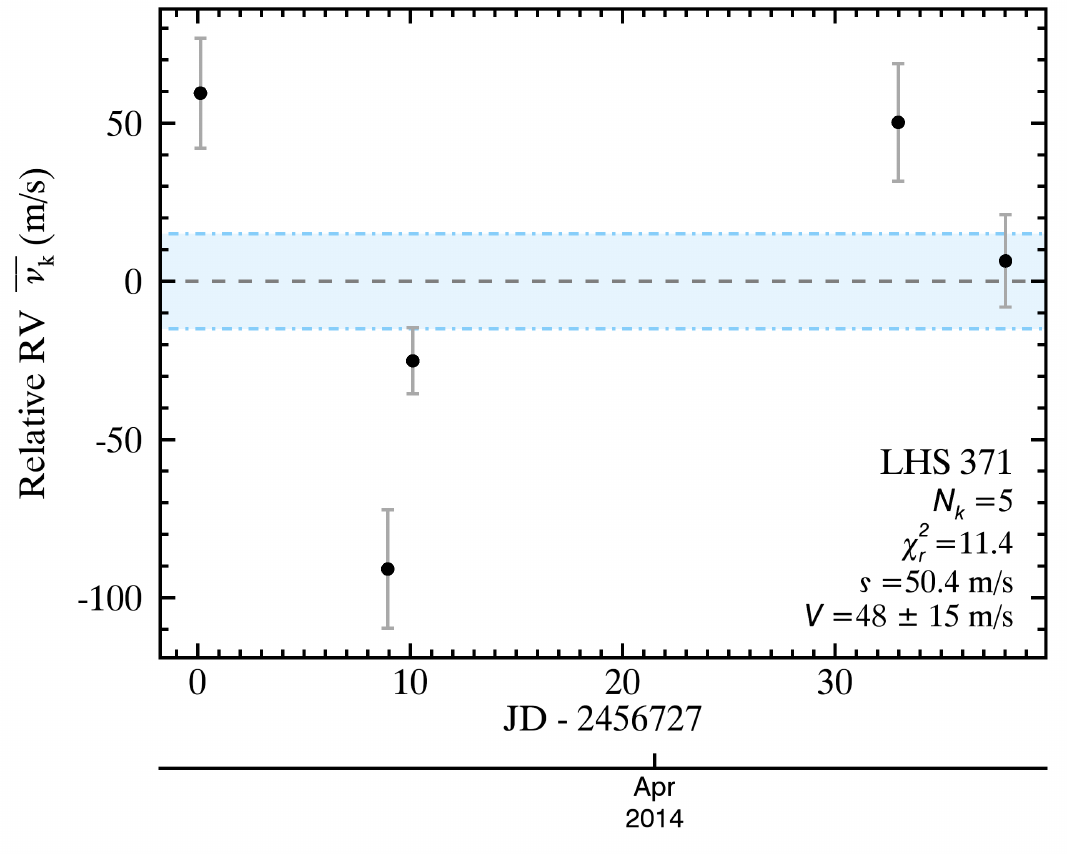}}
	\subfigure[LHS~372]{\includegraphics[width=0.325\textwidth]{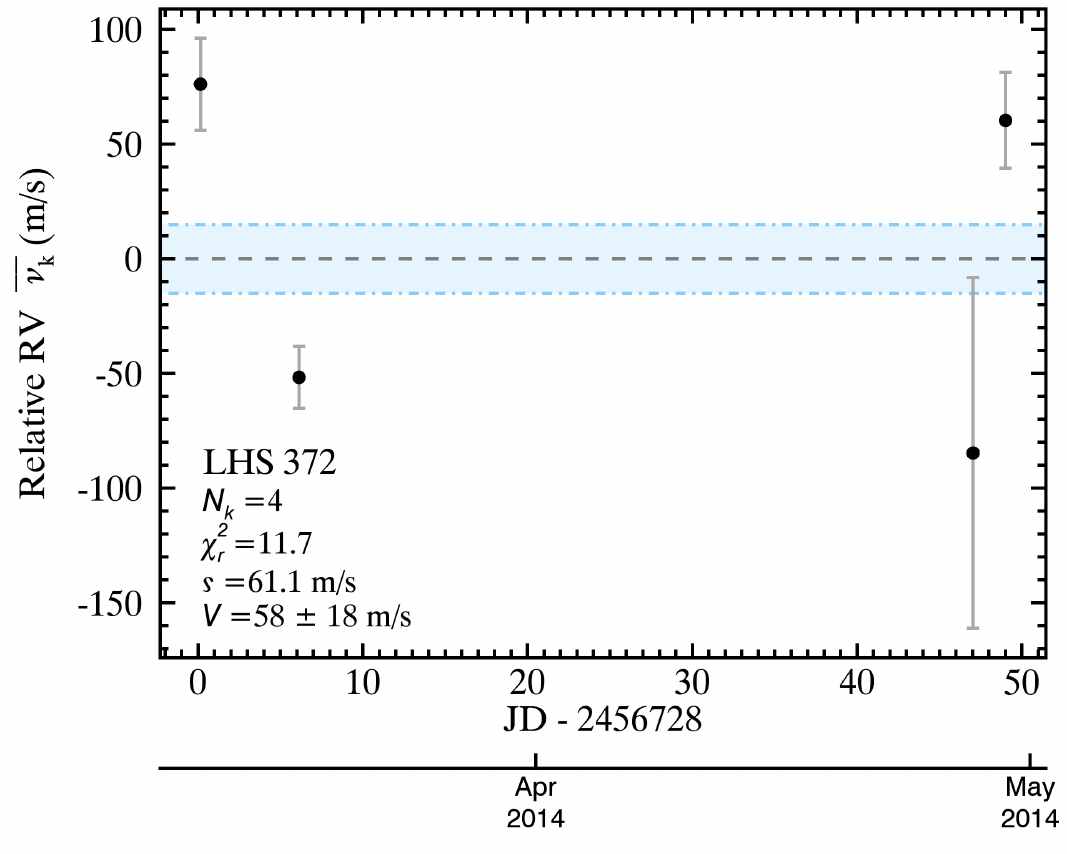}}
	\subfigure[LHS~374]{\includegraphics[width=0.325\textwidth]{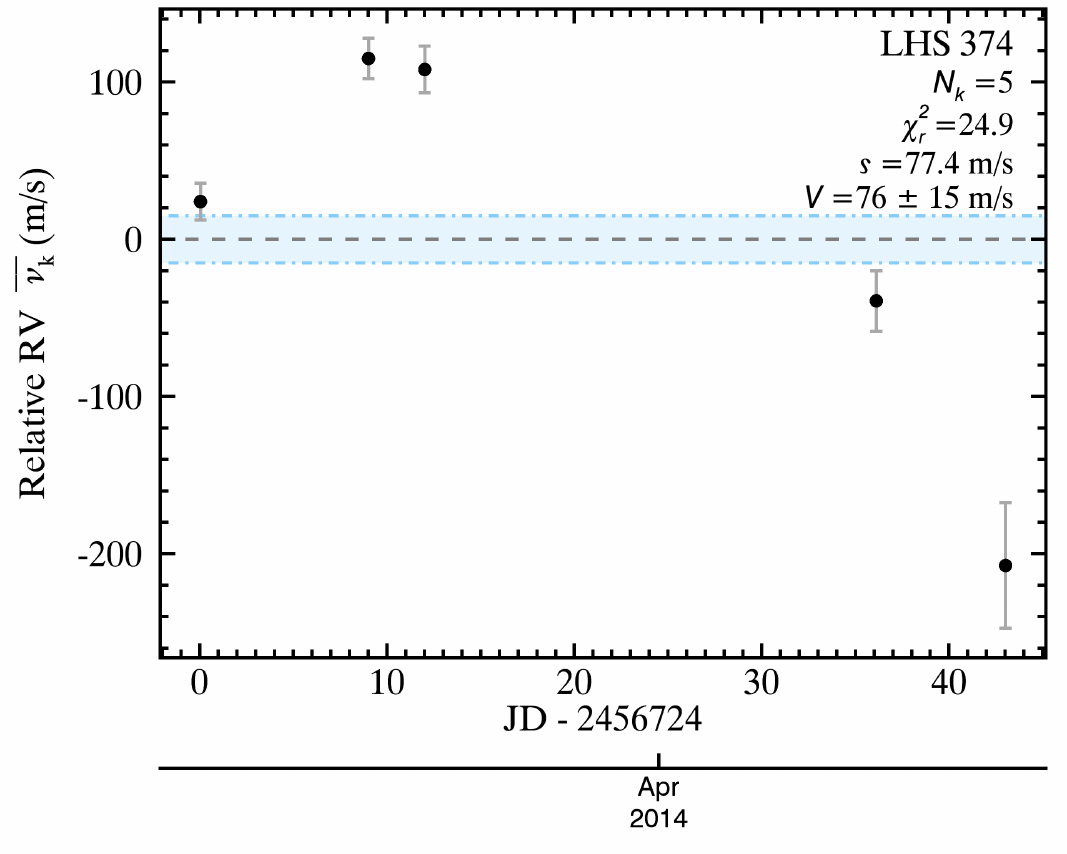}}
	\subfigure[GJ~9520]{\includegraphics[width=0.325\textwidth]{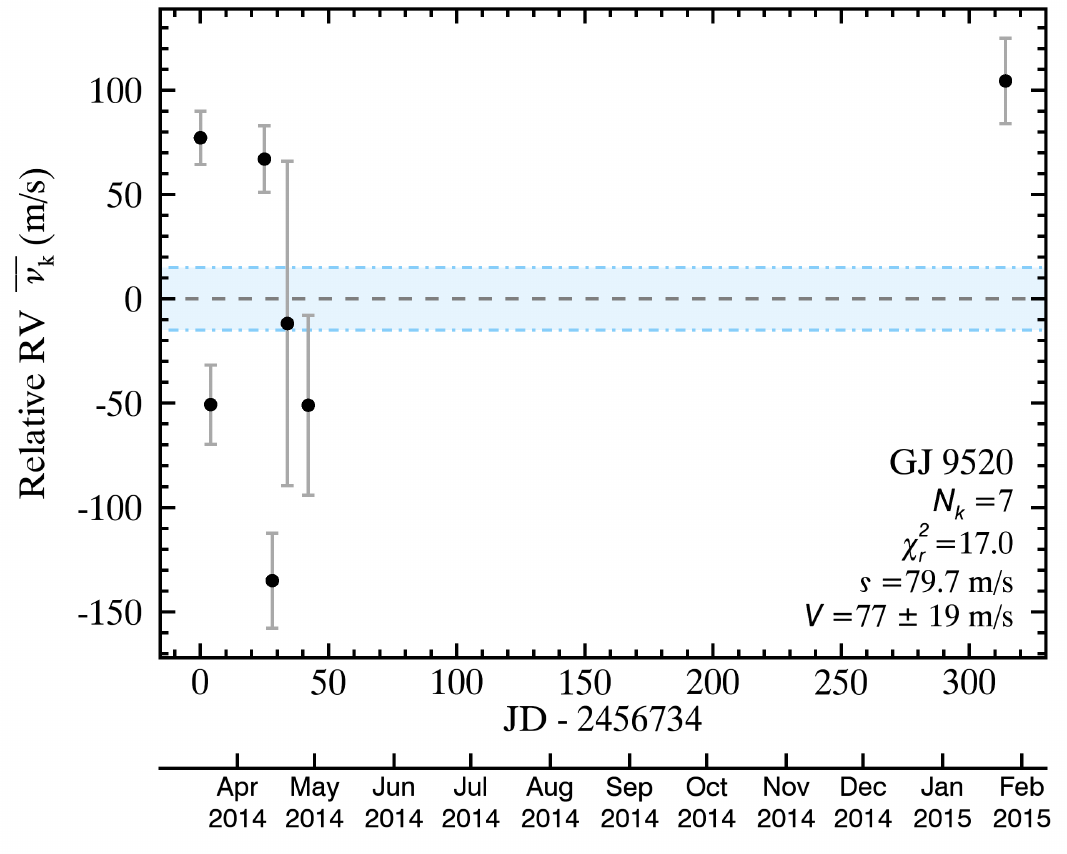}}
	\caption{RV curves of the nearby target sample, part 1 of 2. The formatting is identical to that of Figure~\ref{fig:RV_Curves_1}. See Section~\ref{sec:results} for a discussion of the global survey results, and Section~\ref{sec:indivresults} for a discussion of individual targets}
	\label{fig:RV_Curves_3}
\end{figure*}

\begin{figure*}
	\centering
	\subfigure[GJ~3942]{\includegraphics[width=0.325\textwidth]{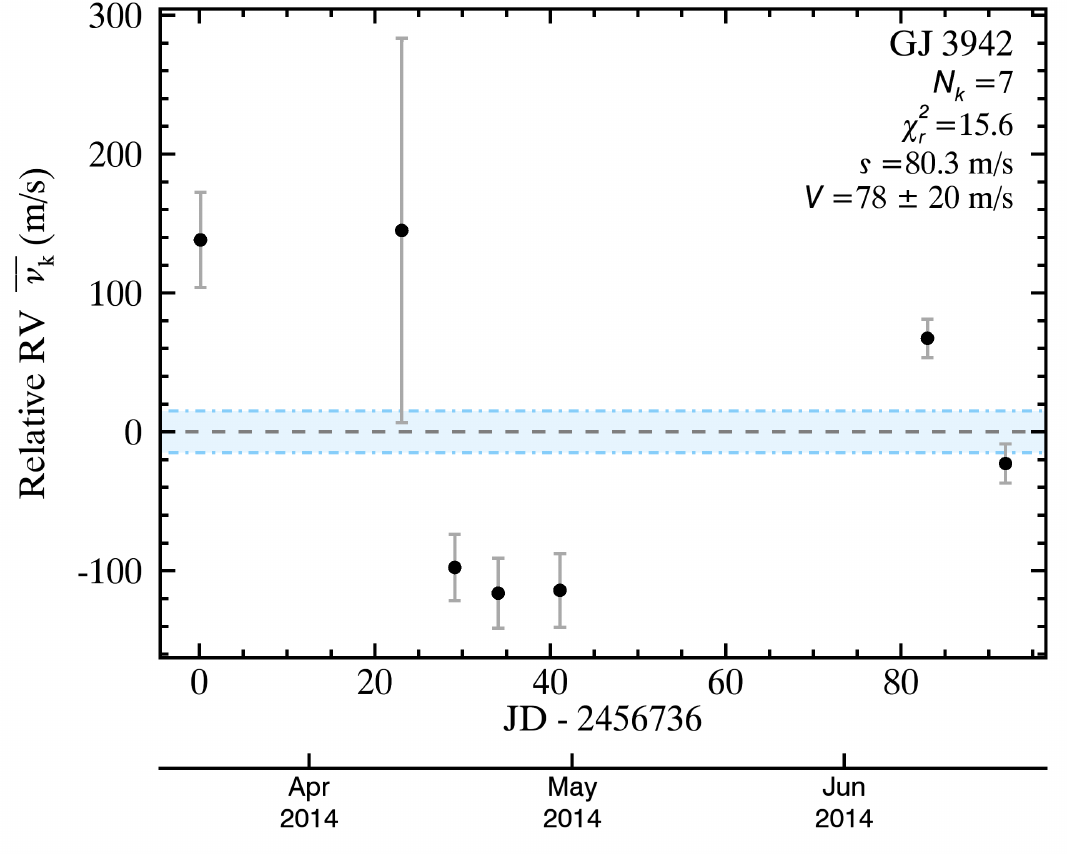}}
	\subfigure[GJ~725~A]{\includegraphics[width=0.325\textwidth]{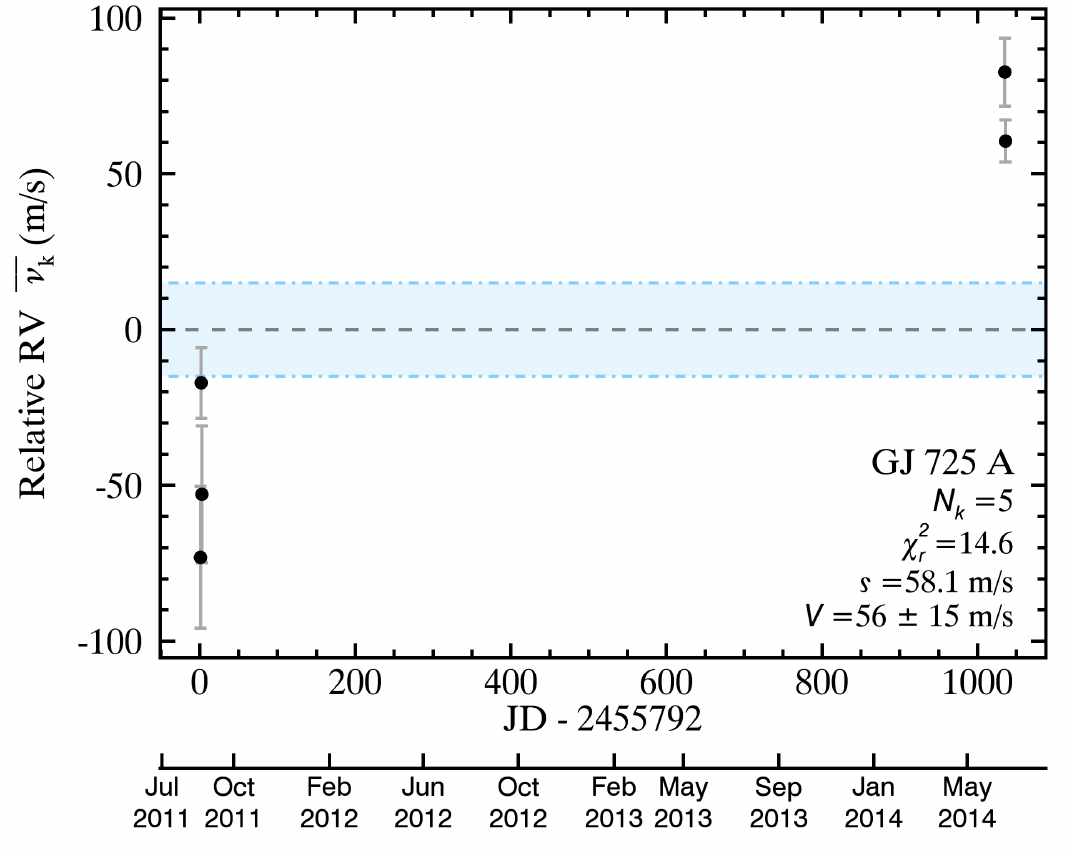}}
	\subfigure[GJ~740]{\includegraphics[width=0.325\textwidth]{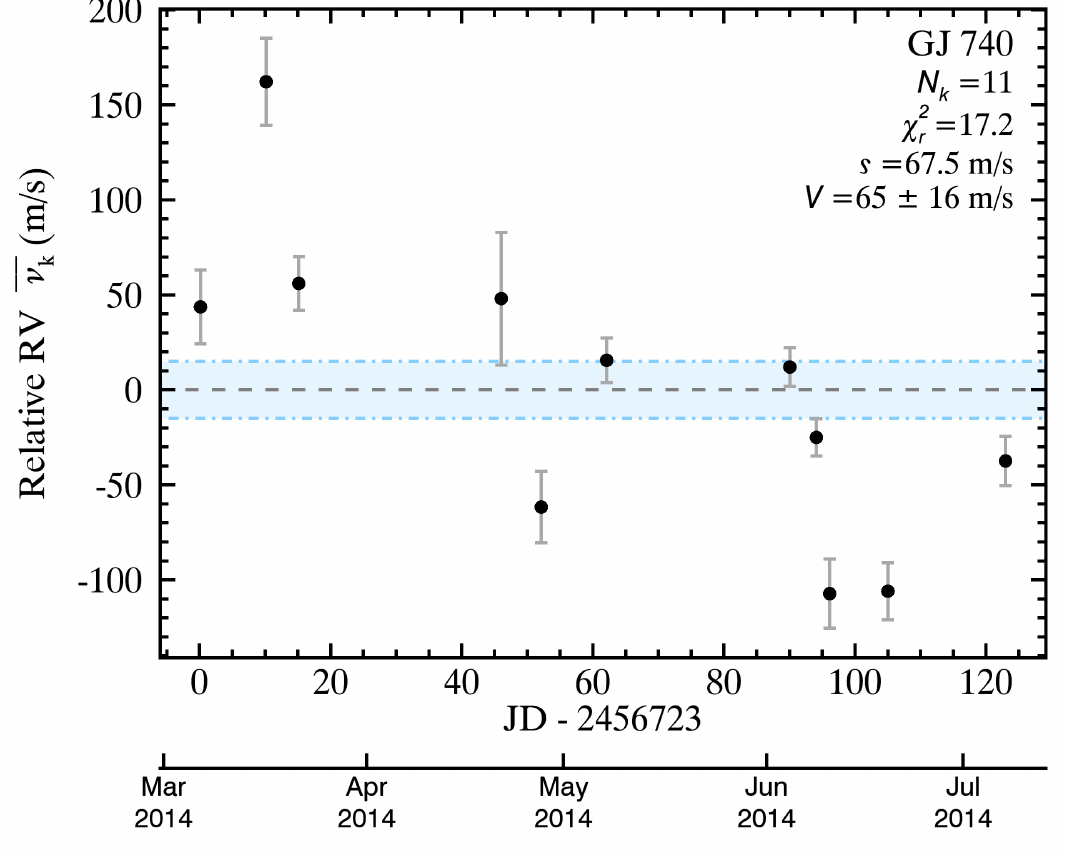}}
	\subfigure[EV~Lac]{\includegraphics[width=0.325\textwidth]{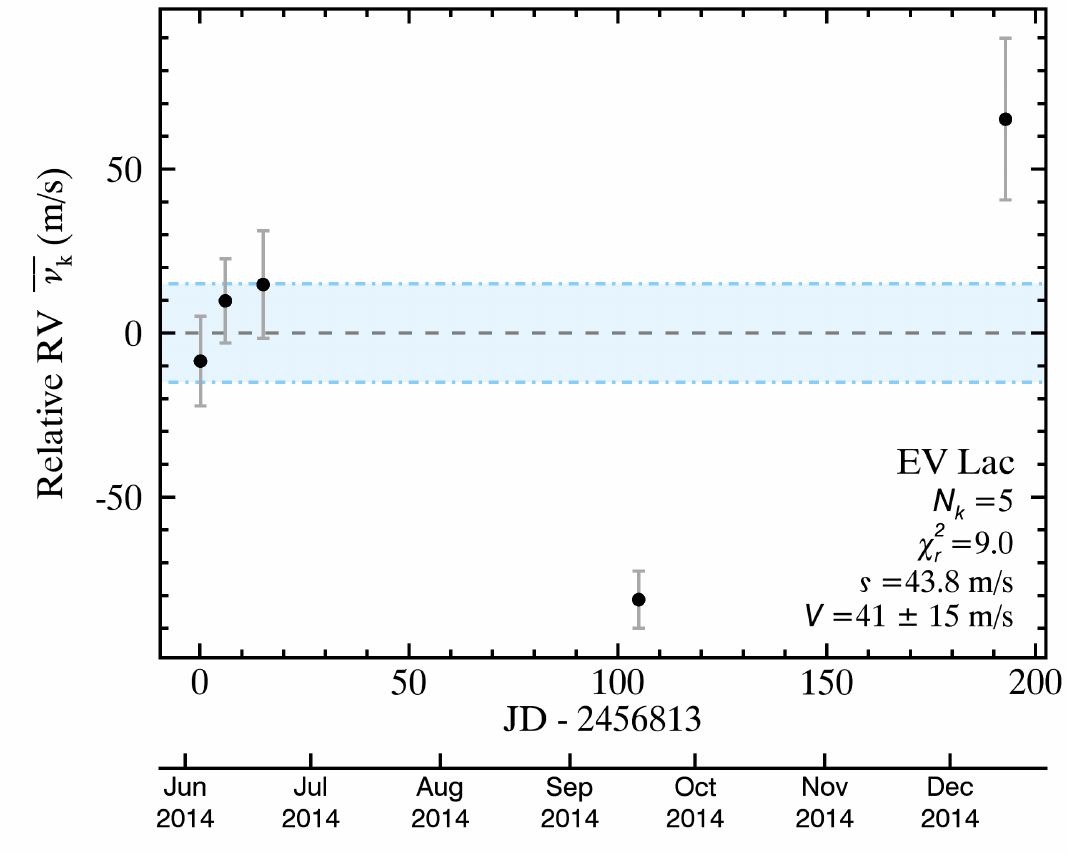}}
	\subfigure[GJ~876]{\includegraphics[width=0.325\textwidth]{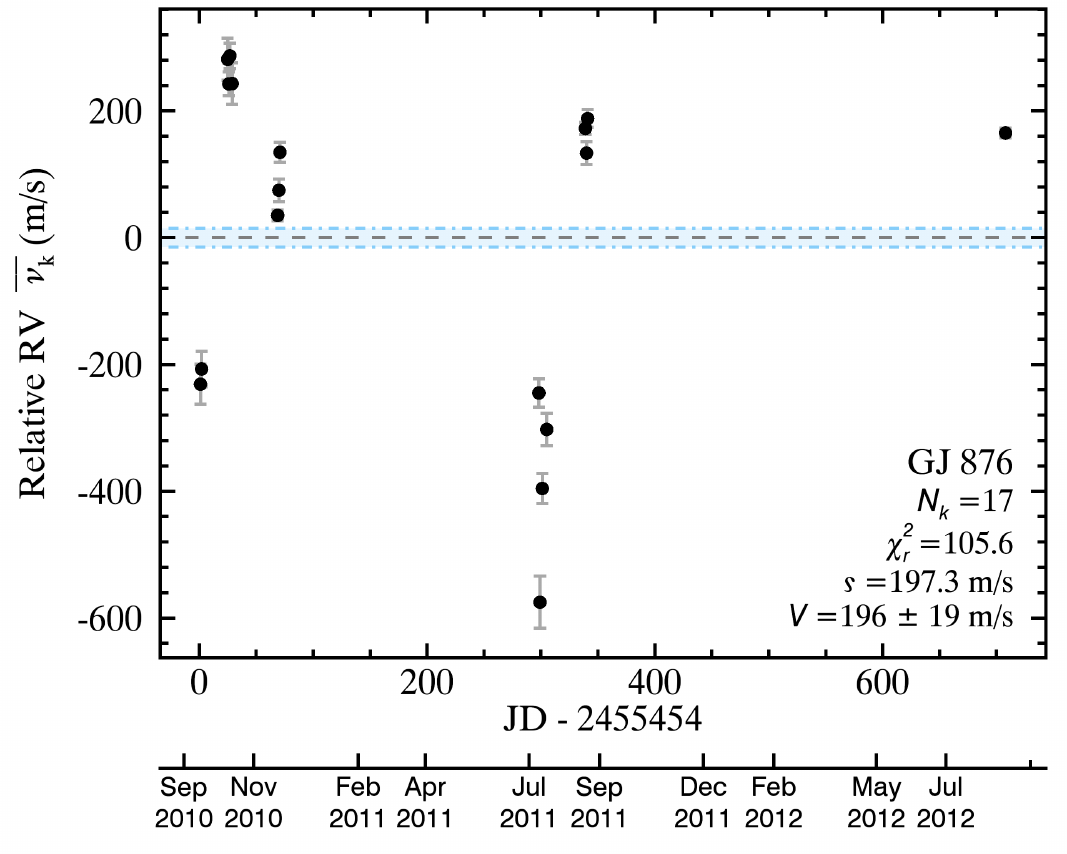}}
	\caption{RV curves of the nearby target sample, part 2 of 2. The formatting is identical to that of Figure~\ref{fig:RV_Curves_1}. See Section~\ref{sec:results} for a discussion of the global survey results, and Section~\ref{sec:indivresults} for a discussion of individual targets}
	\label{fig:RV_Curves_4}
\end{figure*}

The 248 individual per-night RV measurements ($\bar{\nu}_k$, $\sigma_k$) and epochs that were accumulated in this survey are listed in Table~\ref{tab:detrv}, and the associated RV curves are presented in Figures~\ref{fig:RV_Curves_1} through \ref{fig:RV_Curves_4}. The dominant cause of the variation in error bars for the per-night RVs presented in these figures is the varying S/N ratios most often associated with weather or a varying number of total spectra.

We achieved short-term RV precisions (within a single night) as low as $\sim$\,8--15\,\ms\ for higher-S/N observations, which represents a net improvement over previous studies that used similar small facilities (e.g., $\approx$\,45--90\,\ms; \citealp{2012ApJ...749...16B,2015AJ....149..106D}), but that did not benefit from a methane gas cell or \replaced{the}{our} novel RV extraction method\deleted{ presented by P.~Gao et al. (submitted to PASP)}. These precisions are almost comparable to what is achieved using 8\,m-class telescopes with a spectral grasp $\sim$\,10 times larger, although we can only achieve similar S/N on brighter targets (e.g., 5--10\,\ms\ on targets with $4.4 < K_S < 8.0$; \citealt{2010ApJ...713..410B}).

\tabletypesize{\normalsize}
\begin{deluxetable}{lr@{\hspace{1pt}}c@{\hspace{1pt}}lr@{\hspace{1pt}}c@{\hspace{1pt}}lc}
\tablecolumns{8}
\tablecaption{Detailed Radial Velocity Measurements \label{tab:detrv}}
\tablehead{\colhead{Target} & \multicolumn{3}{c}{Red. Julian Date} & \multicolumn{3}{c}{Relative} & \colhead{\added{S/N}}\\
\colhead{Name} & \multicolumn{3}{c}{JD$-2400000$} & \multicolumn{3}{c}{RV (\ms)} & \colhead{Ratio}}
\startdata
\textbf{AG Tri} & 55479.003 & $ \pm $ & 0.003 & 107 & $ \pm $ & 51 & 23\\
& 55480.930 & $ \pm $ & 0.004 & -139 & $ \pm $ & 50 & 16\\
& 55482.015 & $ \pm $ & 0.006 & 10 & $ \pm $ & 27 & 23\\
& 55522.875 & $ \pm $ & 0.003 & 198 & $ \pm $ & 43 & 22\\
& 55523.855 & $ \pm $ & 0.003 & 76 & $ \pm $ & 45 & 26\\
& 55524.871 & $ \pm $ & 0.003 & -76 & $ \pm $ & 77 & 20\\
& 55608.734 & $ \pm $ & 0.004 & 64 & $ \pm $ & 103 & 22\\
& 55611.739 & $ \pm $ & 0.003 & 29 & $ \pm $ & 50 & 25\\
& 55759.126 & $ \pm $ & 0.004 & 26 & $ \pm $ & 48 & 20\\
& 55793.018 & $ \pm $ & 0.004 & -77 & $ \pm $ & 22 & 28\\
\hline
\textbf{AT Mic A} & 55479.735 & $ \pm $ & 0.001 & 34 & $ \pm $ & 25 & 39\\
& 55480.729 & $ \pm $ & 0.004 & 0 & $ \pm $ & 40 & 35\\
& 55482.776 & $ \pm $ & 0.001 & -15 & $ \pm $ & 59 & 28\\
& 55523.713 & $ \pm $ & 0.001 & -97 & $ \pm $ & 118 & 26\\\enddata
\tablecomments{See Section~\ref{sec:results} for more details. \added{Table~\ref{tab:detrv} is published in its entirety in the machine-readable format. A portion is shown here for guidance regarding its form and content.}}
\end{deluxetable}
\tabletypesize{\small}

In Table~\ref{tab:rvs}, we present the distributions of $\varsigma$, $\chi_r^2$, $V$, $S$\added{, $V_{\rm max}\big(N_\varsigma=1\big)$ and $V_{\rm max}\big(N_\varsigma=3\big)$} for the two survey samples. The methane gas cell and RV extraction method that we used allowed us to achieve long-term RV precisions of $S \sim$\,15--50\,\ms, which represents an improvement of a factor $\gtrsim$\,2 over similar NIR surveys that use small observing facilities. \added{It can be noted that we obtain $\chi_r^2$ values above 1 for the majority of targets in our survey. This is expected, as the instrumental stability of CSHELL introduces a systematic RV uncertainty on the combined RV measurements $\bar{\nu}_k$ that is not captured by their error bars \replaced{$\sigma_k^2$}{$\sigma_k$}, which are determined from the weighted standard deviation of single-exposure RV measurements within one night (see Section~\ref{sec:rvcomb}). As a result, a $\chi_r^2$ value above one alone does not imply that a given target is an RV variable. Hence, the survey targets must be compared relative to one another \added{in order }to determine which ones are most likely RV variables. It should also be noted that our survey is expected to have a larger number of RV variables than a blind survey, as it is biased toward young and active stars (see Section~\ref{sec:sample}).}

\begin{figure*}
	\centering
	\subfigure[Reduced chi-square value versus number of nights]{\includegraphics[width=0.495\textwidth]{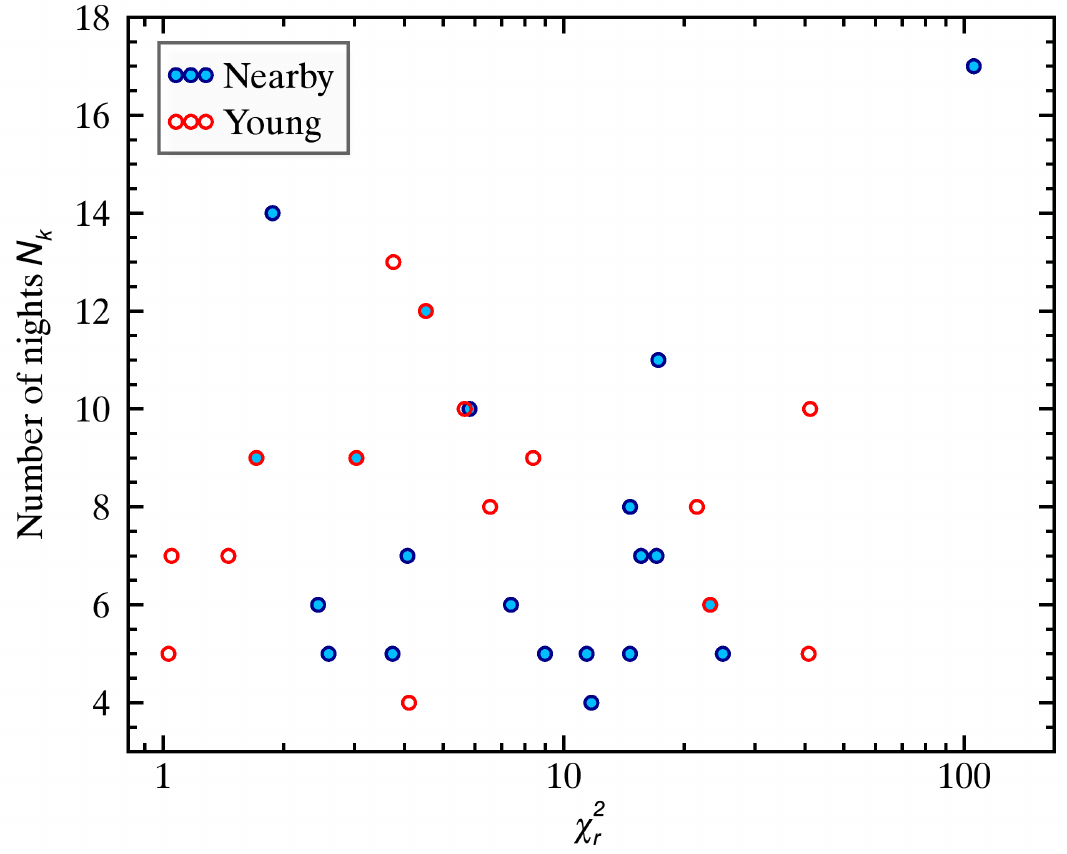}}
	\subfigure[RV scatter versus number of nights]{\includegraphics[width=0.495\textwidth]{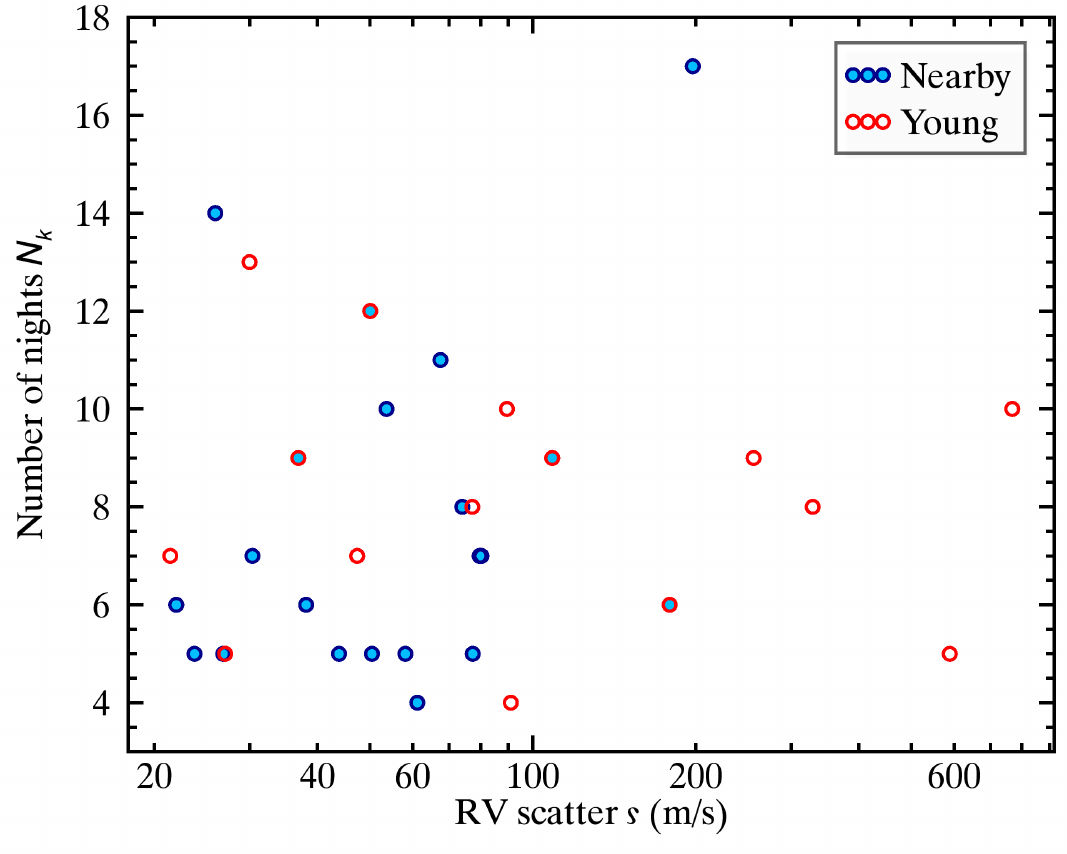}}
	\caption{Reduced $\chi^2$ with respect to zero variation and RV scatter $\sigma$ as a function of the total number of nights for which a target was observed. Young sample targets are displayed as red circles and nearby sample targets as filled blue circles. For more details, see Section~\ref{sec:ensemble}.}
	\label{fig:Chi2RMS_Night}
\end{figure*}

\begin{figure*}
	\centering
	\subfigure[Reduced chi-square value versus $K$-band magnitude]{\includegraphics[width=0.495\textwidth]{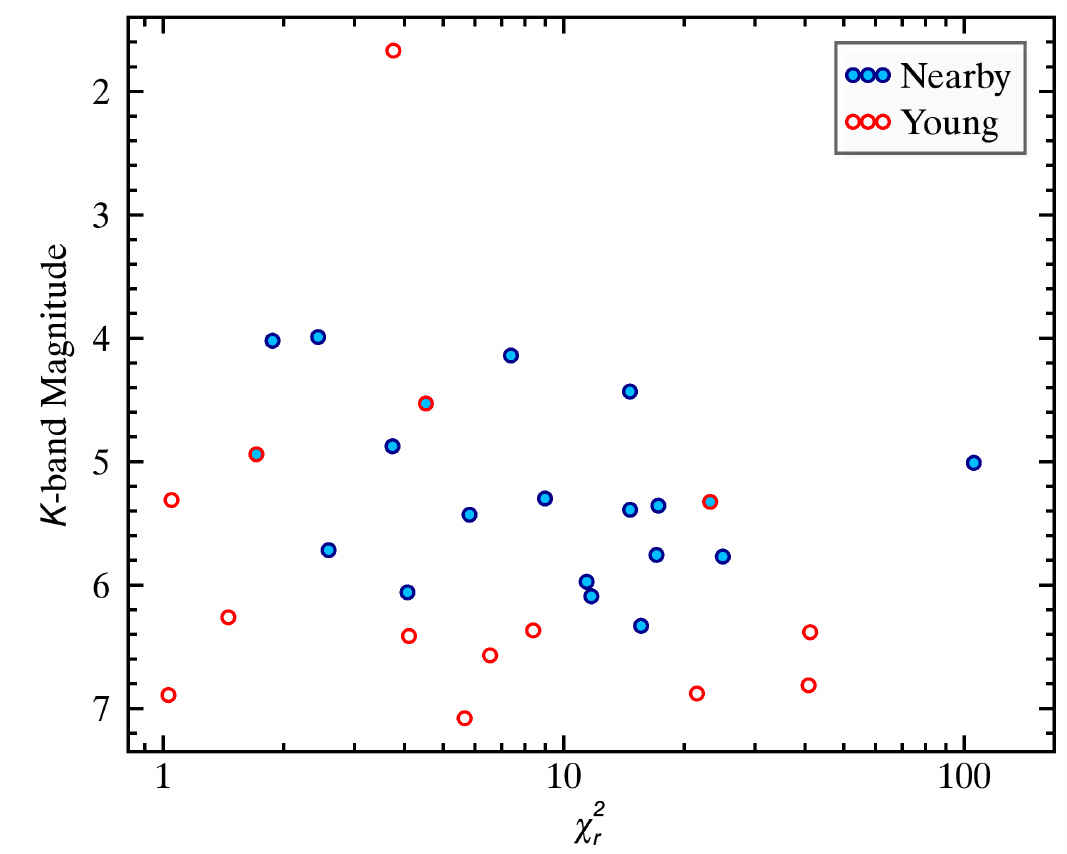}}
	\subfigure[RV scatter versus $K$-band magnitude]{\includegraphics[width=0.495\textwidth]{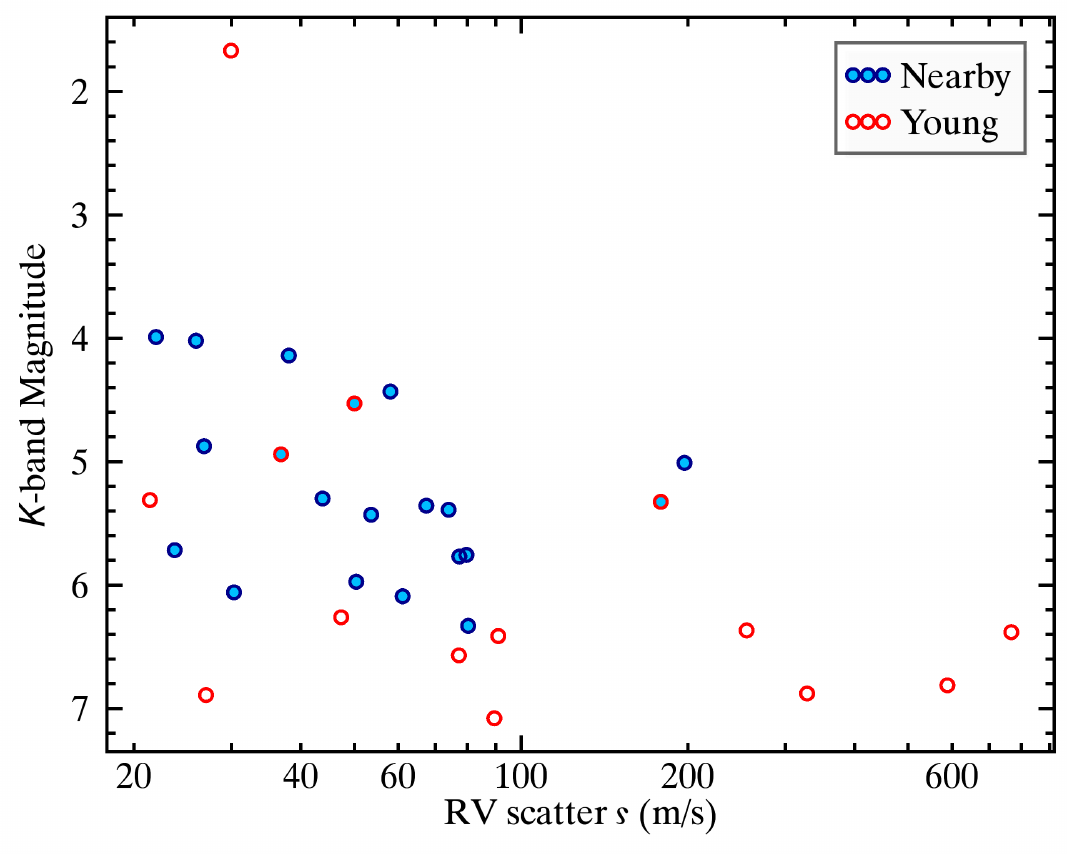}}
	\caption{Reduced $\chi^2$ with respect to zero variation and RV scatter $\varsigma$ as a function of the \emph{2MASS} $K_S$-band magnitude of our targets. The $\chi^2_r$ values do not seem to be correlated with the $K$-band magnitude, which indicates that our RV measurement errors are realistic. The RV scatter $\varsigma$ is however correlated with the $K$-band magnitude, which is a natural effect of the lower S/N that were obtained for fainter targets. The color scheme is identical to that of Figure~\ref{fig:Chi2RMS_Night}. Targets in the young sample have lower S/N both because they were observed with smaller integration times as they were mostly observed before 2014, but also because they have fainter $K$-band magnitudes on average. For more details, see Section~\ref{sec:ensemble}.}
	\label{fig:Chi2RMS_Kmag}
\end{figure*}

In Figures~\ref{fig:Chi2RMS_Night} and \ref{fig:Chi2RMS_Kmag}, we present the distribution of $\chi_r^2$ and $\varsigma$ as a function of the total number of epochs for all of our targets. These figures bring out the absence of a correlation between these quantities, an indication that no significant long-term systematics are affecting our survey results.

\begin{figure*}
	\centering
	\includegraphics[width=0.92\textwidth]{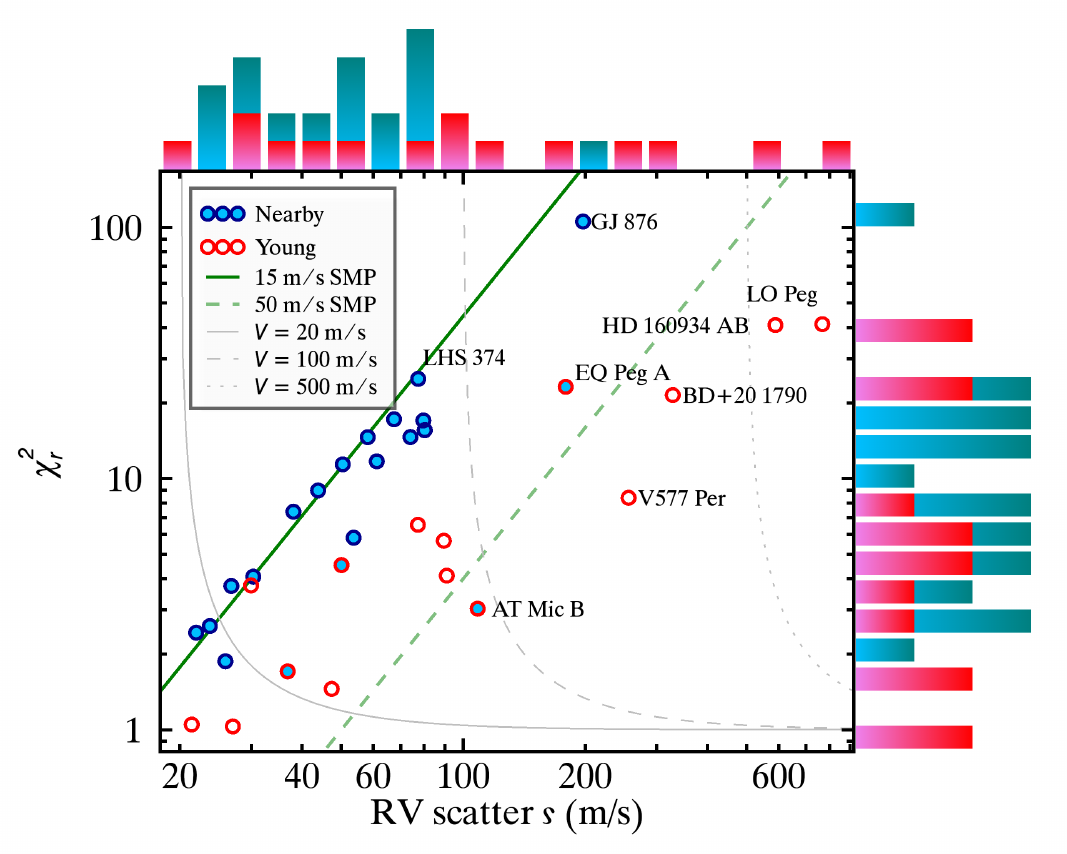}
	\caption{Reduced $\chi_r^2$ with respect to zero variation as a function of the RV scatter $\varsigma$ for targets in the nearby (filled blue circles) and young (red circles) samples. The solid and dashed green lines correspond to single-measurement precisions of 15 and 50\,\ms, respectively. The solid, dashed and dotted grey lines correspond to respective RV variability values $V$ of 20, 100 and 500\,\ms\ ($V = \sqrt{\varsigma^2-S^2}$, see Section~\ref{sec:rvcomb}). Histogram distributions for the nearby (green bars) and young (pink bars) samples are displayed next to each plot axis. For more details, see Section~\ref{sec:ensemble}.}
	\label{fig:Chi2_RMS}
\end{figure*}

In Figure~\ref{fig:Chi2_RMS}, we present the reduced $\chi_r^2$ with respect to zero RV variation as a function of the RV scatter $\varsigma$ for all of our targets. This figure illustrates how the young survey sample has been observed with a typically lower S/N, resulting in typical single-measurement precisions around 50\,\ms\ on average, whereas those of the nearby sample are lower at around 15\,\ms. Targets located \replaced{further to the upper and to the right of}{in the upper right of} the figure (along lines of constant single-measurement precisions), are the most secure RV variables.

A Kolmogorov-Smirnov test yields a 54\% probability that the reduced $\chi_r^2$ values around zero RV variation for the young and nearby samples are drawn from a single random distribution (the nearby sample targets have slightly larger $\chi_r^2$ values on average). This indicates a weak statistical significance that there is any fundamental difference in the RV variability amplitude \replaced{in}{between} the two samples. We recover a larger fraction of candidate RV variable targets in the nearby sample (10/21 $\approx$ 48\%) than in the young sample (4/15 $\approx$ 27\%). However, considering Poisson statistics and the number of targets in each sample, there is a relatively large $\approx$\,23\% chance that this discrepancy is due to pure chance. Both these results could be explained by the fact that we obtained higher S/N observations on average for the nearby sample, if we assume that there is a larger number of RV variables with an amplitude small enough so that they would not be detected in the young sample (see Figure~\ref{fig:Chi2_RMS}).

\begin{figure*}
	\centering
	\subfigure[RV variability versus single-measurement precision]{\includegraphics[width=0.495\textwidth]{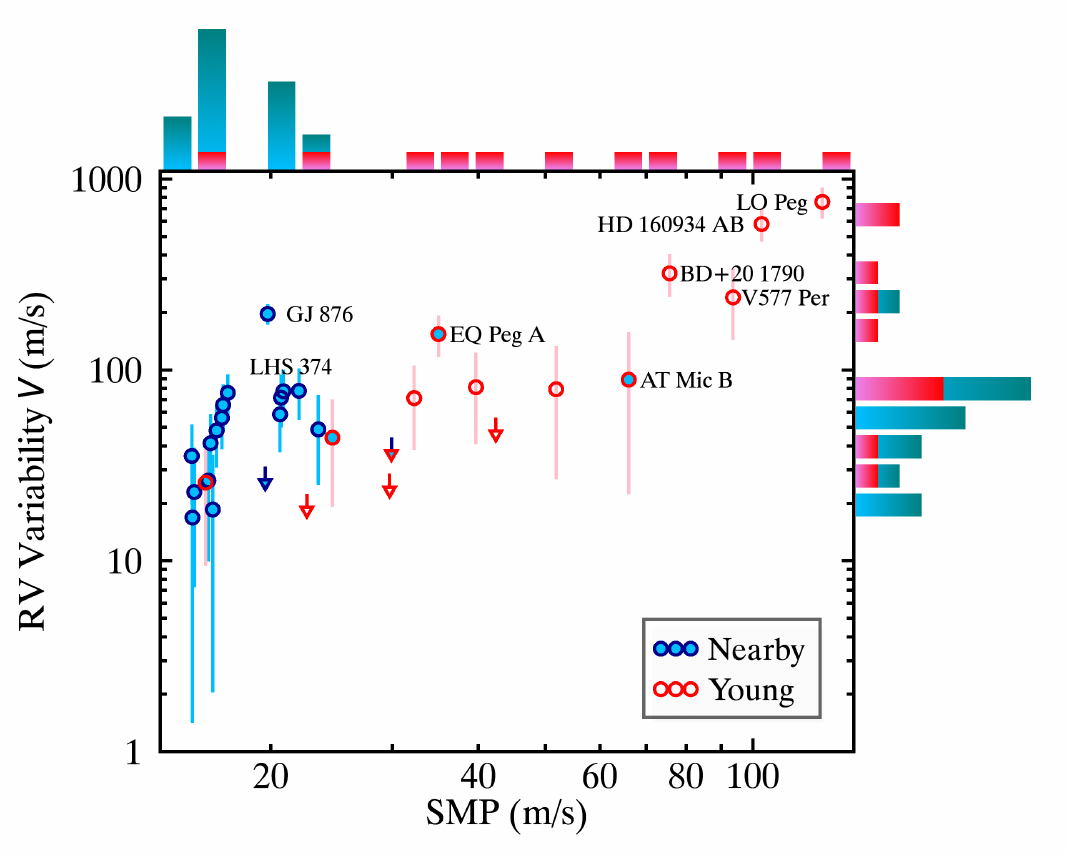}\label{fig:smpvar}}
	\subfigure[Statistical significance of RV variability]{\includegraphics[width=0.495\textwidth]{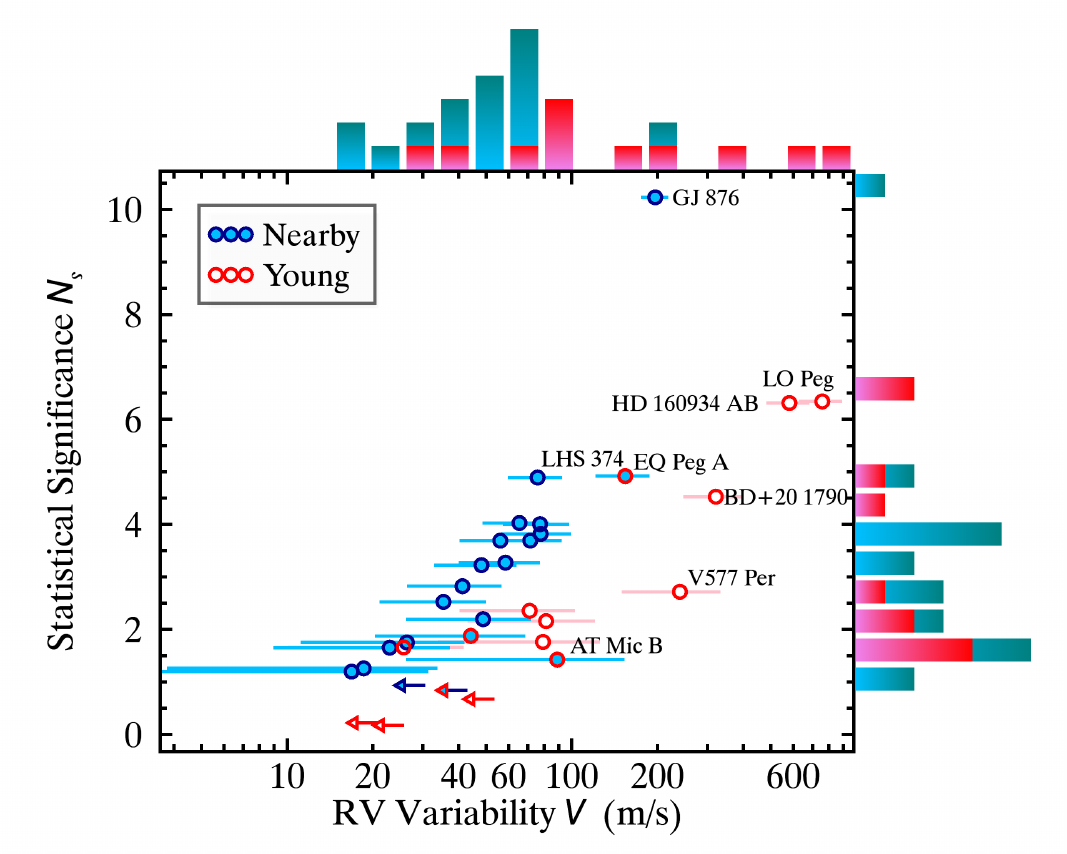}\label{fig:nsigvar}}
	\caption{Panel~a: RV Variability $V$ as a function of single-measurement precision for the young (red circles) and nearby (blue circles) samples.\\ Panel~b: Statistical significance as a function of the RV variability. Targets with an RV variability below 1$\sigma$ are displayed as 1$\sigma$ upper limits (left- or down-pointing arrows). For more details, see Section~\ref{sec:ensemble}.}
	\label{fig:var}
\end{figure*}

In Figure~\ref{fig:var}, we show the RV variability $V$ as a function of the single-measurement precision $S$ and the statistical significance \added{$N_\varsigma$} of $V$. The first distribution outlines the vastly different single-measurement precisions that were obtained for the young and nearby samples, which is an effect of the different S/N observations. Targets located higher up in Panel~b of Figure~\ref{fig:var} are the most probable RV variables, and those located further to the right in the same figure could correspond to more massive and/or close-in companions.

Nineteen of the targets presented in this work were never part of a precise RV follow-up ($S \lesssim$\,100\,\ms) to date. There are, however, 13 targets that already benefitted from precise RV monitoring. These targets are listed in Table~\ref{tab:rvcomp}, where we compare the number of epochs, single-measurement precision and RV scatter of existing optical and NIR surveys to our survey results. For eight of the targets listed in this table, we present a more precise RV follow-up to those already published, and for five of the targets, we present the first precise RV follow-up in the NIR. There is only one case (AT~Mic~B) for which a NIR follow-up already existed at a better precision than the results presented here. \added{These data do not allow performing a significant comparison between the level of RV variability of single stars in the optical versus NIR;  only three such targets (BD+01~2447, GJ~15~A and \replaced{$\epsilon$~Eri}{$\varepsilon$~Eridani}) have precise RV measurements in both regimes, but they are consistent within 1$\sigma$.}

\subsection{Constraints from Non-Detections}\label{sec:nondet}

\begin{figure*}
	\centering
	\subfigure[AU~Mic]{\includegraphics[width=0.665\textwidth]{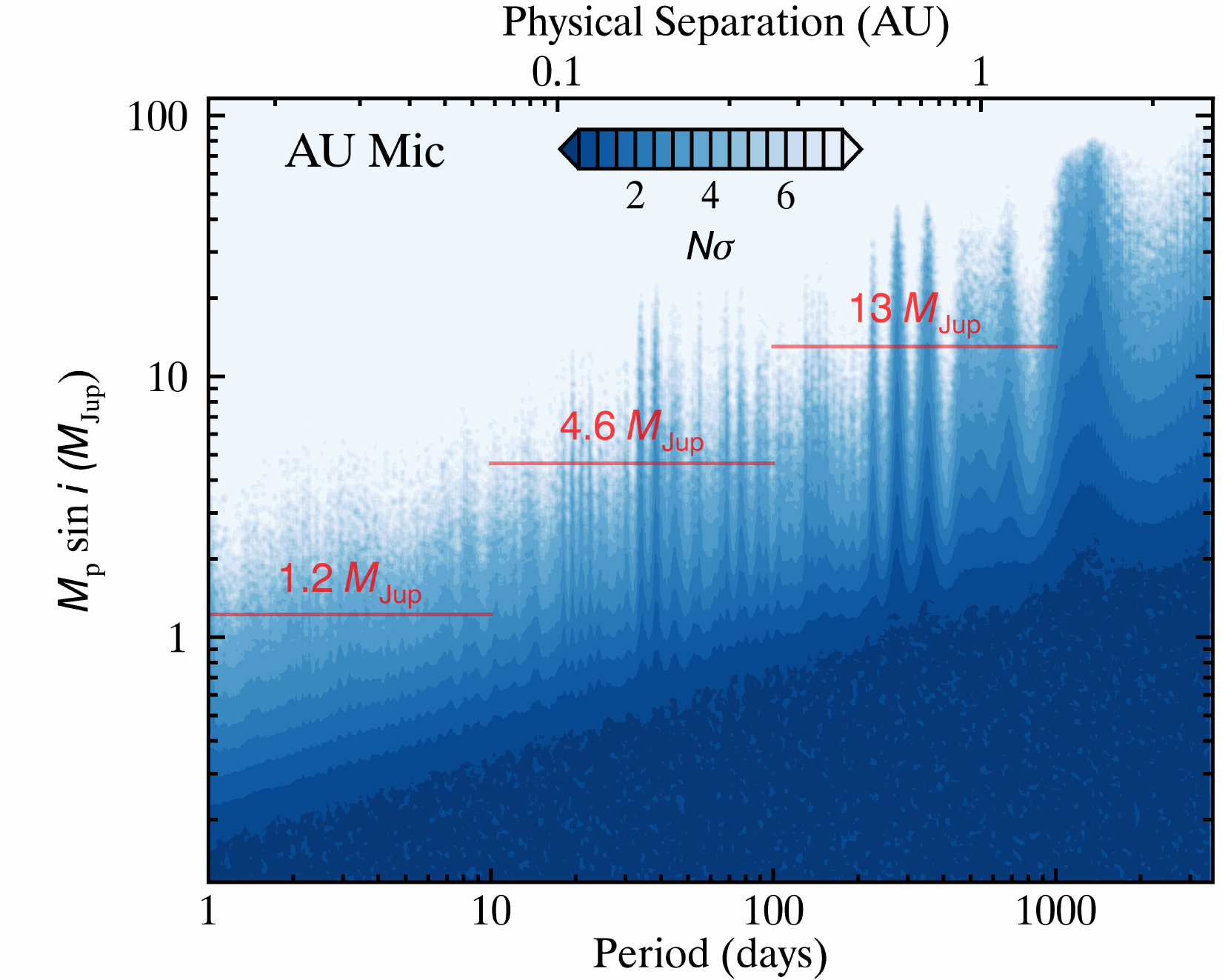}}
	\subfigure[$\varepsilon$~Eridani]{\includegraphics[width=0.665\textwidth]{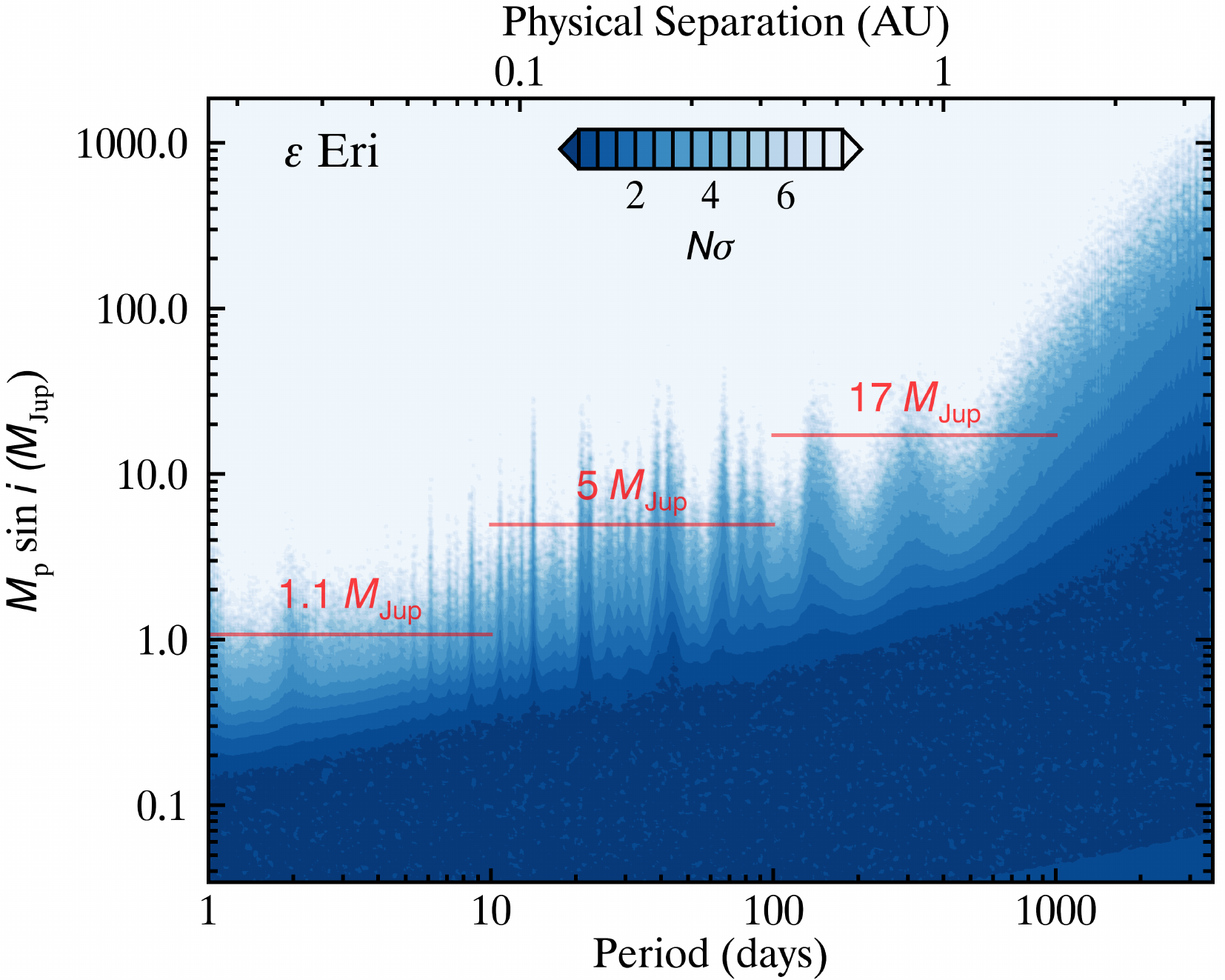}}
		\caption{\added{Rejection statistical significance of synthetic companions from our observed RV variation $V$, RV precision and temporal sampling, as a function of companion mass $M_p \sin i$ and period, for AU~Mic (Panel~a) and $\varepsilon$~Eridani (Panel~b). Lighter-colored regions correspond to companion configurations that are safely rejected with our observations. We display in red the upper mass limits where 95\% of the synthetic companions would have been detected at a 3$\sigma$ significance in three distinct period regimes. See Section~\ref{sec:nondet} for more details. A Figure set is available in the online version displaying similar figures for all of our survey targets.}}
	\label{fig:MassLimit}
\end{figure*}

\added{The upper limit measurements on RV variability $V_{\rm max}$ defined in Equation~\eqref{eqn:vmax} and listed in Table~\ref{tab:rvs} can be translated to upper limits on companion masses as a function of physical separation or orbital period.

In order to estimate these upper limits, we have carried out a $10^8$--steps Monte Carlo simulation, where the orbital parameters of synthetic companions (projected companion mass $M_p \sin i$, eccentricity $e$, period $P$, absolute RV $\gamma$, longitude of periastron $\omega$ and periastron date $T_0$) are drawn from a random distribution. 

In order to properly account for the eccentricity distribution of known exoplanets, the random values for $e$ are sampled from a Beta distribution with $a = 0.867$ and $b = 3.03$ \citep{2013MNRAS.434L..51K}. $P$, $M_p \sin i$ and $T_0$ are drawn from uniform distributions in log space, and $\gamma$ is drawn from a normal distribution with a standard deviation of $V/2$. This is done in order to reflect our uncertainty on the absolute RV of our targets, especially in the cases with only a few RV epochs.

Synthetic RV measurements are then extracted by sampling the RV curve at the same epochs as our individual observations, and assigning the same measurement errors that we have observed on each of them.

The synthetic RV variability term $V_{\rm synth}$ is then calculated for every synthetic companion, from which we derive a detection probability according to its $N_\varsigma$ distance from the measured $V$. This is done by finding the $N_\varsigma$ value in Equation~\eqref{eqn:fnsig} that enforces $V_{\rm max}\big(N_\varsigma\big) = V_{\rm synth}$. This value can then be translated to a probability, assuming that the measurement errors on $V$ follow a normal distribution. A two-dimensional 800x800-elements histogram is then constructed over the variables ($P$, $M_p \sin i$) in logarithmic space, which acts as a marginalization of the other orbital parameters. This two-dimensional probability distribution is then converted back to $N_\varsigma$ values, which can be represented as a contour plot. We display these contour plots in Figure~\ref{fig:MassLimit}, for the cases of AU~Mic and $\varepsilon$~Eridani. The contour data have been smoothed with a 2-pixels filter for visibility.}

\added{We define three regimes of periods $P$ for which we determine mean upper mass limits, i.e. \emph{Hot Jupiters} (1--10\,d), \emph{Warm Jupiters} (10--100\,d) and \emph{Cool Jupiters} (100--1000\,d). Within each one of these ranges, we determine the critical mass $M_p \sin i$ at which 95\% of the synthetic companions would have been detected at a 3$\sigma$ confidence level. These values are reported in Table~\ref{tab:rvs}. These values do not allow us to contradict the existence of any companion detection that was previously reported in the literature.}

\begin{deluxetable*}{llcccccccccccccc}[p]
\tablecolumns{16}
\tablecaption{Survey Results \label{tab:rvs}}
\tablehead{\colhead{Target} & \colhead{Survey} & \colhead{$N_k$} & \colhead{Baseline} & \colhead{Total} &  \colhead{$\varsigma$\tablenotemark{c}} & \colhead{$\chi_r^2$\tablenotemark{d}} & \colhead{$S$\tablenotemark{e}} & \colhead{$V$\tablenotemark{f}} & \colhead{$N_\varsigma$\tablenotemark{g}} & \multicolumn{2}{c}{$V_{\rm max}$\tablenotemark{h} (\ms)} & \colhead{} & \multicolumn{3}{c}{\added{3$\sigma$ Mass Limit (\MJup)} \tablenotemark{i}}\\
\cline{11-12}
\cline{14-16}
\colhead{Name} & \colhead{Sample\tablenotemark{a}} & \colhead{Nights} & \colhead{(days)} & \colhead{S/N\tablenotemark{b}} & \colhead{(\ms)} & \colhead{} & \colhead{(\ms)} & \colhead{(\ms)} & \colhead{} & \colhead{1$\sigma$} & \colhead{3$\sigma$} & \colhead{} & \colhead{HJ} & \colhead{WJ} & \colhead{CJ}}
\startdata
\sidehead{\textbf{RV Variables}}
GJ~876 & N & 17 & 707 & 450 & 197 & 105.6 & 19 & 196 & 10.2 & 203 & 234 & & 2.8 & 14 & 41 \\
HD~160934~AB & Y & 5 & 338 & 160 & 589 & 40.8 & 92 & 581 & 6.3 & 612 & 763 & & 88 & 150 & 1100 \\
LO~Peg & Y & 10 & 314 & 260 & 768 & 41.2 & 120 & 758 & 6.3 & 799 & 994 & & 35 & 220 & 680 \\
\sidehead{\textbf{Candidate RV Variables}}
BD+20~1790 & Y & 8 & 1295 & 200 & 328 & 21.5 & 71 & 321 & 4.5 & 345 & 460 & & 32 & 130 & 820 \\
EQ~Peg~A & Y,N & 5 & 1211 & 410 & 157 & 25.2 & 31 & 154 & 4.9 & 165 & 216 & & 7.2 & 18 & 34 \\
GJ~3942 & N & 7 & 92 & 410 & 80 & 15.6 & 20 & 78 & 3.8 & 84 & 118 & & 4.1 & 10 & 300 \\
GJ~537~B & N & 8 & 326 & 500 & 74 & 14.6 & 19 & 71 & 3.7 & 78 & 110 & & 3.2 & 5.4 & 45 \\
GJ~725~A & N & 5 & 1035 & 380 & 58 & 14.6 & 15 & 56 & 3.7 & 61 & 86 & & 5.6 & 130 & 950 \\
GJ~740 & N & 11 & 123 & 630 & 67 & 17.2 & 16 & 65 & 4.0 & 71 & 97 & & 1.9 & 3.7 & 140 \\
GJ~9520 & N & 7 & 314 & 460 & 80 & 17.0 & 19 & 77 & 4.0 & 84 & 115 & & 4.1 & 11 & 110 \\
LHS~371 & N & 5 & 38 & 400 & 50 & 11.4 & 15 & 48 & 3.2 & 53 & 78 & & 3.8 & 8 & 620 \\
LHS~372 & N & 4 & 49 & 270 & 61 & 11.7 & 18 & 58 & 3.3 & 64 & 94 & & 8.6 & 41 & 650 \\
LHS~374 & N & 5 & 43 & 410 & 77 & 24.9 & 15 & 76 & 4.9 & 81 & 106 & & 6.2 & 12 & 1000 \\
\sidehead{\textbf{Other Targets}}
AG~Tri & Y & 10 & 314 & 240 & 90 & 5.7 & 38 & 81 & 2.2 & 94 & 155 & & 3.6 & 16 & 52 \\
AT~Mic~A & Y,N & 9 & 1365 & 310 & 37 & 1.7 & 28 & 24 & 0.8 & 35 & 80 & & 1.1 & 4.7 & 11 \\
AT~Mic~B & Y,N & 9 & 1488 & 320 & 108 & 3.0 & 62 & 89 & 1.4 & 111 & 212 & & 2.8 & 13 & 31 \\
AU~Mic & Y,N & 12 & 1462 & 480 & 50 & 4.5 & 24 & 44 & 1.9 & 52 & 90 & & 1.2 & 4.6 & 13 \\
BD+01~2447 & Y & 7 & 89 & 290 & 21 & 1.0 & 21 & 5 & 0.2 & 17 & 48 & & 1.4 & 41 & 140 \\
BD--13~6424 & Y & 8 & 313 & 260 & 77 & 6.5 & 30 & 71 & 2.4 & 81 & 131 & & 4.4 & 30 & 84 \\
$\varepsilon$~Eridani & Y & 13 & 339 & 1130 & 30 & 3.8 & 15 & 26 & 1.7 & 31 & 56 & & 1.1 & 5 & 17 \\
EV~Lac & N & 5 & 193 & 490 & 44 & 9.0 & 15 & 41 & 2.8 & 46 & 70 & & 2.5 & 16 & 36 \\
GJ~15~A & N & 14 & 818 & 490 & 26 & 1.9 & 19 & 18 & 0.9 & 25 & 55 & & 0.62 & 2.8 & 8.3 \\
GJ~3305~AB & Y & 4 & 132 & 160 & 91 & 4.1 & 45 & 79 & 1.8 & 94 & 168 & & 12 & 120 & 670 \\
GJ~169 & N & 5 & 130 & 490 & 27 & 3.7 & 14 & 23 & 1.7 & 28 & 50 & & 2.5 & 7.2 & 51 \\
GJ~338~A & N & 6 & 253 & 530 & 22 & 2.4 & 14 & 17 & 1.2 & 22 & 45 & & 1.4 & 2 & 16 \\
GJ~338~B & N & 6 & 320 & 530 & 38 & 7.4 & 14 & 35 & 2.5 & 40 & 63 & & 2.3 & 4.2 & 20 \\
GJ~458~A & N & 7 & 100 & 470 & 30 & 4.1 & 15 & 26 & 1.8 & 31 & 56 & & 1.6 & 4.4 & 82 \\
GJ~537~A & N & 10 & 90 & 500 & 54 & 5.8 & 22 & 49 & 2.2 & 56 & 93 & & 1.7 & 3.2 & 190 \\
LHS~26 & N & 5 & 136 & 380 & 24 & 2.6 & 15 & 19 & 1.3 & 24 & 48 & & 1.7 & 4.4 & 23 \\
TYC~5899--26--1 & Y & 5 & 130 & 200 & 27 & 1.0 & 27 & 5 & 0.2 & 21 & 60 & & 2.1 & 14 & 41 \\
V1005~Ori & Y & 7 & 131 & 250 & 47 & 1.5 & 39 & 26 & 0.7 & 44 & 105 & & 3.6 & 37 & 120 \\
V577~Per & Y & 9 & 338 & 270 & 256 & 8.4 & 88 & 240 & 2.7 & 269 & 413 & & 16 & 43 & 140 \\
\enddata
\tablenotetext{a}{Y: Young, N: Nearby.}
\tablenotetext{b}{Combined signal-to-noise ratio of all observed spectra for a given target, assuming that all data is photon-noise limited.}
\tablenotetext{c}{Standard deviation of the per-night combined RV measurements. See Section~\ref{sec:rvcomb} for more details.}
\tablenotetext{d}{Reduced $\chi_r^2$ value of a zero-variation RV curve. See Section~\ref{sec:rvcomb} for more details.}
\tablenotetext{e}{Typical single-measurement precision of per-night combined RV measurements. See Section~\ref{sec:rvcomb} for more details.}
\tablenotetext{f}{RV variability, defined as $V = \sqrt{\varsigma^2-S^2}$. See Section~\ref{sec:rvcomb} for more details.}
\tablenotetext{g}{Statistical significance of the RV variability $V$, defined as $N_\varsigma = V/S$. See Section~\ref{sec:rvcomb} for more details.}
\tablenotetext{h}{$N$-sigma upper limits on the RV variability term $V$. See Section~\ref{sec:rvcomb} for more details.}
\tablenotetext{i}{\added{Upper mass limit above which a Hot Jupiter (HZ; $P$\,$\sim$\,1--10\,d), Warm Jupiter (WJ; $P$\,$\sim$\,10--100\,d) or Cool Jupiter (CJ; $P$\,$\sim$\,100--1000\,d) would have been detected at 3$\sigma$, 95\% of the time. See Section~\ref{sec:ensemble} for more details.}}
\tablecomments{See Section~\ref{sec:ensemble} for more details on the survey results.}
\end{deluxetable*}
\begin{deluxetable*}{lccccccccccccccc}
\tablecolumns{16}
\tablecaption{Comparison With Previous Work \label{tab:rvcomp}}
\tablehead{\colhead{Target} & \colhead{Survey} & \colhead{} & \multicolumn{4}{c}{Optical} & \colhead{} & \multicolumn{4}{c}{NIR} & \colhead{} & \multicolumn{3}{c}{This Work}\\
\cline{4-7}
\cline{9-12}
\cline{14-16}
\colhead{Name} & \colhead{Sample\tablenotemark{a}} & \colhead{} & \colhead{Ref.\tablenotemark{b}} & \colhead{$N_{\mathrm{data}}$\tablenotemark{c}} & \colhead{$S$ (\ms)\tablenotemark{d}} & \colhead{$\varsigma$ (\ms)\tablenotemark{e}} & \colhead{} & \colhead{Ref.\tablenotemark{b}} & \colhead{$N_{\mathrm{data}}$\tablenotemark{c}} & \colhead{$S$ (\ms)\tablenotemark{d}} & \colhead{$\varsigma$ (\ms)\tablenotemark{e}} & \colhead{} & \colhead{$N_{\mathrm{data}}$\tablenotemark{c}} & \colhead{$S$ (\ms)\tablenotemark{d}} & \colhead{$\varsigma$ (\ms)\tablenotemark{e}}}
\startdata
AG Tri & Y & & $\cdots$ & $\cdots$ & $\cdots$ & $\cdots$ & & 1 & 14 & 55 & 98 & & 10 & 38 & 90\\
AT Mic A & Y,N & & $\cdots$ & $\cdots$ & $\cdots$ & $\cdots$ & & 1 & 14 & 50 & 151 & & 9 & 28 & 37\\
AT Mic B & Y,N & & $\cdots$ & $\cdots$ & $\cdots$ & $\cdots$ & & 1 & 14 & 55 & 207 & & 9 & 62 & 108\\
AU Mic & Y,N & & $\cdots$ & $\cdots$ & $\cdots$ & $\cdots$ & & 1 & 14 & 50 & 125 & & 12 & 24 & 50\\
BD+01 2447 & Y & & 2 & 13 & 80 & 100 & & $\cdots$ & $\cdots$ & $\cdots$ & $\cdots$ & & 7 & 21 & 21\\
BD+20 1790 & Y & & \textbf{3}--5 & 61 & 5.5 & 580 & & $\cdots$ & $\cdots$ & $\cdots$ & $\cdots$ & & 8 & 71 & 328\\
$\epsilon$ Eri & Y & & 6 & 33 & 12.0 & 15.3 & & $\cdots$ & $\cdots$ & $\cdots$ & $\cdots$ & & 13 & 15 & 30\\
EV Lac & N & & $\cdots$ & $\cdots$ & $\cdots$ & $\cdots$ & & 1 & 20 & 50 & 115 & & 5 & 15 & 44\\
GJ 15 A & N & & 7 & 117 & 0.6 & 3.21 & & $\cdots$ & $\cdots$ & $\cdots$ & $\cdots$ & & 14 & 19 & 26\\
GJ 3305 AB & Y & & 8 & 3 & 20 & 550 & & 1 & 5 & 50 & 457 & & 4 & 45 & 91\\
GJ 876 & N & & \textbf{9}--13 & 162 & 2.0 & 162 & & $\cdots$ & $\cdots$ & $\cdots$ & $\cdots$ & & 17 & 19 & 197\\
GJ 725 A & N & & $\cdots$ & $\cdots$ & $\cdots$ & $\cdots$ & & 1 & 18 & 50 & 51 & & 5 & 15 & 58\\
V1005 Ori & Y & & $\cdots$ & $\cdots$ & $\cdots$ & $\cdots$ & & 1 & 6 & 55 & 103 & & 7 & 39 & 47\\
\enddata
\tablenotetext{a}{Y: Young, N: Nearby.}
\tablenotetext{b}{When multiple references are listed, the one in bold has presented the highest overall RV precision; data presented in the following columns are obtained from this reference.}
\tablenotetext{c}{Total number of RV epochs.}
\tablenotetext{d}{Typical single measurement precision.}
\tablenotetext{e}{RV scatter (analogous to $\varsigma$ in this paper).}
\tablecomments{See Section~\ref{sec:ensemble} for more details.}
\tablerefs{(1)~\citealt{2012ApJ...749...16B}; (2)~\citealt{2006PASP..118..706P}; (3)~\citealt{2015AA...576A..66H}; (4)~\citealt{2010AA...513L...8F}; (5)~\citealt{2010AA...512A..45H}; (6)~\citealt{1988ApJ...331..902C}; (7)~\citealt{2014ApJ...794...51H}; (8)~\citealt{2014AA...568A..26E}; (9)~\citealt{2010ApJ...719..890R}; (10)~\citealt{1998ApJ...505L.147M}; (11)~\citealt{2001ApJ...556..296M}; (12)~\citealt{1998AA...338L..67D}; (13)~\citealt{2005ApJ...634..625R}.}
\end{deluxetable*}

\subsection{Effects of Rotational Velocity and Age}\label{sec:rotvel}

To assess the impact of rotational broadening on our achievable RV precision limit, we have constructed a set of synthetic data based on our observations of GJ~15~A, which is a slow-rotating, RV-quiet star (within our survey precision) and benefits from a large number of high-S/N observations. We have used the best fitting parameters that were obtained from the RV pipeline for each individual raw spectrum to remove the effects of the blaze function, gas cell and telluric absorption. We have then deconvolved the remaining individual stellar spectra with the appropriate LSF and used the \emph{add\_rotation.pro} IDL routine\footnote{Written by Russel White in December 2000, then Greg Doppmann in July 2003. Note that this routine is distinct from \emph{lsf\_rotate.pro} from the astrolib library at \emph{http://idlastro.gsfc.nasa.gov/} that was recently shown to contain an error \citep{2015arXiv151008994M}.} to \replaced{artificially add}{produce an artificial} rotational broadening. We convolved the result with the LSF and added back the effects of the gas cell, telluric absorption and blaze function. We generated a synthetic data set in this way for 18 values of projected rotational velocities $v \sin i$ that range from 2 to 30\,\kms.

\begin{figure*}[p]
	\centering
	\subfigure[Effect of $v \sin i$ on synthetic GJ~15~A RV precision]{\includegraphics[width=0.445\textwidth]{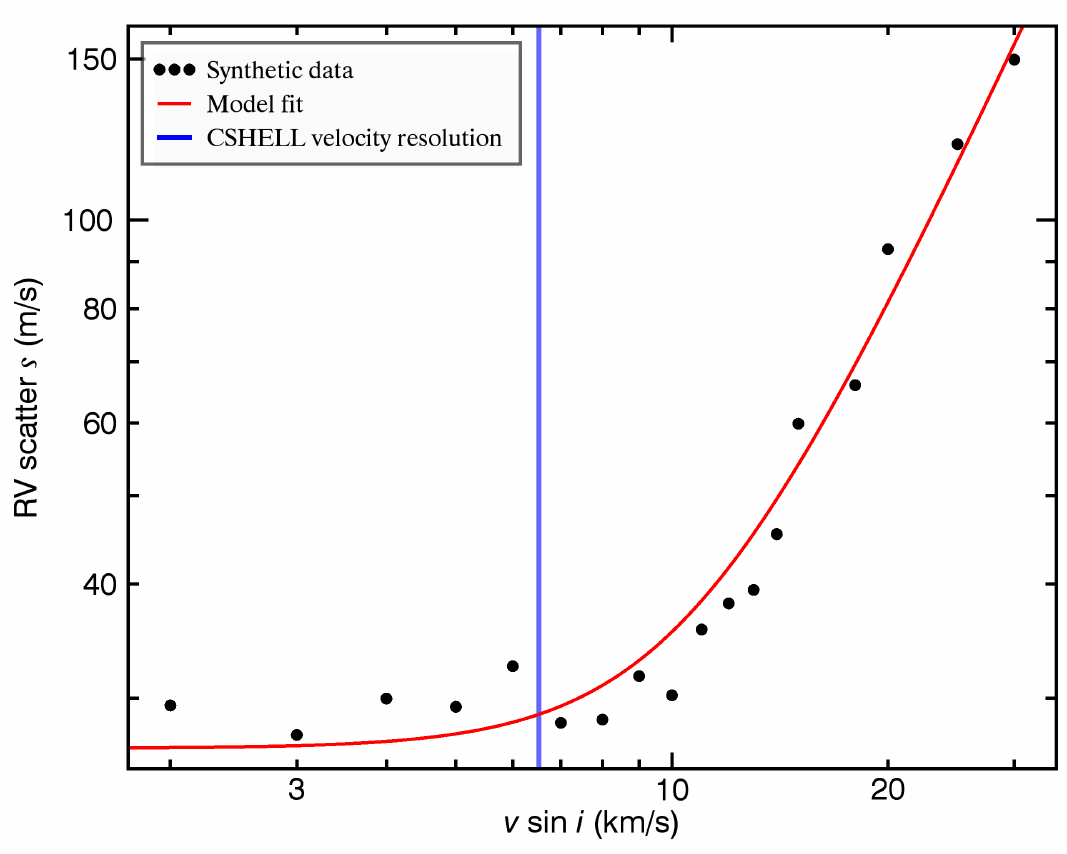}	\label{fig:vsini_rms}}
	\subfigure[RV variability and stellar jitter]{\includegraphics[width=0.645\textwidth]{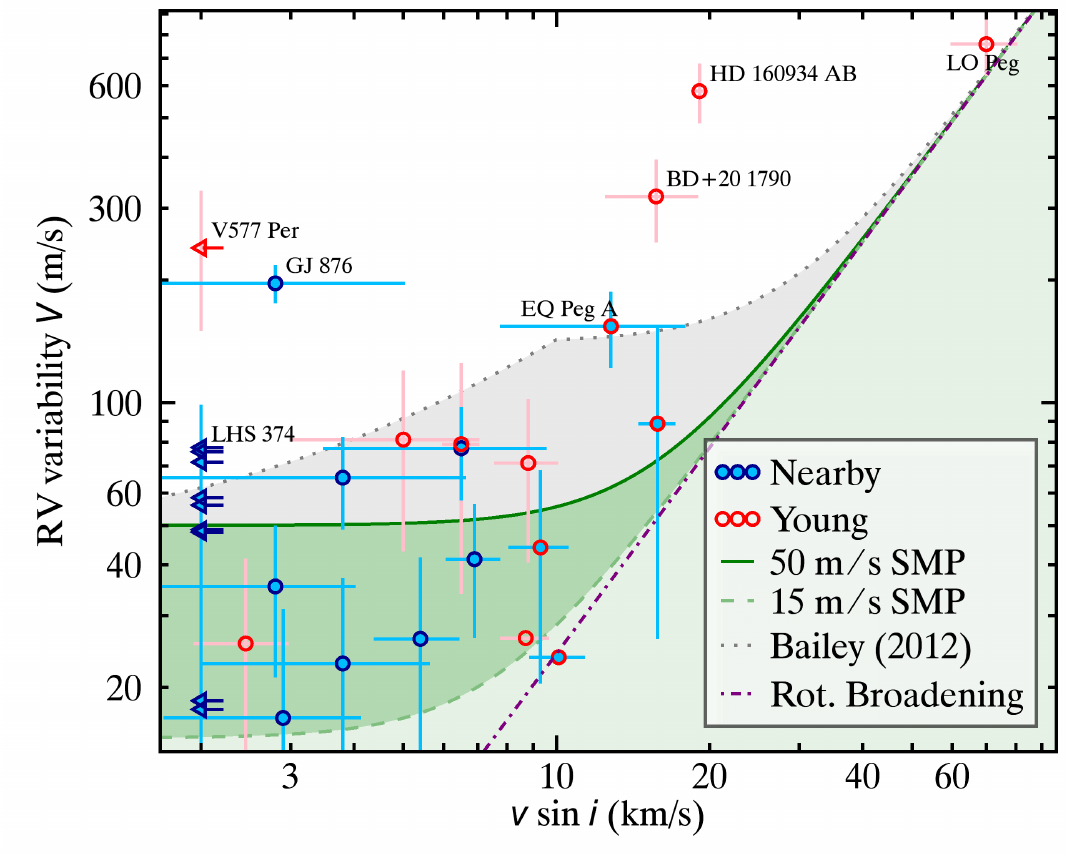}	\label{fig:vsini}}
	\caption{Panel~a: RV scatter $\varsigma$ as a function of the synthetic rotational velocity broadening of GJ~15~A data (black circles). The red line corresponds to a best-fit power law (see Section~\ref{sec:rotvel}). The blue line corresponds to the velocity resolution of a single CSHELL spectrum ($c/R \sim 6.5$\,\kms).\\ Panel~b: RV variability $V$ as a function of the measured projected rotational velocity $v \sin i$ for the nearby (filled blue circles) and young (red circles) samples. Upper limits are displayed with left-pointing arrows. The purple dash-dotted line represents the effect of information loss from rotational velocity alone (extrapolated from the synthetic relation described in Section~\ref{sec:rotvel}). The dashed (solid) green line represents the quadrature sum of a 15\,\ms\ (50\,\ms) single-measurement precision and information loss from $v \sin i$. We display the quadrature sum of a 15\,\ms\ single-measurement precision with the $v \sin i$--jitter relation of \cite{2012ApJ...749...16B} as a dotted grey line. \replaced{We note}{It can be noted} that the targets that we flag as RV variables lie outside of the NIR jitter region defined by \cite{2012ApJ...749...16B}, which is an indication that their RV variability might not be due to stellar activity. For more details, see Section~\ref{sec:rotvel}.}
\end{figure*}

These synthetic data sets were subsequently analyzed with the MATLAB RV pipeline as described in Section~\ref{sec:rvpip}. We show in Figure~\ref{fig:vsini_rms} the resulting RV precision that was achieved as a function of projected rotational velocity. As expected, the RV precision starts decreasing when the projected rotational velocity gets larger than the velocity resolution of CSHELL ($c/R \sim 6.5$\,\kms, where $c$ is the speed of light in vacuum). This loss of RV precision follows a power law as a function of $v \sin i$.

We have thus modelled this effect of $v \sin i$ on the RV precision by using the quadrature sum of a constant term that represents the single-measurement precision, and a two-parameters power law. The resulting fitting function is given by :

\begin{align}
	y = \sqrt{y_0^2+(x/\sigma_x)^{2\beta}}, 
\end{align}

\noindent where $y$ is the RV precision, $x$ is the projected rotational broadening, $y_0$ is the RV scatter caused by all terms except rotational broadening (e.g., single-measurement precision and RV variability), and $\beta$ and $\sigma_x$ are the free parameters of the power-law. We find best-fit values of $\sigma_x = 1.56 \pm 0.05$\,\kms\ and $\beta = 1.70 \pm 0.02$. The best-fit solution is displayed as a red curve in Figure~\ref{fig:vsini_rms}.

In Figure~\ref{fig:vsini}, we compare this relation to our survey results, assuming different single-measurement precisions. \replaced{We note that}{It can be noted that} in several cases with relatively low projected rotational velocities ($v \sin i \lesssim 10$\,\kms), we obtain RV precisions that do not need to include a jitter term increasing with $v \sin i$. It is however possible that a jitter term is the cause of the lower RV precision that we obtain for a few targets located above the $50$\,\ms\ single-measurement precision green solid line.

The NIR jitter--$v \sin i$ relation measured by \cite{2012ApJ...749...16B} \added{for TW~Hydrae members ($10 \pm 3$\,Myr; \citealt{2015MNRAS.454..593B})} is displayed as a grey dotted line in Figure~\ref{fig:vsini}. \added{This shows that the source \replaced{of}{for} the RV variability of targets within the grey region could be explained by jitter, if our targets display stellar spots as significant as the young TW~Hydrae population.}

\deleted{It seems that this contribution would be sufficient to explain the RV variability that we measure for several targets.} \replaced{This does not, however, include}{Such a level of jitter comparable to that of \cite{2012ApJ...749...16B} is not strong enough to reproduce the RV variability of} stars with known or candidate companions (GJ~876, HD~160934~AB), as well as the single stars V577~Per, BD+20~1790, and LHS~374. Those three targets display a level of RV variability that is thus unlikely to be explained by the combined loss of information due to stellar broadening \deleted{of absorption lines} \replaced{and the jitter term proposed by Bailey et al. (2012)}{and stellar jitter}, however additional follow-up will be needed to assess this with certainty.

Survey targets that have a larger projected rotational velocity show larger RV variations, as expected. This correlation is independent of the survey sample (i.e., independent of age), although younger stars are faster rotators on average. \replaced{We note}{It can be noted} that the large RV variability of LO~Peg might be explained by RV information loss due to the rotational broadening of stellar lines, whereas that of EQ~Peg~A would require a significant jitter term at the higher end of what is admitted in the relation proposed by \cite{2012ApJ...749...16B}. Further observations will be required to determine whether EQ~Peg~A can plausibly host a substellar or planetary companion.

\added{A fraction of the low $v \sin i$ measurements that we have compiled from the literature might be spurious, hence in these cases simply comparing $v \sin i$ to the RV variability term $V$ is not a reliable way to determine the source of RV variability with certainty.}

\begin{figure}
	\centering
	\includegraphics[width=0.445\textwidth]{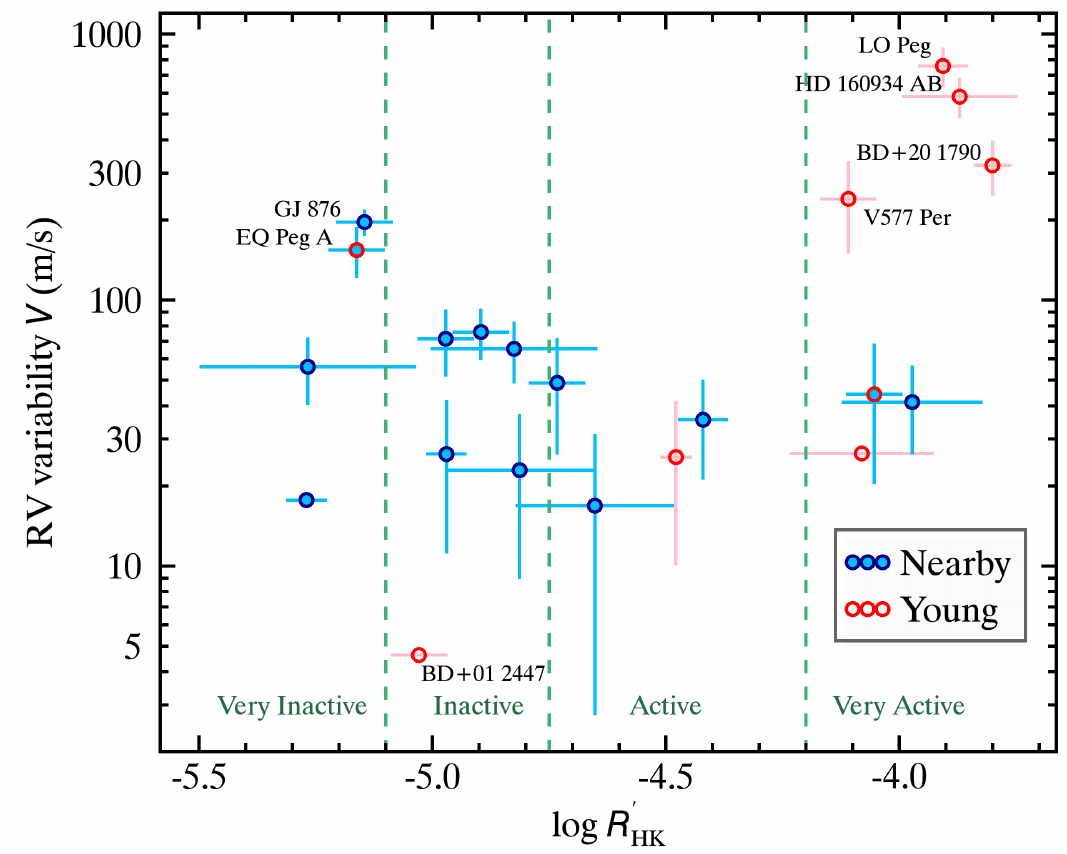}	
	\caption{\added{Activity index $\log{R^\prime_{HK}}$ as a function of RV variability $V$ for our survey samples. The fact that\deleted{ a} three very active targets ($\log{R^\prime_{HK}} > -4.2$) display a low RV variability of $\sim$\,25--50\,\ms\ is a tentative indication for the level of NIR RV jitter. See Section~\ref{sec:rotvel} for more details.}\label{fig:logrhk}}
\end{figure}

\added{In Figure~\ref{fig:logrhk}, we display RV variability as a function of the $\log{R^\prime_{HK}}$ activity index. Targets in the young sample are more active on average than those in the nearby sample, as expected. The fact that we measure a relatively low RV variability for three very active targets ($\log{R^\prime_{HK}} > -4.2$) provides a tentative constraint on the level of jitter in the NIR to $\sim$\,25--50\,\ms\ at $\approx$\,2.3125\,$\mu$m.}


\subsection{Bi-Sector Analysis}\label{sec:bisector}

We measured the bi-sector slopes of CO lines in each of our individual exposures (see e.g., \citealp{2001astro.ph..1377S,2008AA...492..199D}) to investigate the effect of stellar activity on our RV variable targets. We did not identify a correlation between the RV and bi-sector spans in any case.

However, it must be considered that the lack of a correlation might be expected given the moderate resolution ($R \approx 46\,000$) of CSHELL \added{and the observing setup that we have used (i.e., slit spectroscopy)}. Effectively, \cite{2007AA...473..983D} noted that a poor sampling of spectral lines can hinder the measurements of bi-sector spans; e.g. a resolution of $R \approx 50\,000$ would only be able to recover bisector span variations in targets with $v \sin i \gtrsim 6$\,\kms. \added{Slitless} observations at higher resolutions would thus be warranted to guarantee that the RV variability that we measure are not associated with stellar activity.

\section{DISCUSSION OF INDIVIDUAL TARGETS}\label{sec:indivresults}

\subsection{RV Variable Targets}\label{sec:rvvar}

We define targets for which we measure an RV variability with a statistical significance of $N_\varsigma \geq 5$ as likely RV variable\added{s}, and those with $3 \leq N_\varsigma < 5$ as candidate RV variables. All targets that fall in these categories are discussed individually in this section. More follow-up observations will be needed to determine whether \replaced{the}{any} RV variability is due to a companion or to stellar activity\replaced{, although it is expected that the impact of stellar activity is small in the near-infrared regime (Mart\'in et al. 2006; Reiners et al. 2010)}{. Although it is generally expected that the impact of stellar activity is small in the near-infrared regime \citep{2006ApJ...644L..75M,2010ApJ...710..432R}, it has also been shown by \cite{2013AA...552A.103R} that under certain configurations of stellar spots and magnetic fields, the effect of jitter could in fact increase with wavelength.}

\textbf{GJ~876} (HIP~113020) is an M4 low-mass star (\citealt{1995AJ....110.1838R}; mass estimate 0.32\,\Msol; \citealt{2005ApJ...634..625R}) located at $4.69 \pm 0.05$\,pc \citep{2007AA...474..653V}. Its low rotation rate and weak magnetic activity suggest \replaced{an old age ($\gtrsim$\,0.1\,Gyr)}{an age older than $\sim$\,0.1\,Gyr}. Its kinematics place it in the young disk population \citep{2010AA...511A..21C}, however this does not put a strong constraint on its age (i.e., $\lesssim$\,5\,Gyr, \citealt{1992ApJS...82..351L}).

Using the proper motion ($\mu_\alpha\cos\delta = 959.84 \pm 3.36$\,\masyr; $\mu_\delta = -675.33 \pm 1.68$\,\masyr) and parallax measurements of \cite{2007AA...474..653V} along with the systemic RV measurement of $-1.59 \pm 0.18$\,\kms\ \citep{2002ApJS..141..503N} with the BANYAN~II tool \citep{2014ApJ...783..121G,2013ApJ...762...88M}, we obtain a significant probability ($P = 85.1$\%) that this system is a member of the $\beta$~Pictoris moving group, which is comparable to its other bona fide members \citep{2014ApJ...783..121G}. Its $UVW$ space velocity and $XYZ$ galactic position place it at $5.05 \pm 1.80$\,\kms\ and $13.3 \pm 15.9$\,pc from the locus of known $\beta$~Pictoris moving group members. The probability of a random interloper at such a spatial and kinematic distances from the locus of the group (counting both young and old stars) is only of $\approx$\,1.3\% \citep{2014ApJ...783..121G}. However, there are several indications in the literature that this system is old, e.g., \cite{2010AA...515A..98P} measured a low X-ray luminosity of $\log{L_X} = 26.48 \pm 0.13$, \cite{2005AAS...20719103R} measured a large rotation period of $96.7$\,days and a low jitter of $\sim$\,3\,\ms, and \cite{2015AJ....150....6H} showed that it is very quiet with variation amplitudes of only $17.2$\,mmag in the optical. Its absolute magnitude is about $0.7$\,mag brighter than the main sequence, which could be an indication of youth, however this can be explained by its high-metallicity alone (e.g., \citealt{2013AA...551A..36N} measure [Fe/H]\,$\approx$\,0.12--40\,dex). It is therefore most likely that GJ~876 is an old interloper to the $\beta$~Pictoris moving group rather than a member, as its age is not conciliable with that of the group ($24 \pm 3$\,Myr; \citealt{2015arXiv150805955B}) and a star must display both consistent kinematics and a consistent age before it can be considered as a new moving group member \citep{2002AA...385..862S,2013ApJ...762...88M}.

\tabletypesize{\footnotesize}
\begin{deluxetable}{lr@{\hspace{1pt}}c@{\hspace{1pt}}lr@{\hspace{1pt}}c@{\hspace{1pt}}l}
\tablecolumns{7}
\tablecaption{Orbital Solution for GJ~876~\lowercase{b} \label{tab:GJ876}}
\tablehead{\colhead{Parameter} & \multicolumn{3}{c}{\cite{1998ApJ...505L.147M}} & \multicolumn{3}{c}{This Paper}}
\startdata
Orbital period $P$ (days) & $60.85$ & $\pm$ & $0.15$ & $61.23$ & $\pm$ & $0.29$ \\
Eccentricity $e$ & $0.27$ & $\pm$ & $0.03$ & $0.24$ & $\pm$ & $0.10$ \\
Lon. of periastron $\omega$ (deg) & $24$ & $\pm$ & $6$ & $209$ & $\pm$ & $25$ \\
Periastron date $T_0$ (JD$-2.4\times 10^6$) & $50301.0$ & $\pm$ & $1.0$ & $55452.7$ & $\pm$ & $4.6$ \\
Planet's mass $M_p \sin i$ (\MJup) & $2.11$ & $\pm$ & $0.20$ & $2.73$ & $\pm$ & $0.38$ \\
Semimajor axis $a$ (AU) & $0.21$ & $\pm$ & $0.01$ & $0.2041$ & $\pm$ & $0.0007$ \\
\enddata
\tablecomments{See Section~\ref{sec:rvvar} for more details.}
\end{deluxetable}
\tabletypesize{\small}



\cite{1998ApJ...505L.147M} and \cite{1998AA...338L..67D} have identified a 227\,\ms\ RV signal corresponding to GJ~876~b, a Jovian planet ($M_p \sin i \approx 2.1$\,\MJup) on an eccentric ($e = 0.27 \pm 0.03$) $60.85 \pm 0.15$-days orbit at 0.21\,AU around GJ~876. \cite{2001ApJ...556..296M} subsequently discovered GJ~876~c, a $M_p \sin i \approx 0.6$\,\MJup\ planet in 2:1 resonance with GJ~876~b at 0.13\,AU and on a 30.1-days orbit around GJ~876. \cite{2005ApJ...634..625R} then discovered GJ~876~d, a third, $M_p \sin i \approx 5.9$\,\MEarth\ planet on a 1.9-days orbit around GJ~876.

\cite{2010AA...511A..21C} predicted the possible existence of a low-mass ($< 2$\,\MEarth) planet, in 4:1 orbital resonance with GJ~876~b that could explain how the high-eccentricity ($e = 0.14$) of the orbit of GJ~876~d could have survived for more than $\approx$\,1\,Myr, however the existence of such a planet has not been \replaced{yet confirmed}{confirmed yet}. Finally, \cite{2010ApJ...719..890R} confirmed the existence of a fourth planet, GJ~876~e, on a 126.6-days orbit and a minimum mass of $M_p \sin i = 12.9 \pm 1.7$\,\MEarth.

We followed GJ~876 as part of the nearby sample over 17 nights spanning $>$\,700\,days with a typical S/N\,$\approx$\,170 per night, and recovered it as an RV variable target with $V = 196 \pm 19$\,\ms, which is consistent within 1.6$\sigma$ with the RV amplitude measured by \cite{1998ApJ...505L.147M}. Furthermore, we find $V_{\rm max}\big(N_\varsigma=3\big) = 234$\,\ms, which is also consistent with measurements in the literature.

\begin{figure*}
	\centering
	\subfigure[Orbital Fit to GJ~876~b]{\includegraphics[width=0.495\textwidth]{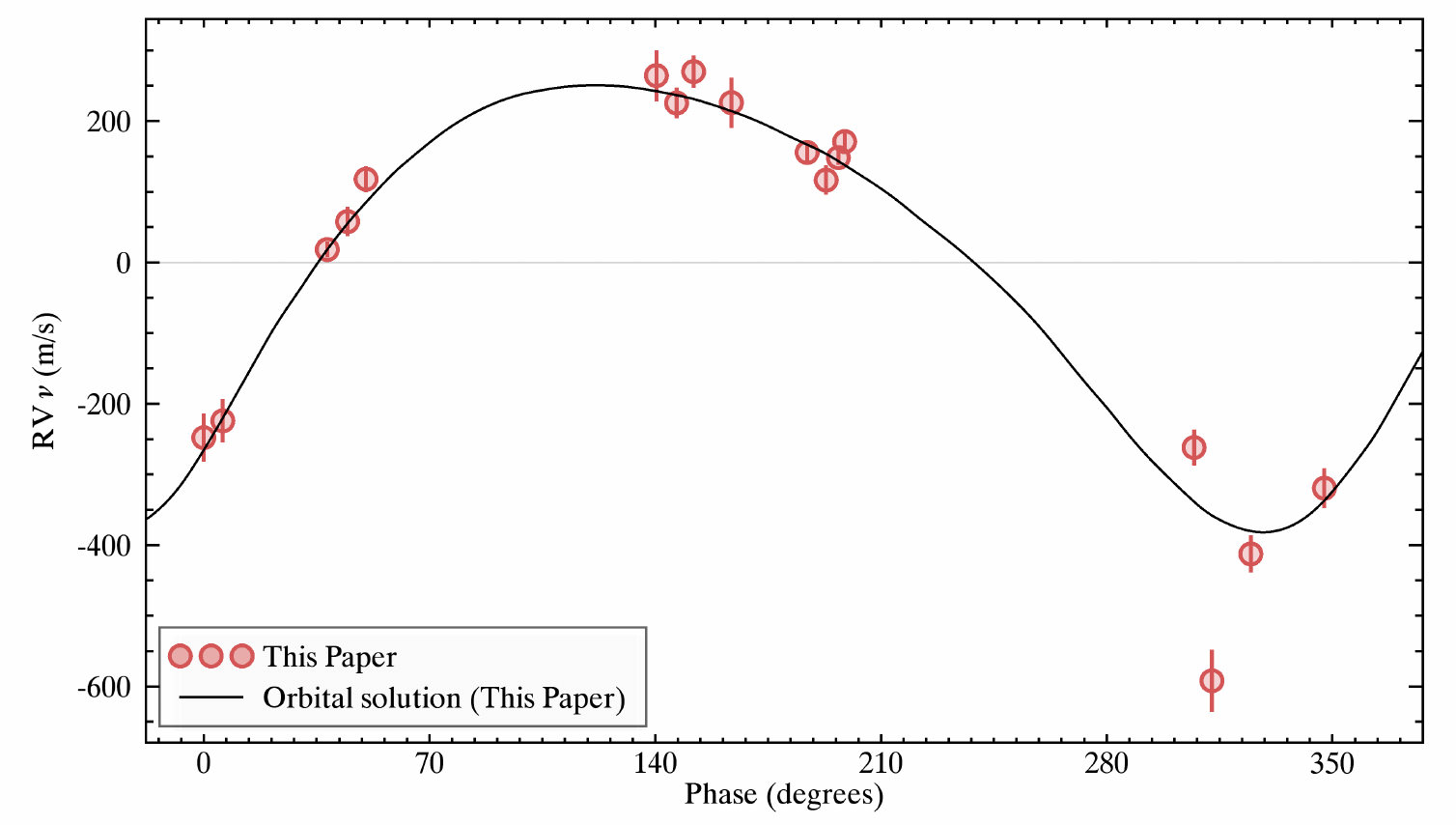}}
	\subfigure[Residuals for GJ~876~b]{\includegraphics[width=0.495\textwidth]{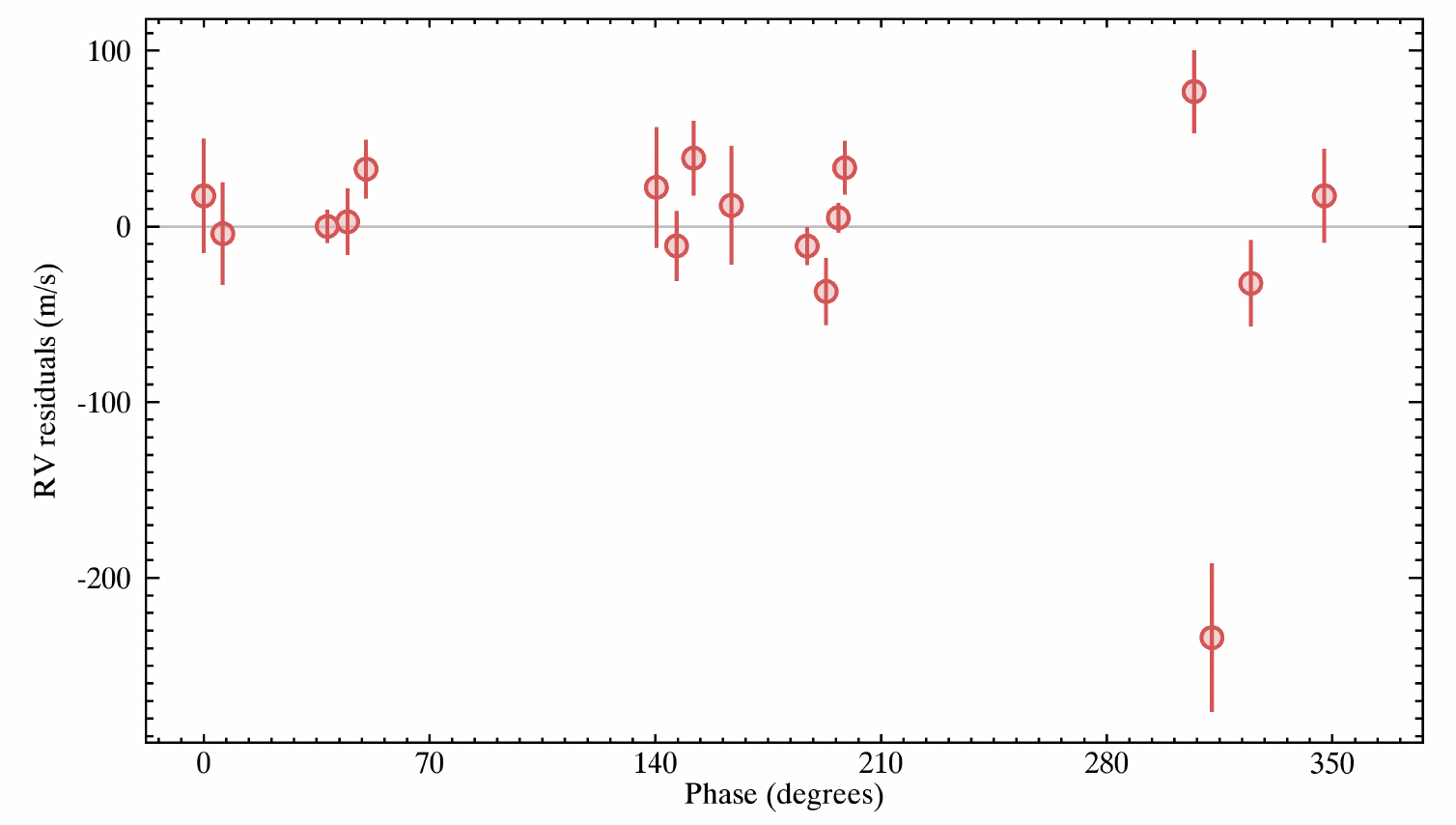}}
		\caption{Panel~a: Orbital fit of GJ~876~b (thick black line) to our RV measurements (red circles). We recover orbital parameters that are fully consistent with those reported in the literature. The outlier data point with a phase of $\sim$\,313\textdegree\ was obtained in bad weather conditions on 2011, July 10 with a seeing above 3\textquotedbl.\\ Panel~b: Residuals after the subtraction of GJ~876~b. Our current data does not allow us to detect the other known planetary companions to GJ~876. For more details, see Section~\ref{sec:rvvar}.}
	\label{fig:Systemic_GJ876}
\end{figure*}

We used the Systemic 2 software\footnote{\url{http://github.com/stefano-meschiari/Systemic2}} \citep{2009PASP..121.1016M,2012ascl.soft10018M} to identify periodic signals in our RV curve that includes 17 epochs spanning $\sim$\,1.9\,yr. We identified a strong signal at $P \approx 61.5$\,days associated to a false alarm probability of only $3.5 \cdot 10^{-4}$\%. Fitting an orbital solution with the Simplex algorithm yielded orbital parameters listed in Table~\ref{tab:GJ876} and the orbital phase curve displayed in Figure~\ref{fig:Systemic_GJ876}. There is one data point (phase\,$\sim$\,313\textdegree\ or 2011\deleted{,} July 10) that is a significant outlier to this orbital fit, however it was obtained in bad weather conditions  with a seeing above 3\textquotedbl.

\deleted{This is however expected, as the planet-planet interaction of GJ~876~b and GJ~876~c produce a libration in their orbital angles, as demonstrated by Bean \& Seifahrt (2009). Laughlin et al. (2005) measured a libration rate of $\dot{\omega} = -41$\,deg/yr for the longitude of periastron of GJ~876~b. Using this to project the measurement of Marcy et al. (1998) to 2011, which corresponds to the center of our observations, yields a predicted $\omega = 211 \pm 6$\,deg, in good agreement with our measurement of $\omega = 209 \pm 25$\,deg.}

The period, planetary mass and eccentricity are remarkably consistent with values associated to GJ~876~b in the literature (e.g., \citealp{1998ApJ...505L.147M}), except for the argument of periastron $\omega$ that is significantly different. \added{This eccentric solution is in fact a combined effect of the RV influence of the two planets GJ~876~b and GJ~876~c on the host star, rather than a physical orbit. It is therefore expected that this artificial value for $\omega$ librates with time. Furthermore, this also means that our orbital parameters can only be meaningfully compared to those of \cite{1998ApJ...505L.147M} who also fitted a single orbit to \replaced{GJ~876~b}{GJ~876~bc}. To our knowledge, our data thus provide the first multi-wavelength confirmation of the planet\added{s} \replaced{GJ~876~b}{GJ~876~bc}, thus confirming that the RV signal cannot be \added{likely }explained by stellar jitter.}

Once the periodic signal of \replaced{GJ~876~b}{GJ~876~bc} is subtracted from our data, our long-term precision does not allow us to detect any additional signal that could be associated with the other known planets orbiting GJ~876 (see Figure~\ref{fig:Systemic_GJ876}). Our analysis however demonstrates that we are able to detect planets with the characteristics of \replaced{GJ~876~b}{GJ~876~bc} using a 3\,m-class telescope and relatively inexpensive equipment.

\begin{figure*}
	\centering
	\subfigure[Orbital Fit to HD~160934~AB]{\includegraphics[width=0.495\textwidth]{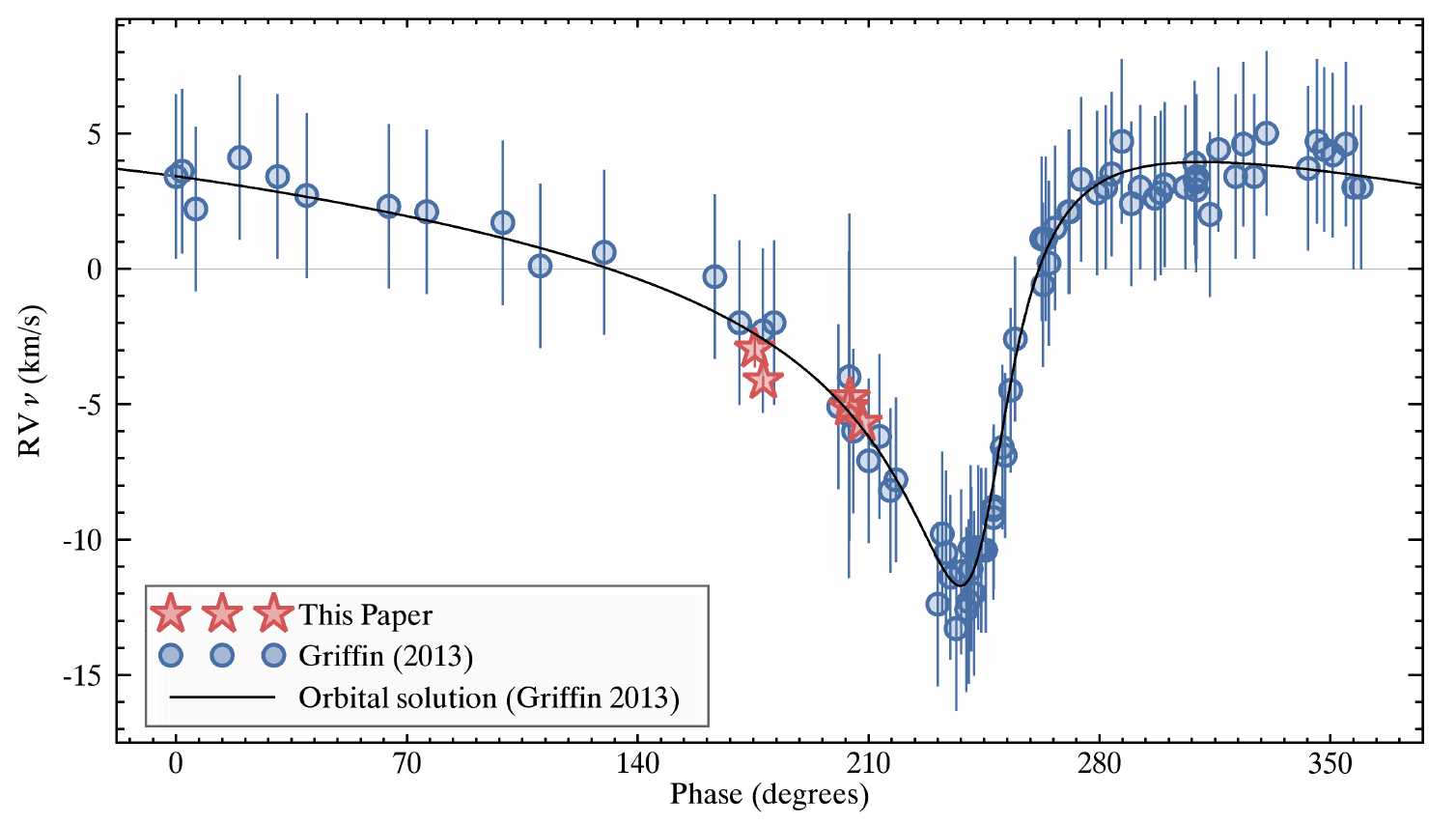}}
	\subfigure[Residuals for HD~160934~AB]{\includegraphics[width=0.495\textwidth]{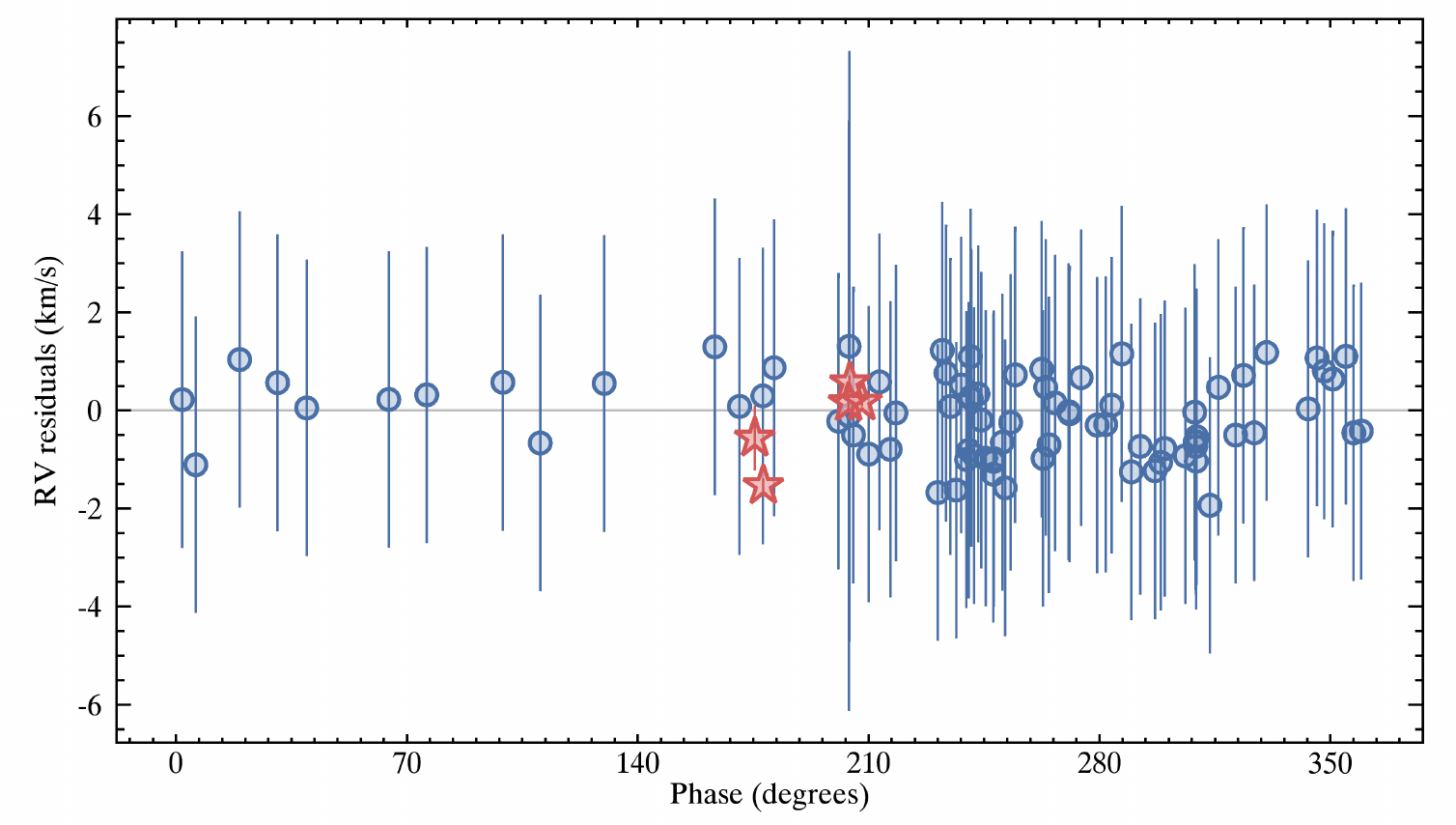}}
		\caption{Panel~a: Comparison of our HD~160934~AB RV measurements (red stars) with those in the literature (blue circles) and the reported orbital solution of the system (thick black line). Our measurements are consistent with the known orbit.\\ Panel~b: Residuals after the subtraction of the known orbit of the HD~160934~AB system. For more details, see Section~\ref{sec:rvvar}.}
	\label{fig:Systemic_HD160934}
\end{figure*}

\begin{figure*}
	\centering
	\subfigure[Orbital Fit to BD+20~1790]{\includegraphics[width=0.495\textwidth]{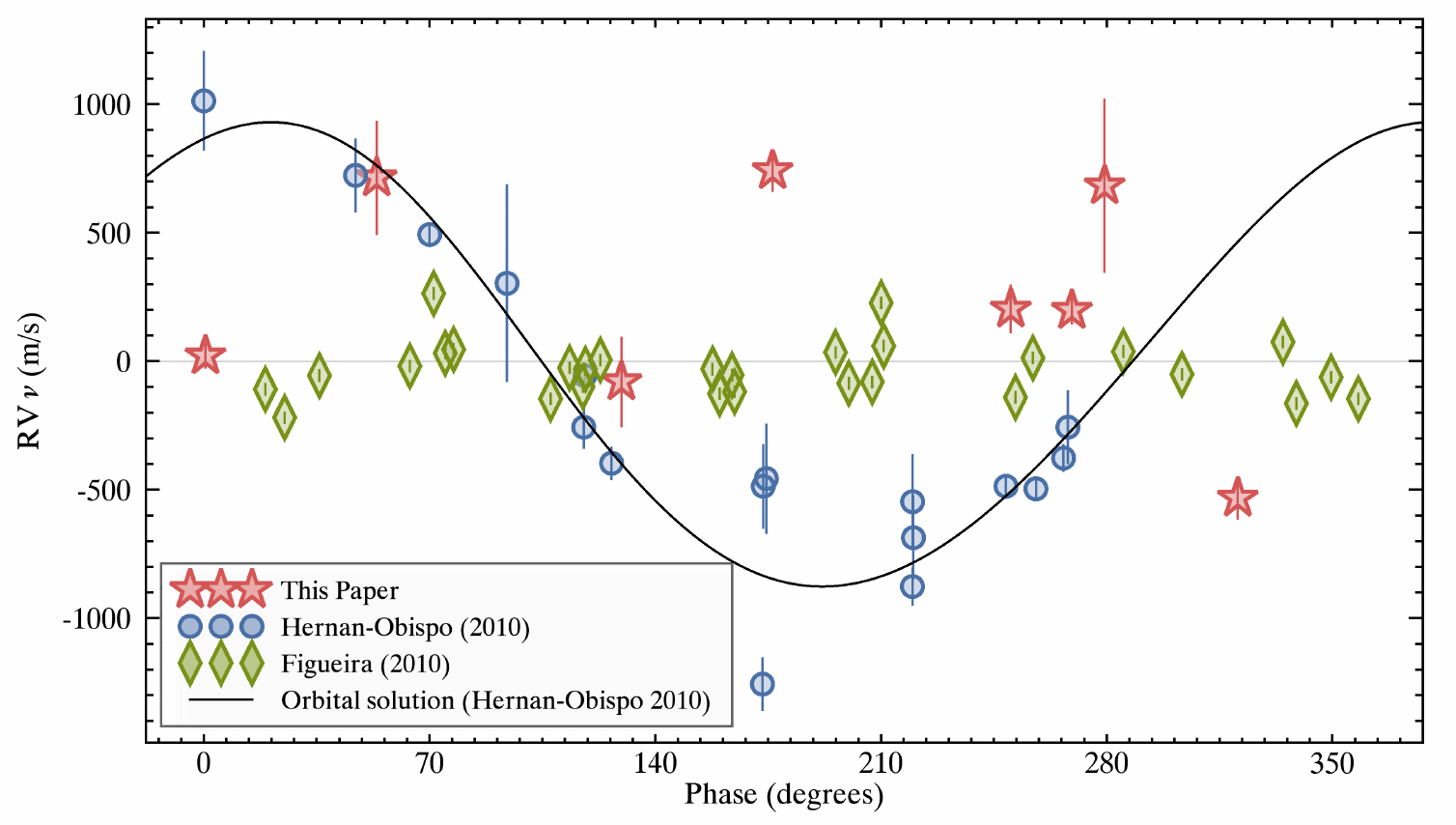}}
	\subfigure[Residuals for BD+20~1790]{\includegraphics[width=0.495\textwidth]{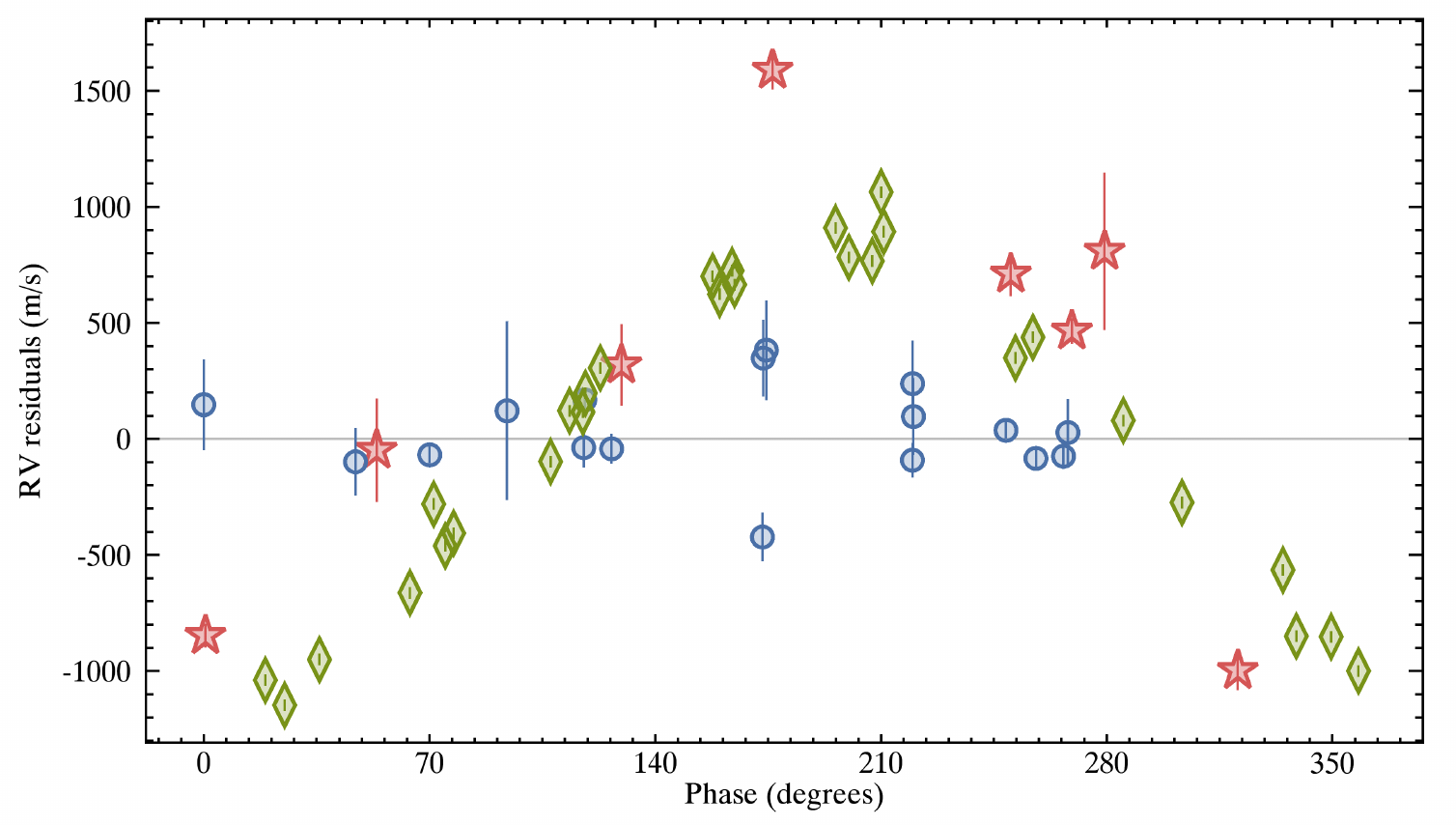}}
		\caption{Panel~a: Comparison of our RV measurements (red stars) with those reported by \citeauthor{2010AA...512A..45H} (\citeyear{2010AA...512A..45H}; blue circles) and \citeauthor{2010AA...513L...8F} (\citeyear{2010AA...513L...8F}; green diamonds), as well as the suggested orbital parameters associated with the purported planet BD+20~1790~b (see Solution~1 with $e = 0.05$ in \citealt{2010AA...512A..45H}). We find an RV scatter that is consistent with that of \cite{2010AA...512A..45H}, suggesting that there might be an RV signal not associated with stellar activity since the amplitude does not depend on wavelength, however our data does not match the orbital solution suggested by \cite{2010AA...512A..45H}, or any of those reported in the literature for the candidate BD+20~1790~b.\\ Panel~b: Residuals after the subtraction of the orbit suggested by \cite{2010AA...512A..45H}. For more details, see Section~\ref{sec:potvar}.}
	\label{fig:Systemic_BD201790}
\end{figure*}

\begin{figure*}
	\centering
	\subfigure[Orbital Fit to GJ~3305~AB]{\includegraphics[width=0.495\textwidth]{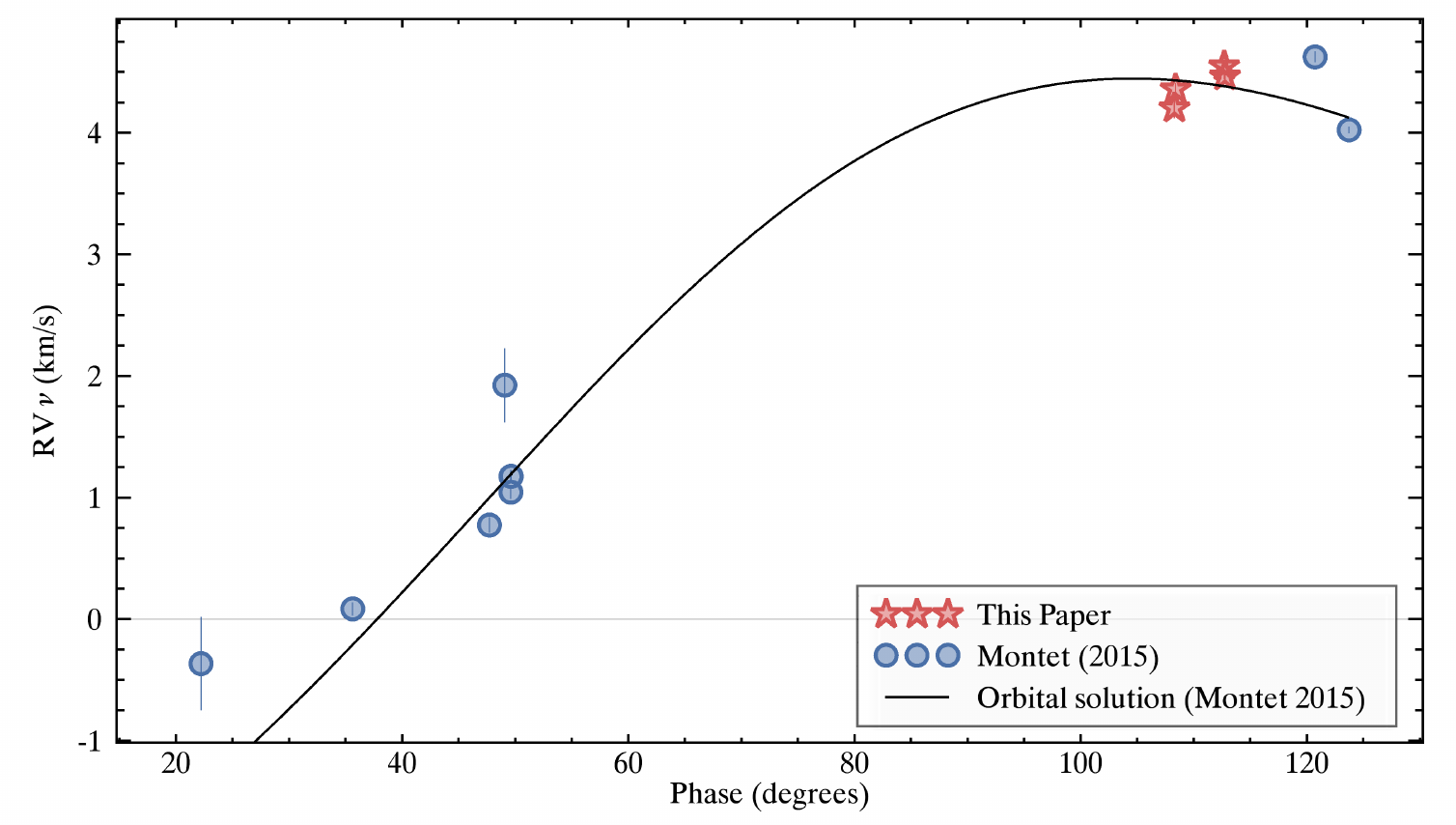}}
	\subfigure[Residuals for GJ~3305~AB]{\includegraphics[width=0.495\textwidth]{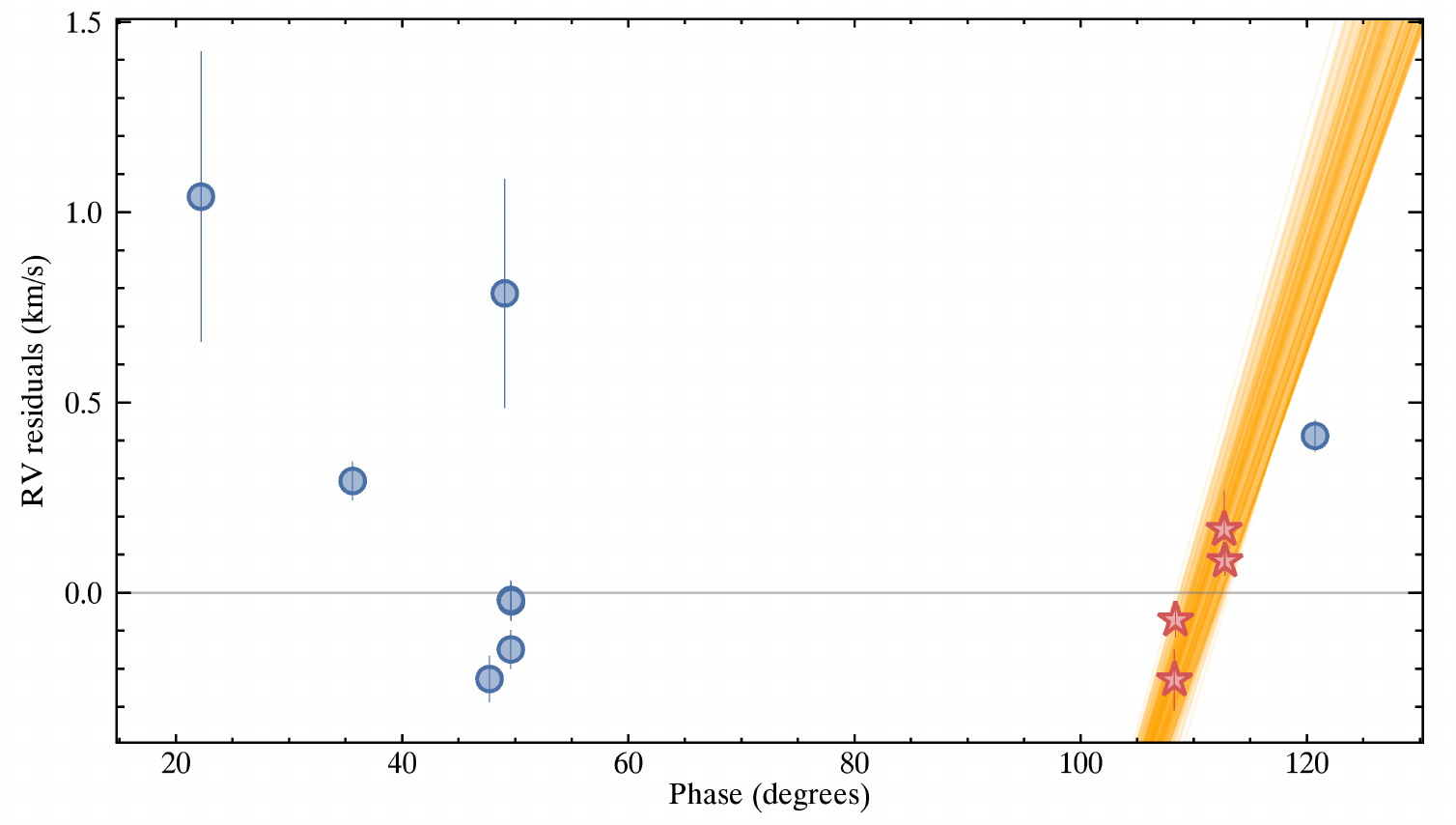}}
		\caption{Panel~a: Comparison of our RV measurements (red stars) with those reported by \cite{2015arXiv150805945M} for GJ~3305~AB (blue circles) as well as their orbital solution. \\Panel~b: Residuals after subtracting the orbital solution of \cite{2015arXiv150803084M}. The linear trend \added{(orange lines)} that we measure in our RV data cannot be explained by the GJ~3305~AB orbital solution of \cite{2015arXiv150803084M}, however it remains to be determined whether it is due to an additional companion or not. For more details, see Section~\ref{sec:lintrend}.}
	\label{fig:Systemic_GJ3305}
\end{figure*}

\textbf{HD~160934~AB} (HIP~86346) is a young and active M0-type low-mass star member of the AB~Doradus moving group \citep{2004ApJ...613L..65Z,2013ApJ...762...88M}, located at $33.1 \pm 2.2$\,pc \citep{2007AA...474..653V}. It has been confirmed as a close (0\farcs12) SB1 binary in an eccentric ($e \approx 0.8$), 17.1--year orbit both by the RV \citep{2006Ap&SS.304...59G} and direct\replaced{ }{-}imaging \citep{2007AA...463..707H} methods. Its individual components have estimated spectral types of M0 and M2--M3 \citep{2006Ap&SS.304...59G} and estimated masses of $0.69$\,\Msol\ and $0.57$\,\Msol\ \citep{2007AA...463..707H}.

\cite{2013Obs...133..322G} used \replaced{novel}{their} RV measurements as well as those reported by \cite{2010AA...514A..97L}, \cite{2010AA...521A..12M} and \cite{2002AJ....123.3356G} to derive an orbital solution for HD~160934~AB. They assumed an orbital inclination of $82.3 \pm 0.8$\,\textdegree, which was obtained from direct imaging data \citep{2007AA...463..707H,2007ApJ...670.1367L,2012ApJ...744..120E} and stellar masses of 0.65 and 0.5\,\Msol, respectively.

We followed HD~160934~AB as part of the young sample for a total of 5 nights spanning 338\,days with a typical S/N of 150 per night and recovered it as an RV variable with $V = 581 \pm 92$\,\ms. We find a strong linear trend of $1853 \pm 163$\,\msyr\ in its RV curve, however the reduced $\chi_r^2$ value remains high ($11.3$) even after subtracting a linear curve.

In Figure~\ref{fig:Systemic_HD160934}, we compare our RV measurements to those reported by \cite{2013Obs...133..322G} and we find that they are consistent with the orbital solution that they propose, however our limited time baseline only allows us to detect a linear trend in our RV data.

\textbf{LO~Peg} (HIP~106231) is yet another young, active K8 low-mass star member of the AB~Doradus moving group \citep{2004ApJ...613L..65Z,2013ApJ...762...88M}, located at $24.80 \pm 0.65$\,pc \citep{2007AA...474..653V}. Measurements of polarization suggest the possibility that a circumstellar envelope remains around LO~Peg or that significant brightness inhomogeneities exist on its surface \citep{2009MNRAS.396.1004P}. This target has the largest rotational velocity of our sample with $v\sin i = 70 \pm 10$\,\kms\ \citep{2005ESASP.560..571G} and a rotation period of 0.42\,days \citep{2010AA...520A..15M}.

We followed LO~Peg as part of the young sample for a total of 10 nights spanning 314\,days with a typical S/N\,$\sim$\,100 and identified it as an RV variable with $V = 758 \pm 120$\,\ms. The RV variability is not well fit by a linear trend and it is thus very unlikely that it can be explained by a massive stellar companion. However, it is possible that the loss of RV information due to rotational broadening of the stellar lines is the only cause of this large RV variation (see Figure~\ref{fig:vsini}). Additional RV follow-up using a larger spectral grasp might be able to \replaced{confirm this}{mitigate this effect}.

\subsection{Candidate RV Variable Targets}\label{sec:potvar}

\textbf{BD+20~1790} is a fast-rotating ($v \sin i = 16 \pm 3$\,\kms; \citealt{2007AJ....133.2524W}), active K5-type \citep{2004AJ....128..463R} member of the AB~Doradus moving group \citep{2006ApJ...643.1160L,2013ApJ...762...88M}, located at $38.8 \pm 2.0$\,pc \citep{2012ApJ...758...56S}.

\cite{2010AA...512A..45H} identified RV variability at an amplitude of $\approx$\,1.8\,\kms\ in the optical. Based on its photometric variability as well as analyses of its bi-sector and spectroscopic indices of chromospheric activity, they interpret the RV signal as the probable signature of a close-in (0.07\,AU), massive (6--7\,\MJup) planet on a 7.8--days orbit rather than the effect of chromospheric activity. They note that two solutions of different eccentricities could fit the RV data ($e = 0.05 \pm 0.02$ or $e = 0.14 \pm 0.04$).

\cite{2010AA...513L...8F} subsequently presented evidence against the interpretation of a planetary companion, by showing that the RV signal correlates with the bi-sector span of the stellar lines, and by obtaining a different RV variation amplitude of 460\,\ms\ with a periodicity of 2.8\,days that correspond to the rotation period of the star.

\cite{2015AA...576A..66H} presented a re-analysis of the RV variations of BD+20~1790 by removing the RV signal due to jitter using a Bayesian method and suggested that the RV variation is due both to stellar activity and a planetary companion. They furthermore suggest that the bisector span--RV correlation reported by \cite{2010AA...513L...8F} was due to flare events and that the correlation disappears in flare-free data. They present new orbital parameters for the candidate BD+20~1790~b that are similar to those reported by \cite{2010AA...512A..45H}, except that they find a more eccentric solution ($e \approx 0.1$ to $e \approx 0.2$).

We observed BD+20~1790 as part of the young sample for a total of 8 nights spanning $\approx$\,3.5\,years with a typical S/N\,$\approx$\,70 per night, and recovered it as a candidate RV variable, with $V = 321 \pm 71$\,\ms. Our RV curve is consistent with variations at an amplitude of $\approx$\,1\,\kms\ (see Figure~\ref{fig:BD2017}), thus providing further indication that the RV variability might not be explained by chromospheric activity alone. However, our data are inconsistent with any of the orbital solutions presented by \cite{2010AA...512A..45H} and \cite{2015AA...576A..66H} (e.g., see Figure~\ref{fig:Systemic_BD201790}).

Using any \added{combination }of our data set or those of \cite{2010AA...512A..45H} and \cite{2010AA...513L...8F}\deleted{ or any of their combinations}, we cannot identify a\added{ny} statistically significant periodicity. More follow-up observations will thus be needed to assess whether the RV variations could be caused by a companion or not. It is unlikely that the RV variability of BD+20~1790 could be explained by the loss of RV due to its fast rotation or to stellar jitter (see Figure~\ref{fig:vsini_rms}).

\textbf{EQ~Peg~A} (GJ~896~A; HIP~116132) is a young, M3.5-type \citep{2014AJ....147...20N,2015AJ....149..106D} flaring low-mass star located at $6.58 \pm 0.16$\,pc \citep{2007AA...474..653V}. \cite{2013ApJ...778....5Z} suggested that it is a member of\added{ the} Octans-Near\added{ association}.

EQ~Peg~A has a stellar companion (EQ~Peg~B) at an angular distance of $\approx$\,5\farcs5 ($\approx$\,36\,AU), which is an M4.0-type flare star \citep{2015AJ....149..106D}.

We followed EQ~Peg~A as part of the young sample for a total of 6 nights spanning $\approx$\,3.3\,years with a typical S/N\,$\approx$\,170 per night, and recovered it as a candidate RV variable with $V = 175 \pm 37$\,\ms. This RV variability cannot be explained by a long-term linear trend that could be produced by EQ~Peg~B. The loss of RV information due to rotational broadening is not important enough to explain this large RV variability, however the addition of a jitter term at the larger end of the distribution measured by \cite{2012ApJ...749...16B} could be sufficient (see Figure~\ref{fig:vsini_rms}). Additional follow-up will be required to address this.

\textbf{GJ~3942} (HIP~79126) is a nearby M0 star \citep{1956AJ.....61..201V} located at $16.93 \pm 0.30$\,pc \citep{2007AA...474..653V}. No precise RV measurements were reported in the literature as of yet.

We followed GJ~3942 as part of the nearby sample for a total of 7 nights spanning 92 days, at a typical S/N\,$\approx$\,150 per night. We identified it as a candidate RV variable with $V = 78 \pm 20$\,\ms. A high-S/N follow-up over a longer baseline will be useful to determine whether this RV signal is physical or not.

\textbf{GJ~537~B} (HIP~68588~B) is a nearby M0 star \citep{2014MNRAS.443.2561G} located at $12.3 \pm 0.9$\,pc \citep{1952gcsr.book.....J}. It is a companion to GJ~537~A at an angular separation of $\approx$\,2\farcs9.

We followed GJ~537~B as part of the nearby sample for a total of 8 nights spanning a total of 326 days with a typical S/N\,$\approx$\,180 per night. We recovered it as a candidate RV variable with $V = 71 \pm 19$\,\ms. Additional follow-up measurements will be needed to determine the nature of this likely RV variability.

\textbf{GJ~725~A} (HIP~91768) is a nearby ($3.57 \pm 0.03$\,pc; \citealt{2007AA...474..653V}), quiet and slowly-rotating M3-type star \citep{2009ApJ...704..975J}.

\cite{2002ApJS..141..503N} have shown that GJ~725~A is stable within 100\,\ms\ on a baseline of $\approx$\,3\,years and \cite{2006ApJ...649..436E} further constrained its RV stability by obtaining a scatter of only 7.4\,\ms\ in their RV measurements over a baseline of $\approx$\,7\,years.

GJ~725~A has a co-moving M3.5-type \citep{2009ApJ...704..975J} companion (GJ~725~B; HIP~91772) at an angular separation of $\approx$\,13\farcs3. \cite{2006ApJ...649..436E} report that they detect a linear RV slope of $6.99 \pm 0.86$\,\msyr\ in the GJ~725~A data over a 7.09\,year baseline, which they interpret as a small portion of its orbit around the center of mass of GJ~725~AB.

We followed GJ~725~A as part of the nearby sample for a total of 5 nights that spanned 2.8\,years with a typical S/N\,$\approx$\,170 per night and identified it as a candidate RV variable with $V = 56 \pm 15$\,\ms. Our data is be well fit by a linear trend with a slope of $35 \pm 5$\,\msyr, 

Using the projected separation of GJ~725~AB ($\approx$\,47\,AU) and assuming typical masses of $\approx$\,0.36\,\Msol\ and $\approx$\,0.3\,\Msol\ that correspond to their respective spectral types of M3 and M3.5 \citep{2005nlds.book.....R,2009ApJ...698..519K}, their orbital period should be $P \approx 400$\,years. This corresponds to a tangential velocity of $v \approx 1.75$\,\kms\ as measured from the center of mass in the case of a circular orbit. In the extreme case where the orbit is seen edge on from the Earth, we could expect a change of RV of up to $\approx$\,25\,\msyr\ per year. Our RV slope measurement is slightly larger than this, which could be an indication that the orbit of GJ~725~AB is eccentric (e.g., $e \gtrsim 0.155$ would be sufficient).

We thus measure an RV slope that is consistent with the orbit of GJ~725~AB as long as it is slightly eccentric, however in the $\approx$\,9\,years that separate our measurements from those of \cite{2006ApJ...649..436E}, it might seem surprising that the RV slope has changed by $\approx$\,28\,\msyr. Assuming that both our measurements are consistent with the orbit of GJ~725~AB indeed puts a much stronger constraint on its eccentricity, at $e \gtrsim 0.435$.

\textbf{GJ~740} (HIP~93101) is a nearby, weakly active M0.5 star \citep{2012AJ....143...93R} located at $10.91 \pm 0.18$\,pc \citep{2007AA...474..653V}. No precise RV measurements were reported in the literature as of yet.

We followed GJ~740 as part of the nearby sample for a total of 11 nights spanning 123\,days with a typical S/N\,$\approx$\,190 per night. We recovered it as a candidate RV variable with $V = 65 \pm 16$\,\ms. A high-S/N follow-up over a longer baseline will be useful to determine whether this RV signal is physical or not. We note a significant linear trend in our RV curve with a slope of $-415 \pm 44$\,\msyr, however we obtain a high reduced $\chi_r^2$ value of $\approx$\,10.2 from a linear fit, which indicates that the scatter is still relatively high even when the linear trend is subtracted. We obtain $\varsigma = 46.3$\,\ms\ and $V = 43.4 \pm 16.3$\,\ms\ ($N_\varsigma = 2.7$) after the subtraction, which would not qualify for an additional statistically significant variation under our criteria.

\textbf{GJ~9520} (HIP~75187) is a nearby M1.5-type \citep{2004AJ....128..463R} star located at $11.41 \pm 0.24$\,pc \citep{2007AA...474..653V}. No precise RV measurements were reported in the literature for this star as of yet.

We followed GJ~9520 as part of the nearby sample for a total of 7 nights spanning 314\,days with a typical S/N\,$\approx$\,170 per night. We recovered it as a candidate RV variable with $V = 77 \pm 19$\,\ms. Additional follow-up will be needed to determine whether this RV signal is physical or not.

\textbf{LHS~371} (HIP~70529) and \textbf{LHS~372} (HIP~70536) form a binary stellar system located at $16.36 \pm 0.40$\,pc \citep{2007AA...474..653V} with respective spectral types of M0 and M1 \citep{2014MNRAS.443.2561G}, and are separated by $\approx$\,45\textquotedbl. No precise RV measurements for any of the two components were reported in the literature as of yet.

We followed both LHS~371 and LHS~372 as part of the nearby sample for a total of 5 and 4 nights that span 38 and 49\,days with typical S/N precisions of $\approx$\,180 and 135 per night, respectively. Both components were identified as candidate RV variables, with $V = 48 \pm 15$\,\ms\ (LHS~371) and $V = 58 \pm 18$\,\ms\ (LHS~372). Subsequent follow-up will be needed to determine whether this RV variation is physical.

\textbf{LHS~374} (HIP~70956) is a slow rotating and \replaced{likely quiet}{chromospherically inactive}, nearby M0 star \citep{2014MNRAS.443.2561G} located at $16.99 \pm 0.42$\,pc \citep{2007AA...474..653V}. No precise RV measurements were reported in the literature for this star as of yet.

We followed LHS~374 as part of the nearby sample for a total of 5 nights spanning 43 days with a typical S/N\,$\approx$\,180 and recovered it as a candidate RV variable with $V = 76 \pm 15$\,\ms. Subsequent follow-up will be needed to determine whether this RV variation is physical.

\subsection{Likely Linear Trends in RV Curves}\label{sec:lintrend}

The presence of a massive companion at a large enough separation can induce a linear variation in our RV curves with a period that possibly exceeds our temporal baseline coverage of a given target. The criteria defined above, which are based on the scatter of RV points around the mean, will be less sensitive to detecting such variations in a given RV curve, compared with one where at least one period is sufficiently sampled. In order to identify such candidate RV variables, we have fit a linear slope to all RV curves presented in this work using the IDL routine \emph{mpfitfun.pro} written by Craig~B.~Markwardt\footnote{See \url{http://cow.physics.wisc.edu/~craigm/idl/idl.html}}. In this section, we focus on the targets for which a linear fit yielded a reduced chi-square of at most 3, corresponding to an non-null RV slope at a statistical significance of at least 3$\sigma$. These criteria have yielded 3 likely RV variable targets :

\textbf{GJ~458~A} is a nearby M0 star \citep{2004AJ....128..463R} located at $15.52 \pm 0.34$\,pc \citep{2007AA...474..653V}. It has an M3-type companion (GJ~458~B, or BD+55~1519~B; \citealt{1997AJ....113.1458H}) at an angular distance of $\approx$\,14\farcs7.

We followed GJ~458~A as part of the nearby sample for a total of 7 nights spanning 100 days with a typical S/N\,$\approx$\,180 per night. We did not recover it as a statistically significant RV variable in terms of RV scatter on our total baseline, however its RV curve is well fit by a linear trend ($\chi_r^2 = 3.0$) with a corresponding slope of $-185 \pm 50$\,\msyr\ (3.7$\sigma$ significance; see Figure~\ref{fig:458A}).

Assuming a mass of $\approx$\,0.58\,\Msol\ for GJ~458~A \citep{2014MNRAS.443.2561G} and a mass of $\approx$\,0.36\,\Msol\ for GJ~458~B that is typical of a field M3 star \citep{2005nlds.book.....R,2009ApJ...698..519K} and using the projected separation of $\approx$\,228\,AU, we \replaced{should}{would} expect a period of $\approx$\,3550\,years for the orbit of the GJ~458~AB system in a case with zero eccentricity. This would be consistent with a maximal RV slope of only $\approx$\,1.2\msyr. Only an well-aligned extremely eccentric orbit ($e \gtrsim 0.95$) could explain this, which is highly unlikely. It is thus probable that we are not measuring the effect of GJ~458~B, but rather possibly that of an unknown, massive companion. It is unlikely that this RV signal is due to stellar jitter, as this would yield a more rapidly varying random RV signal, and GJ~458~A is \replaced{a weakly active}{an inactive}, slow-rotating star \citep{2012AA...537A.147H}.

\textbf{GJ~3305~AB} is a known 0\farcs093 binary low-mass M0-type \citep{2007AA...472..321K} member of the $\beta$~Pictoris moving group located at $29.8 \pm 0.8$\,pc. It has been identified as a 66\textquotedbl\ common proper motion companion system to the F0-type star 51~Eridani \citep{2006AJ....131.1730F}, which itself has a 2\,\MJup\ planetary, T-type companion identified by the method of direct imaging \citep{2015arXiv150803084M}. This system will thus be a very important benchmark to understand stellar and planetary properties at young ages in the near future.

\cite{2015arXiv150805945M} recently led a full RV and astrometric characterization of the GJ~3305~AB pair and their orbital properties. They found a period of $29.16 \pm 0.65$\,yr, a semi-major axis of $9.8 \pm 0.15$\,AU, an eccentricity of $0.19 \pm 0.02$ and individual masses of $0.65 \pm 0.05$ and $0.44 \pm 0.05$\,\Msol\ for A and B, respectively. They compared the observed dynamical masses with evolutionary models to derive an age of $28^{+15}_{-6}$\,Myr for the system, consistent with the age of the $\beta$~Pictoris moving group ($24 \pm 3$\,Myr; \citealt{2015arXiv150805955B}). They, however, obtain a dynamical mass for GJ~3305~B that is discrepant with that of evolutionary models, which they suggest could be explained by the presence of an unresolved companion.

\cite{2012AA...539A..72D} have identified a potential 0\farcs38 companion to GJ~3305~A by direct imaging in a 2009 NACO image in the $L^\prime$ band, however further observations obtained in 2012 revealed that it was not a planetary companion, but rather a speckle or a background star\deleted{ (Delorme et al. 2012)}.

We observed the unresolved GJ~3305~AB pair during a total of 4 nights at a typical S/N$ \approx 80$ per night over a period of 5\,months as part of the young sample. The RV curve of this target is well fit by a linear trend ($\chi_r^2 = 1.8$) with a corresponding slope of $435 \pm 145$\,\msyr\ (3.0$\sigma$ significance; Figure~\ref{fig:RV_Curves_GJ3305}). 

In Figure~\ref{fig:Systemic_GJ3305}, we compare our measurements with those reported by \cite{2015arXiv150805945M} and find that our observed linear trend is inconsistent with the orbital solution that they suggest. \added{To make this more apparent in the figure, we performed a 250-steps Monte Carlo fitting of a linear polynomial relation to our four data points, accounting for their error bars, and displayed the resulting best fits on top of the RV residuals. It can be seen that they are significantly inconsistent with a zero-slope relation that would be expected in RV residuals randomly distributed around the orbital solution. Since our RV measurements are relative, we have applied an arbitrary shift to align our RV data with the known orbit. As a result, the only information contained in our data is this short-timescale slope that is inconsistent with the known orbit. We note that several other RV measurements from the literature also display significant short-timescale variations that are unexplained by the binary orbit, which could be indicative of an additional RV signal.} More follow-up data will be needed to confirm whether there is an additional RV variability to this system that is statistically significant. 

\subsection{Other Noteworthy Targets}

We describe in this section the targets that we did not select as candidate RV variables, but for which relevant information is available in the literature.


\textbf{GJ~15~A}: \cite{2014ApJ...794...51H} reported the detection of a $5.35$\,$M_\oplus$ planet in a $11.443$-day orbit around GJ~15~A. Our data lack the precision and cadence necessary to detect the planet outright. However, by assuming the planet period and ephemeris reported by \cite{2014ApJ...794...51H}, we can place a constraint on the mass of the planet with the data presented here. We used our 10 per-night RV measurements with uncertainties of $< 60$\,\ms, and phased them an $11.443$-day period and ephemeris reported by \cite{2014ApJ...794...51H}. We averaged measurements between phases of 0--$\pi$\,\rad, and subsequently between $\pi$--$2\pi$\,\rad. We then subtracted the results obtained from these two averages and converted this number into a semi-amplitude by making use of the fact that a similar operation carried out on a sinusoidal wave (i.e., subtracting its average between phases of 0--0.5\,\rad\ and that between 0.5--1\,\rad) is equal to $\sqrt{3}\,K$, where $K$ is its semi-amplitude. We used the approximation that our RV measurements are evenly distributed in phase to derive $K = 8$\,\ms. In order to quantify the uncertainty on this measurement, we computed a Monte Carlo simulation of 20 trial periods. For each trial period, we carried out the same phase-averaged measurement of the semi-amplitude, and measured a standard deviation of 12\,\ms. We thus derive a value of $K = 8 \pm 12$\,\ms\ for the RV variation semi-amplitude of GJ~15~A, which corresponds to a 3$\sigma$ upper limit of $< 36$\,\ms\ on its RV variation, or an upper limit of $< 66$\,$M_\oplus$ on the mass of its companion.

\replaced{\textbf{$\epsilon$~Eridani}}{\textbf{$\varepsilon$~Eridani}} is a young K2 star located at $3.216 \pm 0.002$\,pc \citep{2007AA...474..653V}, for which disputed planet candidates have been reported by \cite{1988ApJ...331..902C} and \cite{2002ApJ...578L.149Q}, associated with RV scatters lower than 15\,\ms. 

We followed this target as part of the young survey for a total of 13 nights spanning 339 days, with a typical S/N\,$\approx$\,300 per night. We did not identify it as an RV variable ($V = 26 \pm 15$\,\ms), however we lack the long-term precision that would be needed to determine whether the signal reported by \cite{1988ApJ...331..902C} \added{or \cite{2002ApJ...578L.149Q}} \replaced{is}{are} spurious.

%
%
%
%

\section{CONCLUSION}\label{sec:conclusion}

In this paper we report the \deleted{first }results of a precise NIR RV survey of 32 low-mass stars with spectral types K2--M4, carried out with CSHELL and an isotopologue gas cell at the NASA IRTF, 19 of which were never followed by high precision RV surveys. We used a novel data reduction and RV extraction pipeline \deleted{(see P.~Gao et al., submitted to PASP)} to demonstrate that we can achieve short-term photon-limited RV precisions of $\approx$\,8\,\ms\ with long-term stability of $\approx$\,15\,\ms, which are unprecedented using a small telescope that is easily accessible to the community.

We used the non-detections of our survey to assign upper limits on the masses of close-in companions to our targets, and we provide the first multi-wavelength confirmation of \replaced{GJ~876~b}{GJ~876~bc} and recover orbital parameters that are fully consistent with those reported in the literature. We obtained RV curves for two binary systems (HD~160934~AB, GJ~725~AB) that are consistent with the literature, and report that GJ~740 and GJ~458~A could be bound to unknown, long-period and massive companions. We identified 7 new candidate RV variables (EQ~Peg~A, GJ~3942, GJ~537~B, GJ~9520, LHS~371, LHS~372, and LHS~374) with statistical significances in the 3--5$\sigma$ range. Additional observations will be needed to verify whether these RV variable stars host substellar or planetary companions. 

Comparing our results with the projected rotational velocities of our sample, we showed that the proposed jitter relation of \cite{2012ApJ...749...16B} \added{for young TW~Hydrae members }is not large enough to account for the observed RV variations of LHS~374, BD+20~1790 and V577~Per. The probability that targets in the nearby sample display larger RV variations than those in the young sample is of 54\%; the two samples are thus not significantly different in this regard. We find that very active stars in our survey can display RV variabilities down to $\sim$\,25--50\,\ms, providing a constraint on the effect of jitter in the NIR.

In the near future, iSHELL will be mounted on the IRTF with a methane gas cell similar to that used in this work; the improved spectral grasp ($\approx$\,50 times larger), resolution ($R \approx 70\,000$) and instrumental sensitivity will achieve RV precisions of $\lesssim$\,5\,\ms\ that will allow the detection of super-Earth planets ($\gtrsim$\,13\,\MEarth) near the habitable zone of mid-M low-mass stars in the solar neighborhood. Achieving such precisions on active, very low-mass stars using optical facilities will be \deleted{very }challenging, hence NIR RV techniques will play a key role in characterizing Earth-like planets in the habitable zone of low-mass stars. These will serve as a crucial complement to transiting exoplanet studies, as the combination of both the RV and transit methods will provide a measurement of the mean planet density and put strong constraints on the physical properties of future Earth-like discoveries.

\acknowledgements

\added{We thank the anonymous referee to this paper, who provided valuable comments and suggestions that significantly improved the overall quality of this paper.} We thank \added{Jason Wright, }Julien Rameau, \'Etienne Artigau, David Montes and No\'e Aubin-Cadot for useful comments and discussions, as well as Keeyoon Sung, Sam Crawford, Brian Drouin, Edgardo Garcia-Berrios, Nathan S. Lewis and S. Lin for precious help in the construction and setup of the methane isotopologue gas cell. We thank the IRTF staff for their collaboration and help throughout this project, in particular John Rayner, Lars Bergknut, Bobby Bus and the telescope operators. This work was supported in part through an \replaced{IPAC}{Infrared Processing and Analysis Center (IPAC)} fellowship, a grant from the Fond de Recherche Qu\'eb\'ecois - Nature et Technologie and the Natural Science, a grant from the Engineering Research Council of Canada, an iREx postdoctoral Fellowship, \added{and }a JPL Research and Technology Development Grant. This work was performed in part under contract with the California Institute of Technology (Caltech)/Jet Propulsion Laboratory (JPL) funded by \replaced{NASA}{the National Aeronautics and Space Administration (NASA)} through the Sagan Fellowship Program executed by the NASA Exoplanet Science Institute. This research made use of: the SIMBAD database and VizieR catalog access tool, operated at the Centre de Donn\'ees astronomiques de Strasbourg, France \citep{2000AAS..143...23O}; data products from the Two Micron All Sky Survey (\emph{2MASS}; \citealp{2006AJ....131.1163S,2003yCat.2246....0C}), which is a joint project of the University of Massachusetts and the \replaced{Infrared Processing and Analysis Center (IPAC)}{IPAC}/Caltech, funded by \replaced{the National Aeronautics and Space Administration (NASA)}{NASA} and the National Science Foundation; the Extrasolar Planets Encyclopaedia (\url{exoplanet.eu}), which was developed and is maintained by the exoplanet TEAM; the NASA Exoplanet Archive, which is operated by Caltech, under contract with the NASA under the Exoplanet Exploration Program; the NASA/IPAC Infrared Science Archive (IRSA), which is operated by JPL, Caltech, under contract with NASA; the IRTF, which is operated by the University of Hawaii under Cooperative Agreement NNX-08AE38A with NASA, Science Mission Directorate, Planetary Astronomy Program. This publication uses observations obtained at IRTF through programs number 2010B022, 2011A083, 2011B083, 2012A065, 2012B021, 2014A048, 2014B082 and 2015B043, as well as through engineering time in the 2012A and 2012B semesters. The authors recognize and acknowledge the very significant cultural role and reverence that the summit of Mauna Kea has always had within the indigenous Hawaiian community. We are most fortunate to have the opportunity to conduct observations from this mountain.

\facility{IRTF (CSHELL)}

\bibliographystyle{apj}


\end{document}